\documentclass[11pt,a4paper]{article}

\usepackage{amssymb, amsthm, amsmath, amsfonts, eurosym, newtxmath, nccmath}

\usepackage{graphicx, subfig}

\usepackage{booktabs, dcolumn, float, adjustbox, multirow}

\usepackage{hyperref}
\usepackage[font = small]{caption}

\usepackage[textsize = small]{todonotes} 

\usepackage[normalem]{ulem}

\usepackage{lineno}

\usepackage[backend=biber, style=apa, uniquename=false, sorting=ynt]{biblatex}

\usepackage{geometry, color, setspace, pdflscape, indentfirst}
\usepackage[bottom]{footmisc}

\newcolumntype{L}[1]{>{\raggedright\let\newline\\arraybackslash\hspace{0pt}}m{#1}}
\newcolumntype{C}[1]{>{\centering\let\newline\\arraybackslash\hspace{0pt}}m{#1}}
\newcolumntype{R}[1]{>{\raggedleft\let\newline\\arraybackslash\hspace{0pt}}m{#1}}

\definecolor{linkcolour}{rgb}{0, 0.2, 0.6}
\hypersetup{colorlinks, breaklinks, urlcolor = linkcolour, linkcolor = linkcolour, citecolor = linkcolour}

\newtheorem{assumption}{Assumption}

\usepackage[title]{appendix}

\usepackage[ruled]{algorithm2e}

\renewcommand{\thetable}{\arabic{table}}
\renewcommand{\thefigure}{\arabic{figure}}
  
\DeclareMathOperator{\EX}{\mathbb{E}} 
\DeclareMathOperator{\R}{\mathbb{R}} 

\DeclareMathOperator*{\argmin}{arg\,min}

\newcommand{\open}{“} 

\DeclareFontFamily{U}{mathx}{}
\DeclareFontShape{U}{mathx}{m}{n}{<-> mathx10}{}
\DeclareSymbolFont{mathx}{U}{mathx}{m}{n}
\DeclareMathAccent{\widehat}{0}{mathx}{"70}
\DeclareMathAccent{\widecheck}{0}{mathx}{"71}

\newboolean{showcomments}
\setboolean{showcomments}{true}
\ifthenelse{\boolean{showcomments}}
{ \newcommand{\mynote}[3]{
    \fbox{\bfseries\sffamily\scriptsize#1}
    {\small$\blacktriangleright$\textsf{\textit{\color{#3}{#2}}}$\blacktriangleleft$}}}
{ \newcommand{\mynote}[3]{}}

\bibliography{references}

\begin{document}


\begin{titlepage}
    \title{Behavioral Consequences of Sexual Orientation Disclosure in a Large-Scale Digital Environment\thanks{\ A previous version of this paper circulated under the title \textit{The Cost of Coming Out}. We especially would like to thank Michael Lechner and Franco Peracchi for feedback and suggestions. We are also grateful to Jaime Arellano-Bover, Nora Bearth, Jonathan Chassot, Caroline Coly, Daniel Goller, Eric Guan, Giorgio Gulino, Moritz Janas, Giovanni Mellace, David Neumark, Paolo Pinotti, Mounu Prem, Dario Sansone, Erik-Jan Senn, Andrea Smurra, seminar participants at University of Rome Tor Vergata and SEW-HSG research seminars, and conference participants at the 2nd Rome Ph.D. in Economics and Finance Conference, the SES 2024, the 29th IPDC, and the ESPE 2025 for comments and discussions.}}
    
    \author{Enzo Brox\thanks{\ University of Bern, Center for Research in Economics of Education and University of St.Gallen, Swiss Institute for Empirical Economic Research. Electronic correspondence: enzo.brox@unibe.ch.} \and Riccardo Di Francesco\thanks{\ University of Southern Denmark, Department of Economics, Econometrics and Data Science. Electronic correspondence: rdif@sam.sdu.dk.}}
    
    \date{\today}
    
    \maketitle

    \begin{abstract}
        \noindent Many individuals hesitate to disclose their sexual orientation, anticipating that disclosure may alter how others respond to them. At the same time, concealing one’s identity can entail substantial personal and social costs. Understanding how others react to sexual orientation disclosure is therefore central to evaluating the broader consequences of coming out. This paper uses an innovative data set from a popular online video game together with a natural experiment to causally identify behavioral responses to sexual minority disclosure. We exploit exogenous variation in the identity of a playable character to identify the effects of coming out on players' revealed preferences for that character across diverse regions globally. Our findings reveal a substantial and persistent negative impact of coming out.
                
        \vspace{6pt}
        \noindent\textbf{Keywords:} LGB economics, sexual orientation disclosure, homophobia, revealed preferences, online gaming, videogames.
        
        \noindent\textbf{JEL Codes:} J15, J71, K38 \\

        \bigskip
    \end{abstract}
    
    \setcounter{page}{0}
    
    \thispagestyle{empty}
\end{titlepage}

\pagebreak \newpage

\doublespacing


\section{Introduction} 
\label{sec_introduction}


\noindent Sexual orientation is an important component of identity that shapes social and economic interactions \parencite{badgett2021lgbtq}. However, while legal protections and social acceptance of lesbian, gay, and bisexual (LGB) individuals have expanded in many countries, negative attitudes and unequal treatment persist in many contexts \parencite{badgett2020economic, badgett2023review}. As a result, many individuals anticipate adverse reactions and hesitate to disclose their sexual orientation \parencite{aksoy2023sexual}. 

Yet, concealment can be costly \parencite{akerlof2000economics}. The stress associated with hiding one’s sexual orientation constitutes a \textit{minority stressor} \parencite{meyer1995minority, meyer2003prejudice} and contributes to documented disparities in mental health outcomes between LGB and heterosexual individuals \parencite{pachankis2020sexual}. Therefore, understanding the behavioral and social responses faced by individuals upon coming out is crucial to fostering the groundwork for supportive work and living environments.

In this paper, we use an innovative data source from an online multiplayer video game together with a natural experiment to provide causal evidence on the behavioral costs of sexual orientation disclosure in an anonymous, high-frequency revealed-preference setting. Existing studies examining anti-LGB sentiments, which we review below, usually face at least one of two limitations. First, they rely on a selection-on-observables identification strategy that is likely to suffer from omitted variable bias. Second, they depend on survey data where individuals have the discretion to choose whether to disclose their sexual orientation, thereby introducing additional endogeneity issues \parencite{coffman2017size, ham2024102503}. Our use of video game data effectively avoids these selection concerns, and the natural experiment we leverage allows for a credible identification of the behavioral responses to revealing sexual minority status. 

We use a rich data set from one of the most popular online video games, \textit{League of Legends}. In this game, before a match starts, players are required to choose one playable character. Each character is characterized by game-relevant attributes and a background story that provides details about their history, origin, and relationships with other in-game characters. We leverage an unexpected change in the background story of a playable character, which discloses its sexual orientation minority status, to examine individuals' responses to sexual minority status disclosure. 

Specifically, at the beginning of the $2022$ LGBT Pride Month, the game developers announced that one of their playable characters is gay. This event introduces exogenous variation in the character's identity, providing a unique opportunity to examine reactions to sexual minority disclosure. We demonstrate that the announcement was not anticipated, thereby strengthening the credibility of our identification strategy. We then utilize detailed daily data to track players' revealed preferences for the character over a meaningful period. To isolate the effects of the disclosure on players' preferences from potential confounding influences, we employ synthetic control methods to construct a synthetic character closely resembling the pre-announcement preference history of our treated character \parencite[see, e.g.,][]{abadie2021using, abadie2022synthetic}.

Our findings reveal a substantial and persistent negative impact of coming out on players' preferences for the treated character, with a decline of more than $40\%$ of the pre-treatment average preferences for that character. This result consistently holds across various robustness checks. To further assess the external validity of our findings, we conduct a heterogeneity analysis based on players’ skill levels. The decline in preferences is equally pronounced among both lower- and higher-skilled players, indicating that the behavioral response is widespread rather than confined to a particular segment of the player base. Additionally, we exploit another unique feature of our setting: the online video game is played globally on different regional servers under very comparable circumstances. We make use of the information on the regional servers to compare how preferences for the treated character evolve across diverse regions across the world. The results consistently demonstrate a negative response across regions. 

To strengthen the interpretation of the estimated effects as responses to sexual minority disclosure, it is crucial to ensure that players' decisions to switch from the character are not influenced by other factors. We address and eliminate several alternative channels, thereby enhancing the plausibility that the observed behavior reflects a response to the disclosure itself rather than confounding gameplay or design factors. First, we rule out the possibility that shifts in characters' relative strengths could explain our estimated effect. Second, we show that players' skills have no correlation with the choice to drop the character, thus mitigating the concern that gameplay factors are the driving force behind the players' observed behavior. Additionally, we demonstrate that players are not leaving the game after the disclosure but are shifting their focus to other characters. Third, we provide evidence that switching to other characters does not affect the performance of the players involved, highlighting that the decision to abandon the character is unlikely to be driven by performance considerations. Fourth, we dismiss the possibility that the release of a new character after the disclosure explains the estimated effect.

Finally, we exploit the presence of other playable characters with sexual minority status to show that LGBT Pride Month, which started on the day of the announcement, is unlikely to explain our findings. To do so, we introduce a theoretical framework that formalizes the existence of two \open simultaneous treatments"---the disclosure of the character’s sexual orientation and the start of LGBT Pride Month.\footnote{\ See, e.g., \textcite{roller2023differences} for a discussion on \open simultaneous treatments" and methodologies for disentangling their effects under a difference-in-differences identification strategy.} We outline sufficient assumptions that enable us to separate the impacts of these treatments on players’ preferences for the character. The empirical results support the interpretation that the estimated effects are driven by the character’s disclosure.

The observed decline in character usage may reflect negative preferences or prejudice toward sexual minorities \parencite{bharadwaj2017mental, badgett2023review}, but it may also arise from reduced identification following disclosure of an identity that differs from that of many players \parencite{mcpherson2001birds}. Both mechanisms imply reduced demand conditional on disclosure, even when material payoffs remain unchanged.\footnote{\ In models of identity utility \parencite{akerlof2000economics}, individuals derive utility not only from material outcomes but also from alignment with others’ identities; disclosure may therefore reduce perceived similarity and, in turn, demand, even absent animus \parencite{oh2023does}. Relatedly, homophily models of in-group favoritism \parencite{currarini2009economic, bekes2025cultural} imply that disclosure can shift a character from an implicitly majority-aligned identity to an out-group identity, lowering utility through either out-group aversion or in-group preference.} As in much of the discrimination literature, reduced-form evidence cannot distinguish between these mechanisms. Our contribution is therefore to identify the causal behavioral response to identity disclosure, capturing economically meaningful shifts in revealed preferences.


While our setting is unconventional, it provides unique advantages for causally identifying behavioral responses to sexual minority disclosure. First, video games offer a highly controlled research environment in which behavior can be observed in real time, often in ways that are difficult to capture using traditional data sources \parencite{palacios2025beautiful}.\footnote{\ A similar argument underlies several studies that use data from the professional sports industry to examine, for example, discrimination \parencite{Price2010, pope2018awareness}, incentive effects \parencite{Pope2011}, and other behavioral patterns \parencite{Brox2025}.} Second, they allow us to leverage objective measures of behavior and identity, unlike survey-based studies, which rely on self-reported identity and are vulnerable to reporting incentives and social desirability biases \parencite{coffman2017size, ham2024102503}. Third, our setting provides data on high-frequency longitudinal revealed preferences over an economically meaningful time horizon. We can track character selection decisions at a daily frequency, enabling a granular assessment of both the immediate and persistent behavioral responses to disclosure. Such temporal resolution is rarely available in traditional social or labor market environments. Fourth, online gaming platforms offer the benefit of anonymity, minimizing social desirability bias and increasing the likelihood that observed behavior reflects individuals’ genuine choices rather than socially desirable responses.\footnote{\ Moreover, a growing body of work documents prejudice, harassment, and exclusionary behaviors in online gaming communities. Surveys and observational studies show that hostile, sexist, and sexually prejudiced behaviors are common in online multiplayer games, with negative comments about sexual orientation frequently observed \parencite{tang2020investigating, gillin2024attitudes, marques2024positive}. This evidence motivates online gaming environments as particularly salient settings for studying behavioral responses to sexual minority disclosure in a setting characterized by large-scale interaction and anonymity.} 

Despite these advantages, our study's findings should be interpreted with three important caveats. First, the sample of participants in our setting is unlikely to be representative of the general population. Our data are fully anonymized and do not contain individual demographic information. However, external evidence consistently indicates that the League of Legends player base is predominantly male (around $80\%$) and relatively young, with the majority of players between 16 and 34 years old \parencite{wang2019personality, gandhi2024beliefs}. Our estimates therefore most plausibly reflect behavioral responses within a population of young male players rather than the broader population. At the same time, this group is typically underrepresented in traditional online and telephone surveys. Observing revealed behavior in large-scale digital data thus provides a complementary perspective on social preferences within a demographic that is otherwise difficult to study using standard survey instruments. Second, our analysis focuses on responses to a male character revealing his sexual minority status. Given prior evidence of differential labor market discrimination against gay men and lesbian women \parencite{badgett2023review}, it is plausible that reactions would vary if a female character disclosed the same identity. Third, our study examines preferences for a fictitious character rather than an actual human being. As a result, factors such as the character's perceived prestige or popularity may influence individuals' responses, making our setting most analogous to the context of supporting a well-known public figure, such as a politician or professional athlete.\footnote{\ In male-dominated domains such as professional sports, public disclosure of sexual orientation remains relatively rare, consistent with the view that identity revelation in such contexts may entail substantial social or reputational costs.}

Our paper contributes to three distinct strands of the literature. First, it relates to a growing literature studying the economics of LGBTQ+ individuals \parencite[see, e.g.,][]{badgett2021lgbtq, badgett2023review}. The current body of research primarily focuses on measuring discrimination against LGB individuals by comparing their labor market outcomes with those of non-minority individuals with similar observable characteristics, mostly documenting wage penalties for gay and bisexual men and wage premiums for lesbian women \parencite[see, e.g.,][]{badgett1995wage, martell2021labor, carpenter2017does, carpenter2023orient}.\footnote{\ The only study finding a wage premium for gay men is that of \textcite{carpenter2017does}. \textcite{tampellini2024pride} does not find a wage penalty but finds a lower probability of being full-time employed and a higher probability of being a victim of work-related violence for gay men. There is also evidence that transgender workers face earning and employment penalties \parencite{geijtenbeek2018penalty, carpenter2022economic}.} However, despite their valuable contributions, these studies suffer from the endogeneity and selection issues discussed above that hinder the ability to draw causal inferences from their findings.\footnote{\ A related literature studies whether changes in laws and norms affect labor market outcomes and attitudes towards LGB individuals \parencite[see, e.g.,][]{burn2018not, sansone2019pink, ofosu2019same, burn2020relationship, delhommer2020effect, deal2022bound, deal2023heterogeneity}. \textcite{broockman2016} show that a randomized intervention that encourages actively taking the perspective of others can reduce transphobia.} To tilt towards a more causal interpretation, other studies use correspondence designs to probe into hiring discrimination against LGB individuals, consistently revealing that LGB job candidates are less likely to be invited for interviews or offered job opportunities \parencite{weichselbaumer2003sexual, drydakis2009sexual, Tilcsik2011, ahmed2013gay, drydakis2014sexual}. Nevertheless, a significant challenge lies in communicating that an individual belongs to a sexual orientation minority group, since such information is not typically included in job applications. This raises the question of whether the observed results are affected by the choice and nature of the signal used \parencite{bertrand2017field}.

We make several contributions to this literature. First, our use of data from an online video game allows us to overcome the endogeneity issues of previous studies and to identify behavioral responses to sexual minority disclosure. Second, to the best of our knowledge, our study is the first to investigate the immediate reactions to coming out. Due to the substantial anecdotal and scientific evidence that individuals often conceal their sexual orientation status \parencite[see, e.g.,][]{badgett2020economic} and the prevalent evidence of the associated costs \parencite{meyer2003prejudice, pachankis2020sexual}, several studies have explicitly focused on investigating the determinants and incentives of coming-out decisions.\footnote{\ \textcite{seror2021legalized} investigate the impact of same-sex marriage legalization on coming-out decisions using a revealed preference mechanism inferred from data on Catholic priests' vow of celibacy. \textcite{grodmadzki2026} explore spillover effects of coming-out decisions using Twitter data, discovering positive externalities, as exposure to peers coming out is associated with a higher probability of individuals coming out themselves. \textcite{aksoy2023sexual} conduct a lab experiment demonstrating that individuals strategically hide their sexual orientation in anticipation of discrimination in prosocial behavior, a result consistent with that of \textcite{kudashvili2022minorities}.} Our paper complements these studies by focusing on behavioral responses to sexual minority disclosure, thereby shedding light on how disclosure may affect subsequent social and economic interactions.

Second, our findings relate to the broader literature on identity-based behavior. Building on the seminal framework of \textcite{becker1957economics}, a large empirical literature, primarily focused on race and gender, documents differential treatment across labor and other markets \parencite[see, e.g.,][]{arnold2022measuring, kuhn2013gender}. Much of this evidence relies on audit and correspondence designs that exploit observable identity characteristics to isolate causal effects \parencite[see, e.g.,][]{ayres1995race, oreopoulos2011}. Sexual orientation, however, differs from most commonly studied traits in that it is often concealable. Individuals may anticipate discrimination and strategically withhold disclosure \parencite{aksoy2023sexual}, rendering identity endogenous and complicating standard empirical approaches. When identity is revealed, both discrimination models and homophily frameworks predict sorting and demand shifts along group lines, albeit through distinct mechanisms. Our setting leverages an exogenous disclosure shock in a context where identity would otherwise remain hidden, thereby allowing us to identify the net behavioral response to sexual orientation revelation. In doing so, we provide causal evidence on how identity disclosure reshapes demand.


Third, our study contributes to the emerging literature on social behavior in online gaming environments by providing causal evidence on how sexual orientation disclosure affects revealed behavior. Despite the economic size of the industry and the substantial fraction of time devoted to video games, the scientific literature on social interactions within these platforms is only slowly increasing.\footnote{\ \textcite{aguiar2012recent} report that, according to the American Time Use Survey, almost $10\%$ of leisure time is spent on gaming and computer use. Global video game revenues exceeded $\$180$ billion in $2020$.} Existing work documents harassment, exclusion, and prejudice in multiplayer games \parencite[see, e.g.,][]{tang2020investigating, gillin2024attitudes, marques2024positive}, but causal evidence on behavioral responses in these environments remains scarce. Our setting allows us to exploit a plausibly exogenous sexual orientation shock and trace its impact on actual choice behavior at scale, providing novel empirical evidence on how individuals adjust their behavior following sexual orientation disclosure in digital communities.

Our paper further complements the growing literature that uses video games as empirical laboratories for economic and social behavior. Early work established the credibility of gaming environments as sources of meaningful behavioral data and as part of broader patterns of social and cultural participation, showing that behavior in virtual environments reflects stable underlying preferences rather than purely recreational noise \parencite{borowiecki2018did, borowiecki2015video, borowiecki2017cultural}. Building on this foundation, recent studies exploit large-scale game data to study team performance and diversity \parencite{parshakov2018diversity}, human--AI collaboration \parencite{dell2023super}, racial discrimination \parencite{correll2002police}, consumer responses to identity labeling \parencite{parshakov2023lgbtq}, and belief-based utility theories \parencite{gandhi2024beliefs}.\footnote{\ \textcite{parshakov2023lgbtq} are most closely related to our study. However, our institutional setting, data and design allow us to causally trace behavioral responses to identity revelation over a meaningful time horizon and to rule out several alternative channels.} Our study complements this literature by providing causal evidence on behavioral responses to sexual orientation disclosure in a naturally occurring virtual environment, leveraging a large-scale natural experiment, objective measures of revealed preferences, and the high degree of anonymity characteristic of online gaming platforms.

The rest of the paper unfolds as follows. Section \ref{sec_context} describes the key elements of League of Legends that are relevant to our study and outlines the natural experiment we leverage to identify the effects of coming out. Section \ref{sec_data_and_methodology} introduces our data. Section \ref{sec_methdology_results} presents our main results. Section \ref{sec_mechanisms} examines the underlying mechanisms driving the estimated effects. Section \ref{sec_conclusion} concludes.


\section{Context}
\label{sec_context}
\noindent In this section, we explore the contextual framework that enables us to measure reactions to the disclosure of sexual orientation. Specifically, we turn our attention to the online video game \textit{League of Legends} as our data source and the natural experiment we leverage to credibly identify the causal effects of coming out.

The next subsection describes the key elements of League of Legends that are relevant to our study. Our main analysis does not rely on in-game information, but instead focuses on the pre-match phase. Therefore, we do not provide an exhaustive account of how matches unfold, but rather emphasize the details that inform our research. Then, we discuss the coming-out event we exploit and its implications for identification purposes. 

\subsection{League of Legends}
\label{subsec_league_legends}
\noindent League of Legends is a globally prominent multiplayer online game developed and published by Riot Games. Originally launched on October $27^{th}$, $2009$, the game attracted a vast player base, averaging $180$ million monthly active players as of $2022$ and peaking at $14$ million players in a single day. It has also achieved substantial financial success, generating $\$1.8$ billion in revenue in that same year.\footnote{\ See, e.g., \href{https://prioridata.com/data/league-of-legends}{https://prioridata.com/data/league-of-legends}.} 

League of Legends demands significant time and cognitive engagement from its user base. In our sample alone---which follows a subset of players over a seven-month period---users collectively spent $647{,}966$ hours in matches. Valuing this gameplay time at the 2022 U.S. federal minimum wage of $\$7.25$ per hour implies an aggregate opportunity cost of $\$4.7$ million. This figure represents a conservative lower bound of time spent on the game, as it excludes substantial time spent on pre- and post-match activities.
 
In League of Legends, players are divided into two teams of five players each to compete in matches with the aim of destroying the opposing team's base. Players in each team coordinate to sort themselves into one of five roles, typically based on pre-selected role preferences and informal agreement in the pre-game virtual lobby. These roles represent crucial strategic positions, each requiring specific playstyles and contributing differently to the team's final objective. 

Before a match begins, players must select a playable character to control during the match.\footnote{\ The roster of playable characters in League of Legends expands over time as new characters are periodically released. At the time of our analysis, players could choose from $161$ characters.} In our analysis, we measure players' revealed preferences for a specific character by quantifying how frequently they select that character for their matches. Our objective is to investigate whether these preferences undergo any shifts following the disclosure of the character's sexual orientation. Thus, we devote the rest of this section to exploring the design of characters in League of Legends and the process through which players select their characters for matches.

Each character has a unique set of skills and abilities and is specifically designed to excel in one or two of the distinct roles that players can take on within the team.\footnote{\ These roles are labeled \textit{top}, \textit{jungle}, \textit{mid}, \textit{bottom}, and \textit{support}.} Additionally, characters are crafted with a rich background that adds a narrative dimension to the game but does not have any impact on the game's mechanics. This is achieved through the creation of detailed biographies and short stories that provide players with a deeper understanding of the character's history and motivations, thus offering players the opportunity to connect with their chosen characters on a more personal level.\footnote{\ The list of all characters along with details about their abilities and histories is available at \href{https://www.leagueoflegends.com/en-gb/champions/}{https://www.leagueoflegends.com/en-gb/champions/}.}

The character selection process occurs in a virtual lobby where players can communicate with their teammates through a chat function. In a random order that alternates between teams, players take turns selecting their characters for the match. Once a player chooses a character, their selection becomes visible to all players participating in the match, including the opposing team. Each character can be chosen by only one player, making it unavailable for selection by others. Once all players have selected their characters, the match begins.

When making their character selection, players consider various factors. First, they consider the role they are assigned to fulfill in the game. Each role has its own set of responsibilities and playstyle requirements, and players aim to choose a character that aligns with their designated role. Second, players take into account their personal mastery of specific characters, opting for those they are most skilled and comfortable with. Third, players may also consider their personal preferences, such as the playstyle and background story of the character, adding a subjective element to the selection process.

\subsection{Coming-Out Event}
\label{subsec_coming_out_event}
\noindent Every year in June, \textit{LGBT Pride Month}, a dedicated time to honor and celebrate the LGBT community, takes place. Since $2018$, Riot Games has actively participated in this month-long celebration by integrating new content into League of Legends during the month of June. This includes the introduction of in-game cosmetics, such as character skins, as well as emotes that allow players to express themselves in the game. It is important to note that while these additions enhance the visual and expressive elements of the game, they do not alter the game's mechanics or the characteristics and abilities of the League of Legends characters.\footnote{\ We check this in Section \ref{subsec_graves_performance}, where we demonstrate that characters' strength was unaffected by LGBT Pride Month.}

Riot has also supported the representation of the LGBT community in the game by unveiling the sexual minority status of specific characters.\footnote{\ For instance, in $2018$, the character \textit{Neeko} was confirmed as lesbian, marking her as the first openly LGBT character in the game. Furthermore, in $2021$, the characters \textit{Diana} and \textit{Leona} were revealed to be lesbians, while the character \textit{Nami} was declared to be lesbian and polyamorous.} In our paper, we primarily focus on the character \textit{Graves}, chosen due to the existence of a well-defined announcement regarding his sexual orientation.

At the beginning of the $2022$ LGBT Pride Month, Riot Games released a short story featuring Graves and \textit{Twisted Fate}. This story officially unveils Graves' sexual orientation, establishing him as a gay character.\footnote{\ The story unveiling Graves' sexual orientation also subtly hints at Twisted Fate's pansexuality, although this is not explicitly stated. We investigate whether this implied revelation has captured the players' attention in Supplementary Information \ref{app_why_not_tf}. We highlight the relatively low attention directed towards Twisted Fate from players, who were primarily focused on Graves and the explicit establishment of his sexual orientation. As a result, we concentrate our analysis on Graves and his disclosure for a more credible identification of the effects of coming out.} The following quotes provide two pivotal passages of the narrative:\footnote{\ The short story was published on Riot Games’ official League of Legends Universe website and promoted as part of the 2022 Pride Month content campaign, making it an official element of the game’s narrative canon.} The whole story is available at \href{https://universe.leagueoflegends.com/en_SG/story/the-boys-and-bombolini/}{https://universe.leagueoflegends.com/en\_SG/story/the-boys-and-bombolini/}.

\begin{quote}
  \textit{I do not have terrible taste in men. I have good taste in terrible men.} \hfill (Graves)
\end{quote}

\begin{quote}
    \textit{[\dots] asked Fate with a tinge of poorly concealed jealousy, despite Graves having been gay for the better part of four decades.} \hfill (Storyteller)
\end{quote}

\noindent This \textit{coming-out event} closely approximates an ideal experiment where individuals randomly disclose their sexual minority status, thus providing a unique setting to identify the effects of coming out on players' preferences for Graves.\footnote{\ It is crucial to distinguish between the \textit{coming-out event} and the disclosure of Graves' sexual orientation. The coming-out event encompasses both Graves' disclosure and the start of LGBT Pride Month. While this is not a concern for identification, it requires careful interpretation of the findings. To maintain clarity, we generally refer to the effects of the coming-out event in our analysis. Further discussion on this topic is deferred to Section \ref{subsec_coming_out_vs_lgbt_month} and Supplementary Information \ref{app_anatomy_coming_out_event}.}

\begin{figure}[b!]
    \centering
    \includegraphics[scale=0.4]{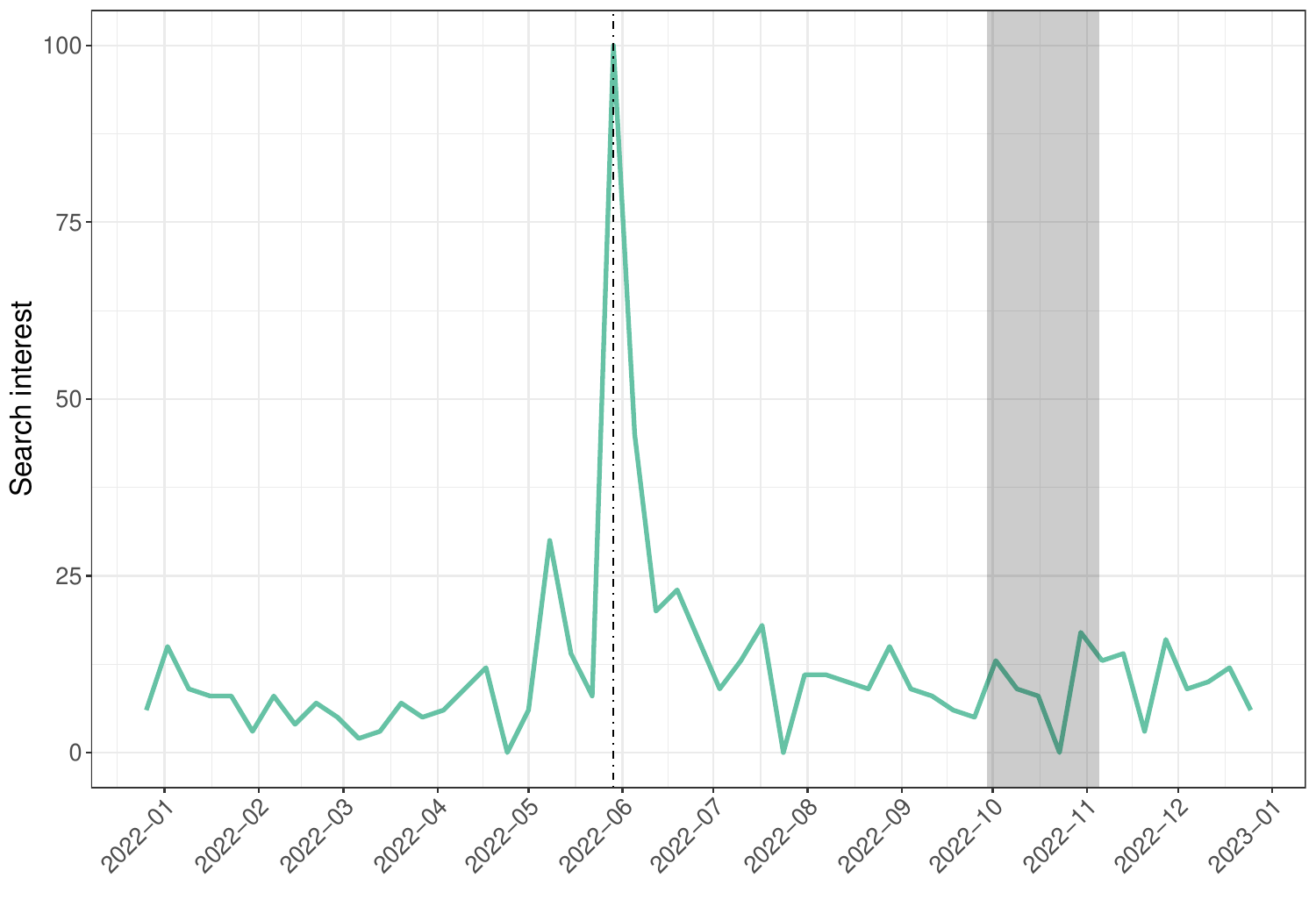}
    \caption{Google search interest over time for the query \textit{\open Graves gay."} The dashed vertical line denotes the week of disclosure, and the shaded area highlights the League of Legends World Championship.}
    \label{fig_google_interest_time_graves}
\end{figure}  

To ensure the credibility of our identification, it is crucial that the disclosure was not anticipated by players. Figure \ref{fig_google_interest_time_graves} displays the Google search interest for the query \textit{\open Graves gay."} We observe minimal interest in this search term throughout the year $2022$, with a remarkable spike occurring during the week of the coming-out event. This pattern supports our assumption of no anticipation and strengthens the credibility of our identification strategy. What is particularly interesting is that this surge in interest surpasses the level observed during the $2022$ League of Legends World Championship (held from September $29^{th}$ to November $5^{th}$), despite Graves being among the top-eight most played characters during the tournament. This finding emphasizes the substantial impact and attention that the coming-out event received from players.


\section{Data}
\label{sec_data_and_methodology}

\noindent We obtain our data by accessing the Riot Games API, which provides us with valuable information about League of Legends matches.\footnote{\ The API is publicly available and can be accessed using a standard developer key issued by Riot Games, allowing data collection without relying on web scraping.} The game operates on multiple servers located worldwide. Within each server, we focus on players in the highest tier of the ranked system. By targeting this specific group of players, we aim to minimize behavioral noise arising from infrequent or weakly engaged players who may be less exposed to in-game information and whose character choices are more idiosyncratic, thereby reducing the risk of attenuation bias.\footnote{\ Players can choose between \textit{draft} or \textit{ranked} matches. Both game modes share the same mechanics and objectives. However, while draft matches are more casual and do not have consequences for players' rankings or ratings, in ranked matches players earn or lose points based on the outcome of the match to determine their position within the ranked system. By focusing on ranked matches we aim to further reduce the risk of attenuation bias.} Furthermore, this focus increases the chances that players in our data set are aware of the coming-out event, thereby strengthening the credibility of our identification strategy.

Specifically, we identify the players within the top tier of each server as of July 2022. For each of these players, we gather raw data on the full history of matches they engaged in during the period January-July $2022$.\footnote{\ A potential concern is that our sampling frame might exclude players who exited the game immediately following the disclosure. Several features of our data mitigate this concern. First, because we collect full match histories, the data set also includes all teammates and opponents appearing in those matches---many of whom are not part of the July top-tier population. This substantially broadens the effective sample and further reduces concerns that post-treatment exit removes players from observation. Second, as shown in Figure \ref{fig_daily_players_series} in Appendix \ref{app_figures_tables}, daily active player counts exhibit no decline around the disclosure date, either for the full sample or when distinguishing between prior Graves users and non-prior users. If anything, engagement remains stable or increases during the post-treatment period, consistent with the descriptive statistics reported below.} This approach inherently includes, to a large extent, players who were not in the top tier as of July $2022$ but have been matched with our focal players during the time span we consider.\footnote{\ Our data are fully anonymized and do not contain demographic information. As a result, we cannot directly observe or report the gender, age, or other personal characteristics of the players in our estimation sample.}


To clean the raw data, we apply a multi-step procedure. First, we aggregate server-specific regions into four primary macro-regions---Latin America, North America, Europe, and Korea---and exclude servers where match coverage is sparse, namely Japan, Oceania, Russia, and Turkey.\footnote{\ The Latin America macro-region includes servers from North Latin America, South Latin America, and Brazil. Europe combines the North/East Europe and West/South Europe servers. Korea and North America are standalone servers. Further information on server structure and geographic coverage is available at \href{https://wiki.leagueoflegends.com/en-us/Servers}{https://wiki.leagueoflegends.com/en-us/Servers}.} Second, we eliminate duplicate observations, that arise when the same match appears multiple times due to its inclusion in the histories of multiple players. We also drop corrupted matches, defined as those with fewer than ten unique participants, a duration exceeding two hours, or missing data. Finally, we assign each character \open main” and \open auxiliary” roles based on historical usage patterns and exclude matches in which characters were used in atypical roles.\footnote{\ Characters in League of Legends are explicitly designed to fulfill a primary and a secondary role. We use observed historical play patterns to empirically assign each character their main and auxiliary roles. Excluding matches where characters are used outside these roles ensures that our measure of character usage reflects meaningful player choices, rather than experimental or off-meta behavior that could introduce noise into the analysis.} After these steps, our clean sample comprises $136{,}399$ unique players participating in $146{,}451$ matches.

From the clean data, we construct a character-level balanced panel that tracks the daily usage of each character across all major regions. Our final sample includes $161$ characters tracked over $194$ days. To gauge players’ revealed preferences for characters, we construct a metric called \textit{pick rate}, which measures the frequency with which players choose a specific character in their games each day in each region. Our primary objective is to investigate whether the disclosure of Graves' sexual orientation influences the pick rate of this character. We further compute daily \textit{win rates}, measuring the fraction of matches in which each character is on the winning team relative to the total number of matches that character participates in on a given day. This serves as a proxy for character's strength, allowing us to test whether shifts in character popularity are driven by changes in effectiveness rather than preferences alone.

We further construct an additional player-level daily panel that captures individual behavior over time. To ensure meaningful variation, we restrict the sample to players who participated in at least 10 matches both before and after the coming-out event, resulting in $7{,}421$ unique players. For each player, we compute the number of matches played and hours spent in-game each day, along with their daily win rates.\footnote{\ Hours spent in-game capture only the duration of matches. This substantially underestimates total time spent on the game, as it excludes time devoted to matchmaking, character selection, and other pre- and post-match activities.} We also record the frequency with which players select specific characters and roles. This data set enables us to examine whether individual behavior adjusts in response to the coming-out event.

\begin{table}[t!]
    \centering
    \caption{\label{table_summary_stats} Pre- and post-treatment summary statistics.}
    \vspace{-0.25cm}
    \footnotesize
    \begin{tabular}{lccc}
        \toprule
        & \textbf{Pre-treatment} & \textbf{Post-treatment} & \textbf{Mean difference} \\
        \midrule
        \multicolumn{4}{l}{\textbf{Graves metrics}} \\
        Graves pick rate (\%) & 18.56 & 12.32 & -6.24 \\
         & (3.81) & (3.13) & \\
        Graves win rate (\%) & 50.19 & 49.04 & -1.15 \\
         & (4.82) & (4.21) & \\

        \addlinespace
        \multicolumn{4}{l}{\textbf{Player activity}} \\
        Win rate (\%) & 51.51 & 50.47 & -1.04 \\
         & (42.08) & (40.47) & \\
        N. daily matches & 2.03 & 2.27 & +0.24 \\
         & (1.62) & (1.75) & \\
        Time spent in-game (hours) & 0.89 & 1.01 & +0.11 \\
         & (0.72) & (0.78) & \\

        \addlinespace
        \multicolumn{4}{l}{\textbf{Role selection patterns}} \\
        Top (\%) & 18.51 & 18.86 & +0.35 \\
         & (36.63) & (36.67) & \\
        Jungle (\%) & 20.93 & 20.83 & -0.11 \\
         & (38.58) & (38.09) & \\
        Mid (\%) & 19.68 & 19.18 & -0.50 \\
         & (36.89) & (36.10) & \\
        Bottom (\%) & 20.13 & 21.21 & +1.09 \\
         & (38.08) & (38.35) & \\
        Support (\%) & 20.74 & 19.91 & -0.83 \\
         & (38.84) & (37.77) & \\

        \addlinespace
        \bottomrule
    \end{tabular}

    \footnotesize
    \renewcommand{\baselineskip}{11pt}
    \textit{Notes.} Sample averages are reported for each variable, followed by standard deviations in parentheses. Graves' pick and win rates are computed from the character-level data; remaining statistics are computed from the player-level data. Time spent in-game only captures time within matches, excluding character selection and matchmaking time. The final column reports differences in means between the pre- and post-treatment periods.
\end{table}

Table \ref{table_summary_stats} presents descriptive statistics on Graves usage and player behavior before and after the coming-out event. Graves’ pick rate declines from $18.56\%$ to $12.32\%$---a drop of more than $30\%$---suggesting a substantial shift in player preference for Graves.\footnote{\ Table \ref{table_most_played_champions} in Supplementary Information \ref{app_figures_tables} shows that, conditional on Graves's primary role, it ranks as the second most frequently selected character before the disclosure. However, after the treatment, its ranking drops to fifth place.} This decline is not accompanied by a deterioration in Graves' win rate, which remains essentially flat. We further see no evidence that players disengaged from the game following Graves' disclosure. On the contrary, both the average number of daily matches and hours spent in-game slightly increase post-treatment, ruling out attrition or general disengagement.\footnote{\ Figure \ref{fig_player_activity} in Supplementary Information \ref{app_figures_tables} illustrates the distribution of players’ total and average daily activity, measured in both number of matches played and hours spent in-game.} Finally, players’ role selection patterns remain largely stable across the pre- and post-treatment periods. The lack of major shifts suggests that players who stopped using Graves likely substituted toward other characters within the same role (\textit{jungle}). We investigate the possibility of intra-role spillovers in greater detail in our robustness checks and find that our main results are unaffected.


\section{Results}
\label{sec_methdology_results}
\noindent In this section, we present our main results. First, we present our main findings. We then investigate heterogeneity across players with different levels of in-game performance. Finally, we explore the possibility of regional variations in attitudes toward the LGB community by replicating our analysis across different macro-regions. Supplementary Information \ref{app_methodology} provides a formal review of the synthetic control estimators employed to isolate the effects of the coming-out event on players’ preferences for Graves, while Supplementary Information \ref{app_robust} reports a series of robustness checks that support the reliability of our estimates.

\subsection{Main Results}
\label{subsec_main_results}
\noindent A simple comparison of Graves’ pick rates before and after the disclosure may not accurately reflect the impact of the coming-out event on players' preferences for that character, as other factors could have changed during that period. To address this issue, we construct a synthetic control unit \parencite[see, e.g.,][]{abadie2003economic, abadie2010synthetic, abadie2015comparative, abadie2021using, abadie2022synthetic} by weighting other characters to approximate the pick rates of Graves before the disclosure. This method allows us to isolate the effects of the coming-out event on players' revealed preferences for Graves and gain insight into how these preferences would have behaved in the absence of the disclosure. To mitigate the potential for spillover effects, we exclude Twisted Fate and other four characters (\textit{Diana}, \textit{Leona}, \textit{Nami}, and \textit{Neeko}) that were already members of the LGB community prior to the coming-out event from the donor pool.\footnote{\ Nevertheless, even if included in the donor pool, the estimator assigns these characters zero weight.} 

\begin{figure}[b!]
    \centering
    \includegraphics[scale=0.4]{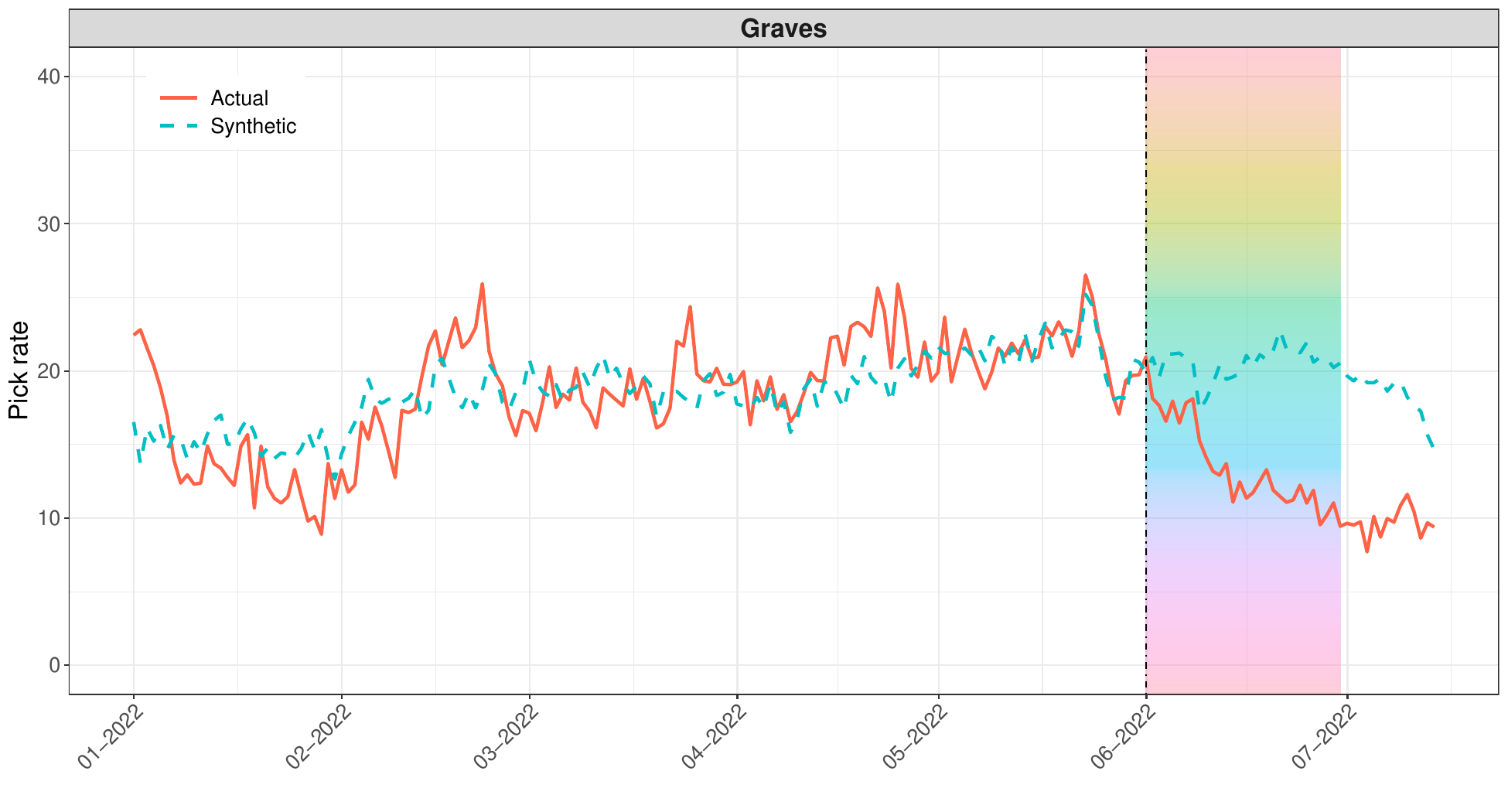}
    \caption{Graves' daily pick rates and synthetic control estimation results. The dashed vertical line denotes the day of disclosure, and the rainbow area highlights LGBT Pride Month.}
    \label{fig_graves_pick_rates_pooled}
\end{figure}

Figure \ref{fig_graves_pick_rates_pooled} displays Graves' actual and synthetic pick rate series.\footnote{\ Figure \ref{fig_graves_pick_rates_pooled_weights} in Supplementary Information \ref{app_figures_tables} displays the identities and the contributions of the characters in the donor pool with non-zero estimated weights. Figure \ref{fig_graves_pick_rates_pooled_donors} plots the corresponding pick rate series.} Overall, our analysis suggests a substantial negative impact of the coming-out event on players' preferences for Graves. Before the disclosure, the synthetic control estimator closely approximates the trajectory of Graves' pick rates, providing support for the estimator's ability to predict the counterfactual series. However, starting from June $1^{st}$, $2022$, the two series diverge substantially, with Graves' pick rates consistently dropping below those of the synthetic control. This gap persists over time, extending even beyond the conclusion of LGBT Pride Month. The estimated average effect, reported in the first column of Table \ref{table_estimation_results_main} in Supplementary Information \ref{app_robust}, is $-7.11$ percentage points and is statistically different from zero at the $5\%$ significance level.
This implies a decline of $38.32\%$ of the pre-treatment average preferences for Graves. 

We conduct a series of robustness checks, detailed in Supplementary Information \ref{app_robust}, all of which reinforce our main finding. First, our results are robust to alternative estimation strategies and donor pool compositions: we obtain consistent estimates using a regularized synthetic control estimator and when restricting the donor pool to characters unlikely to act as substitutes for Graves. This restriction directly addresses potential spillovers arising from substitution behavior: if players who stop selecting Graves mechanically substitute toward closely related characters, including those characters in the donor pool would violate a no-interference assumption by transmitting treatment effects into the synthetic counterfactual. Second, we assess predictive accuracy by artificially backdating the treatment \parencite[see, e.g.,][]{abadie2022synthetic}; discrepancies between the actual and synthetic series emerge only on the true disclosure date and closely match those in the main analysis, demonstrating the estimator’s ability to forecast the counterfactual series and supporting a no-anticipation assumption. Third, a leave-one-out analysis \parencite[see, e.g.,][]{abadie2021using} confirms that no single donor character drives the results. Fourth, we implement a within-role placebo design by replicating the synthetic control analysis for the four most frequently selected jungle characters (excluding Graves) in the pre-treatment period. None of these placebo units exhibits a comparable and sustained post-disclosure divergence from its synthetic counterpart, and we do not observe a synchronized decline in jungle pick rates around June 1, 2022. Overall, these patterns indicate that the estimated effect for Graves is not driven by broader shocks to role preferences, but is instead character-specific.

\subsection{Player Skills Heterogeneity}
\label{subsec_player_heterogeneity}
\noindent To examine whether responses to the coming-out event differ by player ability, we conduct a heterogeneity analysis based on a proxy for player skill derived from pre-treatment in-game performance. Specifically, we compute each player’s win rate prior to the treatment date, restricting the sample to the same $7{,}421$ users selected for our player-level data set to ensure the reliability of this metric. We then classify these players into above- and below-median skill groups based on this metric. Using this stratification, we replicate our main synthetic control analysis separately for each group. The estimation proceeds identically to our baseline specification of the previous subsection.

This stratification allows us to investigate whether the decline in Graves’ pick rate is concentrated among particular subpopulations---such as relatively lower-performing players within the top-tier ranked population---or whether it reflects a broader behavioral response. Consistent estimates across skill groups would strengthen the external validity of our findings, mitigating concerns that the observed response is driven by a narrow or non-representative segment of the player base.

Figure~\ref{fig_player_skills_median} in Supplementary Information \ref{app_figures_tables} plots the distribution of overall win rates and total matches for players in each skill group. Although the win rate distributions overlap substantially, they differ in their means, with players classified as higher-skilled exhibiting slightly higher win rates on average. In contrast, the distribution of total matches is nearly identical across groups, indicating that conditioning on pre-treatment win rate does not mechanically segment players by engagement level. This is important for interpretation: it implies that our stratification holds constant (approximately) overall game activity while isolating variation in a meaningful proxy for player quality.

\begin{figure}[b!]
    \centering
    \begin{minipage}[t]{0.48\textwidth}
        \centering
        \includegraphics[width=\textwidth]{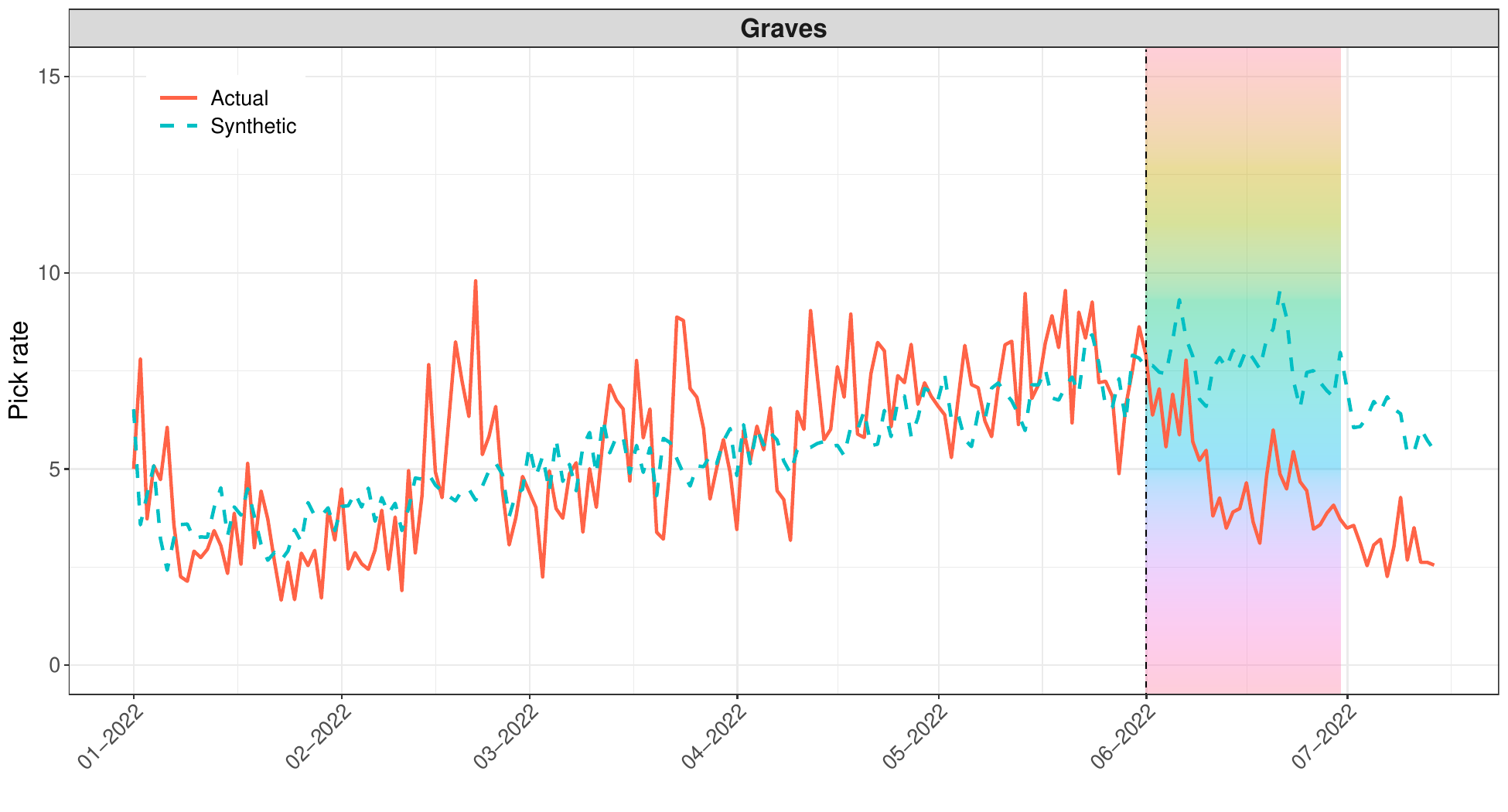}
        \caption*{\footnotesize \textit{(a)} Players below median skill}
    \end{minipage}%
    \hfill
    \begin{minipage}[t]{0.48\textwidth}
        \centering
        \includegraphics[width=\textwidth]{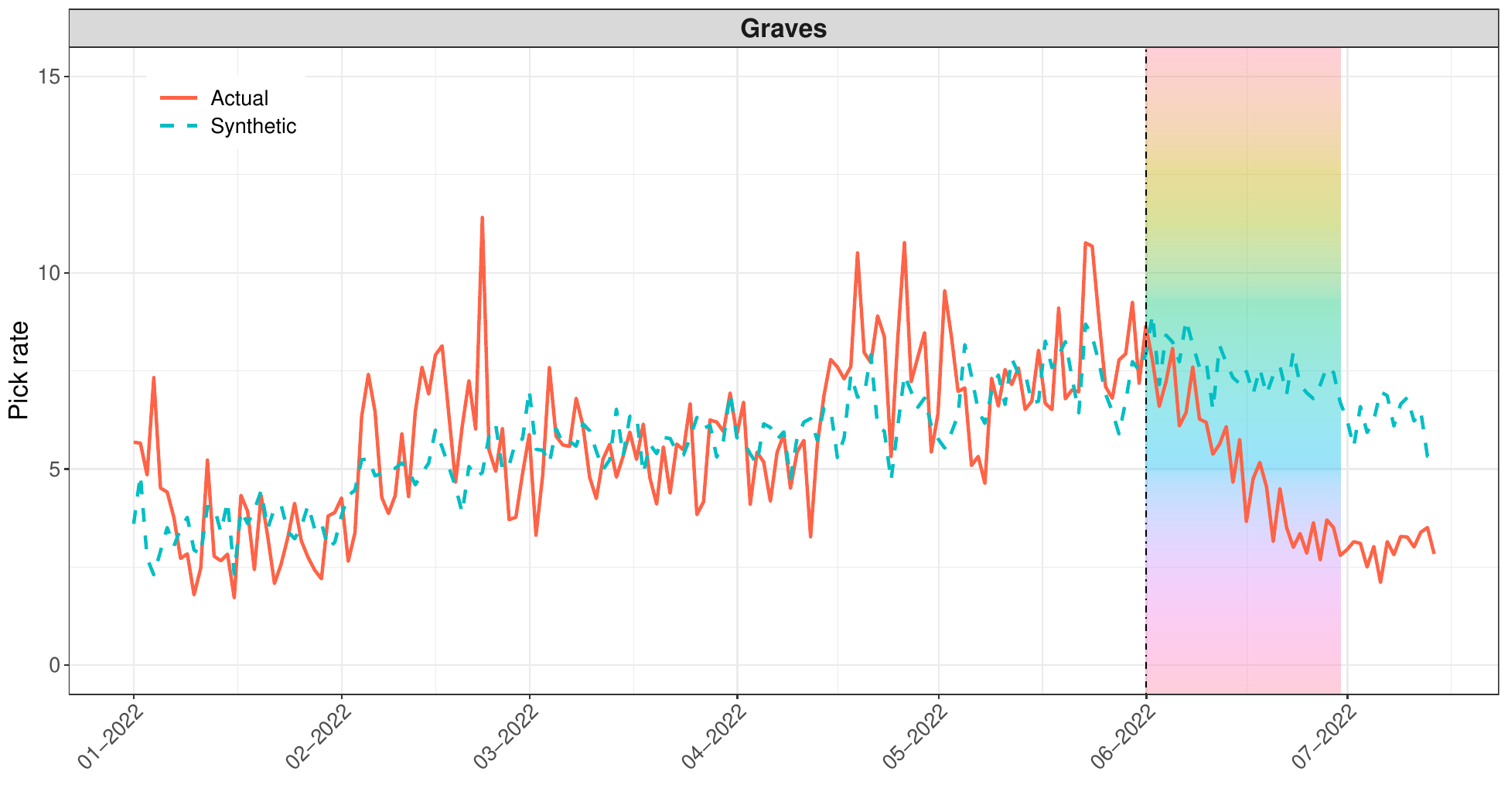}
        \caption*{\footnotesize \textit{(b)} Players above median skill}
    \end{minipage}

    \caption{Graves' daily pick rates and synthetic control estimation results by skill group. The dashed vertical line denotes the day of disclosure, and the rainbow area highlights LGBT Pride Month.}
    \label{figure_sc_result_player_heterogeneity}
\end{figure}

Figure \ref{figure_sc_result_player_heterogeneity} presents the results. We observe a sustained decline in Graves’ pick rate immediately following the coming-out event in both the below- and above-median skill groups. The timing and persistence of the divergence from the synthetic counterfactual are remarkably similar across the two groups and closely mirror the patterns documented in our pooled specification.\footnote{\ Supplementary Information \ref{app_robust} shows that these findings are robust to backdating and leave-one-out exercises.} These findings indicate that the estimated effect is not concentrated among players of a particular ability level but instead reflects a more widespread behavioral response within the top-tier ranked population.\footnote{\ While qualitatively similar, the pre-treatment pick rates of Graves are lower in both skill groups relative to the main specification. This difference arises from the additional sample restriction imposed here, whereby we include only players who participated in at least 10 matches both before and after the coming-out event to ensure a meaningful proxy for skill. The main specification, by contrast, includes the full player base, potentially encompassing more casual users with higher Graves usage. As such, differences in baseline levels should be interpreted as compositional.}

\subsection{Regional Heterogeneity}
\label{subsec_regional_heterogeneity}
\noindent Previous research has documented substantial cross-country variation in social norms, legal environments, and the prevalence of individuals openly identifying as LGB \parencite[see, e.g.,][]{badgett2020economic, badgett2021lgbtq}.\footnote{\ An OECD report shows that even within a set of relatively comparable OECD countries, the size of the LGB communities differs by a factor of four \parencite{oecd2019}.} To explore regional heterogeneity in behavioral responses to sexual minority disclosure, we exploit the geographic information associated with each match. Specifically, we divide the sample according to the macro-regions defined in Section~\ref{sec_data_and_methodology} and apply our synthetic control methodologies separately within each region.

\begin{figure}[b!]
    \centering
    \includegraphics[scale=0.45]{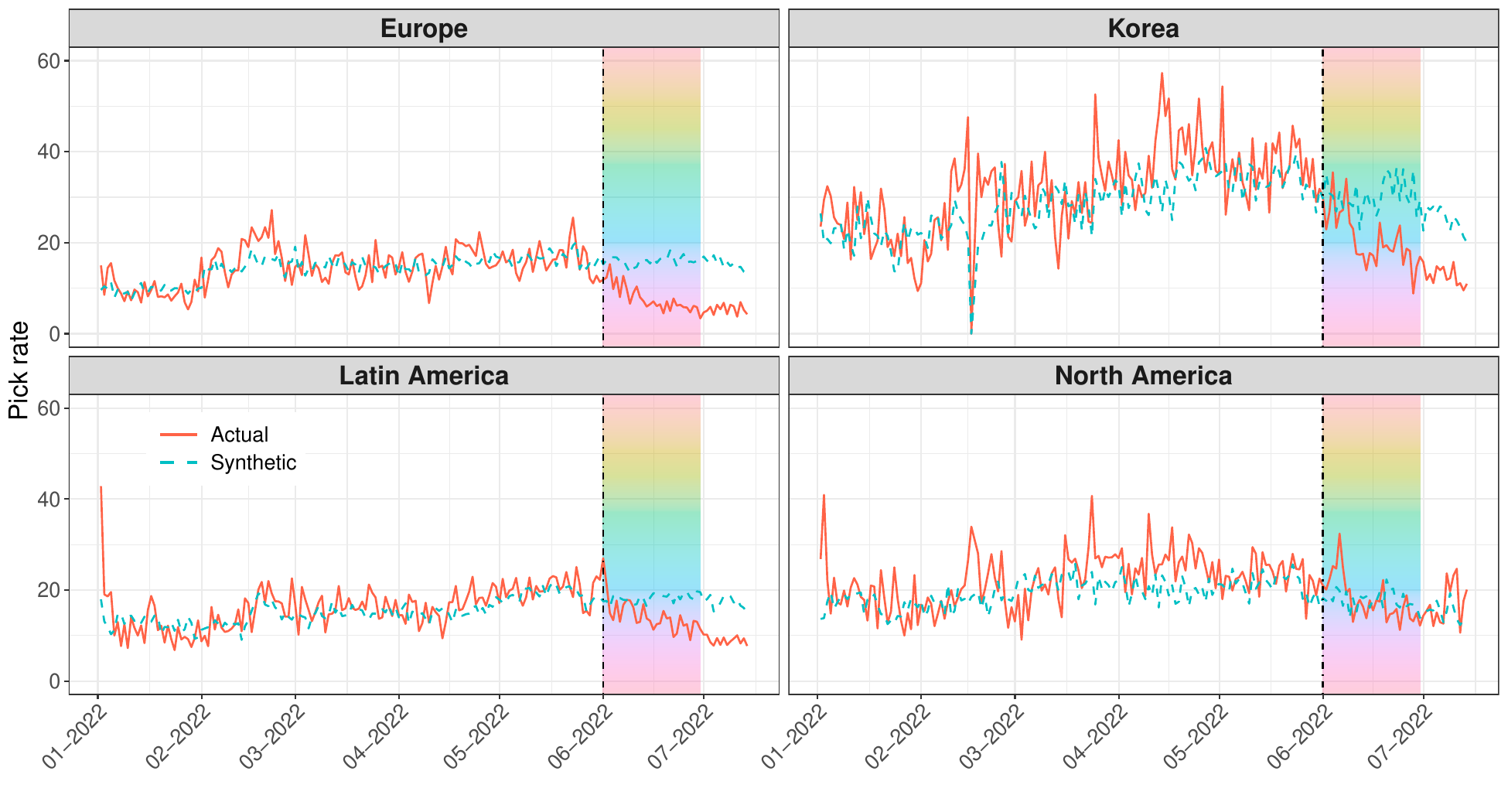}
    \caption{Graves' daily pick rates and synthetic control estimation results by region. The dashed vertical line denotes the day of disclosure, and the rainbow area highlights LGBT Pride Month. Figure \ref{fig_graves_pick_rates_regional_weights} in Supplementary Information \ref{app_figures_tables} displays the identities and the contributions of the characters in the donor pool with non-zero estimated weights.}
    \label{fig_graves_pick_rates_regional}
\end{figure}

Figure \ref{fig_graves_pick_rates_regional} displays the results. The synthetic control estimator closely approximates the trajectory of Graves' pick rates for matches in Europe and Latin America before the disclosure, exhibiting pre-treatment root mean squared errors comparable to those of the pooled specifications. However, discrepancies arise in Korean and North American matches, where the pre-treatment root mean squared error is two to three times higher than that achieved with European and Latin American matches. In Europe, Korea, and Latin America, we estimate negative and persistent effects of the coming-out event on players' preferences for Graves. Point estimates for the average treatment effect in these regions, reported in Table \ref{table_estimation_results_regional} in Supplementary Information \ref{app_figures_tables}, are consistently negative and are statistically significant at the 5\% level in our main specification. In North America, where the pre-treatment root mean squared is relatively large, we are unable to determine the sign of the impact, as the confidence intervals for the estimated average effect consistently encompass zero.

The estimated average effect varies substantially across regions. In our main specification, Europe exhibits the largest relative decline (-57.80\% relative to pre-treatment average preferences), while Korea and Latin America show smaller declines of roughly -32\%, indicating heterogeneous behavioral responses to the disclosure event. However, differences in magnitude should not be interpreted as direct measures of underlying social attitudes. Servers may differ along several dimensions that affect character selection independently of preferences toward sexual minorities. For example, regional differences in competitiveness could influence the degree of subjectivity in character choice. In addition, information about the disclosure may have diffused unevenly across regions.\footnote{\ Another factor may relate to how information about the disclosure diffused across regions. Although the announcement was officially localized into multiple in-game languages, players may also have learned about it through external channels such as social media, online forums, or streaming platforms, where exposure and salience can differ across linguistic and regional communities.} We therefore carefully interpret these estimates as evidence of negative behavioral responses in several regions, rather than as a basis for comparing the intensity of underlying attitudes across regions.


\section{Mechanisms}
\label{sec_mechanisms}
\noindent In Section \ref{sec_methdology_results}, we established evidence of a substantial negative impact of the coming-out event on players' revealed preferences for Graves. However, the players' decision to switch from this character may reflect factors other than a response to the disclosure itself. This section discusses why such alternative explanations are unlikely, thereby strengthening the interpretation that the observed behavior reflects a response to the disclosure rather than confounding gameplay or design factors. Methodological details and supporting analyses are reported in Supplementary Information \ref{app_mechanisms}.

\subsection{Graves' Strength}
\label{subsec_graves_performance}
\noindent Essential for interpreting the estimated effects as responses to the disclosure is the fact that Graves' strength remained unaffected by the coming-out event, as any change in character relative strengths could explain why players' preferences shift away from Graves. To address this concern, we employ our synthetic control methodologies to examine the potential impact of the coming-out event on Graves’ strength. We measure characters’ strength using daily win rates. 

As detailed in Supplementary Information \ref{subapp_graves_performance}, our analysis reveals that the coming-out event had no impact on Graves' strength, thus dismissing the possibility of a shift in his strength as an explanation for the results of Section \ref{sec_methdology_results}. Moreover, players have real-time access to detailed statistics on each character’s performance from widely used online platforms, meaning they would have been aware that no abilities or attributes had changed during the period under study. These factors suggest that the negative impact of the coming-out event estimated in Section \ref{sec_methdology_results} is unlikely to be driven by actual or presumed changes in character relative strengths.

\subsection{Engagement and Performance Across Usage Groups}
\label{subsec_players_skills}
If the observed decline in Graves’ usage were driven by gameplay considerations rather than the disclosure itself, we would expect systematic differences in engagement or performance across players with different pre-disclosure usage patterns.

To address this concern, we compare engagement and performance outcomes across players with different pre-treatment preferences for Graves. As detailed in Supplementary Information \ref{subapp_players_skills}, we classify players into \textit{prior users}---those who selected Graves in at least $5\%$ of their pre-disclosure matches---and \textit{non-prior users}, and compare engagement levels and performance before and after the coming-out event. Both groups play a similar number of daily matches, and neither shows signs of disengagement following the disclosure; if anything, match volumes increase slightly, suggesting players shifted to other characters rather than leaving the game. We also find no meaningful differences in win rates between or within groups, indicating that differences in pre-disclosure
usage of Graves are not associated with differential performance or skill.

Overall, our findings provide no evidence that the post-disclosure decline in Graves’ usage is driven by gameplay considerations, thereby strengthening the interpretation that the observed behavior reflects a response to the disclosure rather than changes in in-game incentives.

\subsection{Prior Usage and Performance}
\label{subsec_players_performance}
\noindent To ensure that our estimates capture behavioral responses to the disclosure rather than strategic considerations, it is crucial to assess whether shifting away from Graves to other characters impacts players' performance. If there are performance costs, our estimates could be biased toward zero, as players might continue using Graves for strategic considerations. Moreover, if players switch characters primarily for convenience or strategic reasons, our analysis might capture a different phenomenon unrelated to the disclosure.

To investigate this possibility, we analyse whether players’ performance changed after moving away from Graves. As detailed in Supplementary Information \ref{subapp_players_performance}, we focus on prior Graves users and classify them into \open treated" and \open control" groups depending on the magnitude of their post-disclosure reduction in pick rate. We then employ difference-in-differences identification and estimation strategies \parencite{callaway2021difference} to compare the daily win rates of treated and control players before and after the disclosure. 

Overall, we find that shifting away from Graves to other characters has no impact on players' performance. None of the estimated effects is statistically different from zero, indicating that transitioning to other characters entails no performance-related costs. This reinforces the view that the decision to abandon Graves is unlikely to be driven by strategic or gameplay considerations. 

\subsection{New Substitute Character}
\label{subsec_belveth}
\noindent On June $9^{th}$, $2022$, a new character (\textit{Bel'Veth}) was released.\footnote{\ Figure \ref{fig_belveth_pick_rates} in Supplementary Information \ref{app_figures_tables} displays Bel'Veth's daily pick rates.} Because Bel'Veth's primary role overlaps with that of Graves, the new character can be considered a close substitute. Therefore, part of the post-June~$9^{th}$ decline in Graves’ usage could, in principle, reflect players experimenting with the new character or seeking potential competitive advantages, rather than a response to his disclosure, thereby challenging our interpretation of the estimated effects.

If this substitution channel were the primary driver of the observed drop, we would expect a positive correlation between the magnitude of a player’s reduction in Graves usage and their subsequent adoption of Bel’Veth. However, the analysis in Supplementary Information~\ref{subapp_belveth} shows that this is not the case. Players with substantial post-disclosure reductions in Graves usage are less likely to select Bel’Veth than those with moderate reductions, and even less likely than those who made no change at all. This pattern suggests that the release of Bel’Veth cannot account for the behavioral shift documented in Section~\ref{sec_methdology_results}.


\subsection{Coming Out versus LGBT Pride Month}
\label{subsec_coming_out_vs_lgbt_month}
\noindent As described in Section \ref{subsec_coming_out_event}, the disclosure of Graves' sexual orientation coincided with the start of LGBT Pride Month. This means that the coming-out event encompasses two \open simultaneous treatments" \parencite[see, e.g.,][]{roller2023differences}, namely the announcement of Graves' homosexuality and the introduction of visual and expressive elements in League of Legends that support the LGBT community. It is therefore plausible that the findings presented in Section \ref{sec_methdology_results} may, to some extent, be influenced by the presence of LGBT Pride Month, which might elicit negative reactions from certain players, leading them to shift their preferences away from LGB characters. While this alternative perspective does not undermine the validity of our identification strategy, it does raise questions about our interpretation of the estimated effects as solely stemming from Graves' disclosure.

Supplementary Information \ref{app_anatomy_coming_out_event} introduces a theoretical framework that formalizes the existence of two simultaneous treatments, discusses the interpretation challenges it poses, and outlines sufficient conditions for separating the effects of the coming out from those of Pride Month. Here, we provide the main intuitions behind our approach, directing the reader to the supplementary information for technical details.

To examine the potential impact of LGBT Pride Month on players' preferences for LGB characters, we leverage the existence in our data set of other four characters already acknowledged as part of the LGB community before the coming-out event. These characters are subject only to a part of our treatment, specifically being part of the LGB community while LGBT Pride Month is ongoing, whereas Graves experiences both the disclosure of his sexual orientation and LGBT Pride Month.  We create a composite LGB unit by averaging their daily pick rates and apply our synthetic control methodologies to measure the effect of Pride Month on players’ preferences for LGB characters. Under the assumption that Pride Month’s influence is homogeneous across all LGB characters, the difference between the Graves and composite LGB effects identifies the impact of his disclosure. Intuitively, if the estimated impact of LGBT Pride Month on players' preferences for LGB characters is small relative to the estimated impact of the coming-out event on players' preferences for Graves, this suggests that the findings of Section \ref{sec_methdology_results} must be primarily attributed to Graves' disclosure.

The results, reported in Supplementary Information~\ref{subapp_coming_out_vs_lgbt_month}, show that Pride Month had no measurable effect on players’ preferences for LGB characters. Under the homogeneity assumption discussed above, these findings support the interpretation that the estimated effects presented in Section \ref{sec_methdology_results} are primarily driven by Graves' disclosure rather than being influenced by the broader context of LGBT Pride Month.


\section{Conclusion}
\label{sec_conclusion}

\noindent Disclosure of sexual orientation can have important social and economic implications. Understanding how individuals respond to such disclosure is therefore relevant for evaluating the broader consequences of coming out. Prior research documents that differential treatment and social exclusion of LGB individuals in educational, health, social, and political settings can lead to losses in human capital with detrimental economic effects \parencite{badgett2020economic}, as well as severe long-term health consequences \parencite{boden2016bullying}. Examining behavioral responses to sexual orientation disclosure contributes to this broader discussion by providing evidence on how disclosure may shape subsequent interactions.


In this study, we utilize a comprehensive data set sourced from the widely popular online video game \textit{League of Legends} and exploit exogenous variation in the identity of a playable character to credibly identify behavioral responses to sexual minority disclosure. Players in the game select a playable character before each match, and each character is characterized by game-relevant attributes and a background story. Leveraging an unexpected revelation during the 2022 LGBT Pride Month---wherein game developers disclosed the sexual orientation of a playable character---we examine how players adjust their revealed preferences following this disclosure. By tracking character selection over a meaningful period, we document a substantial and persistent decline in demand, with preferences decreasing by around $38\%$. These findings highlight that sexual orientation disclosure can meaningfully affect subsequent behavior in interactive environments and contribute to understanding how disclosure may shape social and economic interactions.

To strengthen the interpretation of the estimated effects as responses to the disclosure, we address and eliminate several alternative channels. First, we rule out the possibility that shifts in characters' relative strengths could explain our estimated effect. Second, we show that players' skills have no correlation with the choice to drop the character. Third, we provide evidence that switching to other characters does not affect the performance of the players involved and that the release of new characters in the post-treatment period is unlikely to serve as primary explanation behind the results. Fourth, we introduce a theoretical framework that formalizes the existence of two \open simultaneous treatments" and use information on other playable characters that belong to the LGB group to rule out that LGBT Pride Month serves as the main explanation for the observed result.

{While we do not distinguish between out-group animus or reduced identification and in-group preference, both mechanisms imply economically meaningful behavioral costs of identity disclosure. More broadly, the results highlight that identity revelation can shift demand in high-visibility environments in ways that persist beyond short-lived attention shocks, even absent material payoff changes. Our findings have broad implications for understanding how disclosure of sexual orientation can shape subsequent interactions and behavior. Raising awareness about behavioral responses to sexual orientation disclosure may contribute to designing interventions that promote inclusive and supportive communities \parencite{pope2018awareness}.

Beyond societal implications, our findings are relevant for platform and content design. Then decline in Graves’ usage following the disclosure suggests that inclusive representation can generate immediate behavioral responses among some players. However, our design does not allow us to assess longer-run outcomes such as engagement, retention, brand value, or community norms, which are likely central to platform strategy.\footnote{\ Riot Games has continued to release official Pride Month content in 2023, 2024, and 2025, maintaining a public commitment to inclusive representation.} From a welfare perspective, simmediate demand responses must be weighed against potential longer-run benefits of fostering inclusive environments and attracting more diverse user populations. Understanding how identity-based representation affects platform growth and long-term engagement remains an important direction for future research.


\newpage
\begin{appendices}

\renewcommand{\appendixname}{Supplementary Information}
\renewcommand{\appendixtocname}{Supplementary Information}
\renewcommand{\appendixpagename}{Supplementary Information}

\renewcommand\theequation{\Alph{section}.\arabic{equation}}
\renewcommand\thetable{\Alph{section}.\Roman{table}}
\renewcommand\thefigure{\Alph{section}.\Roman{figure}}

\doublespacing

\section{Additional Figures and Tables}
\label{app_figures_tables}

\setcounter{equation}{0}
\setcounter{table}{0}
\setcounter{figure}{0}

\begin{figure}[H]
    \centering
    \includegraphics[width=\textwidth]{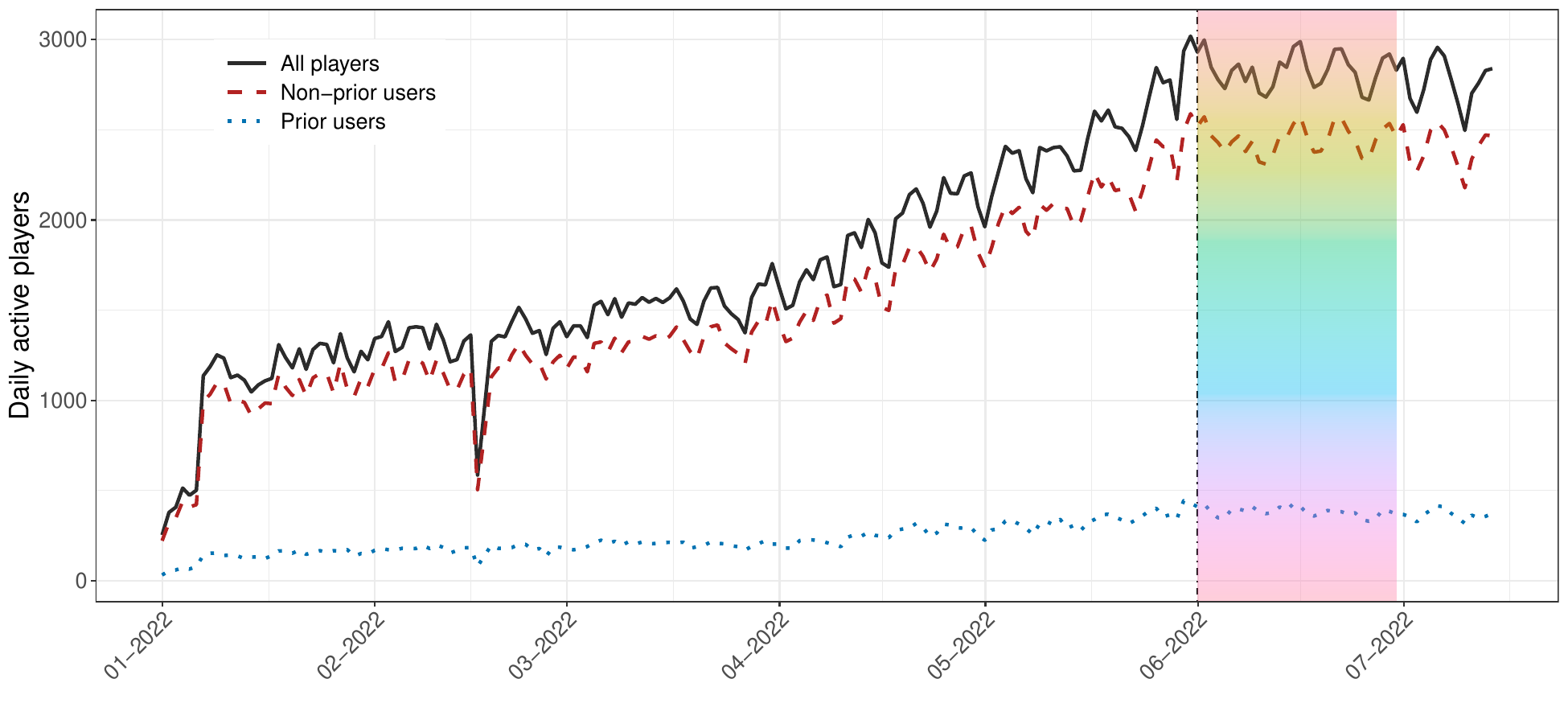}
    \caption{Daily number of active players. The solid black line shows the total number of active players, while the dashed red and dotted blue lines report activity separately for players with high and low pre-treatment Graves usage, respectively. The dashed vertical line denotes the day of disclosure, and the rainbow area highlights LGBT Pride Month.}
    \label{fig_daily_players_series}
\end{figure}

\begin{figure}[H]
    \centering
    \includegraphics[scale=0.5]{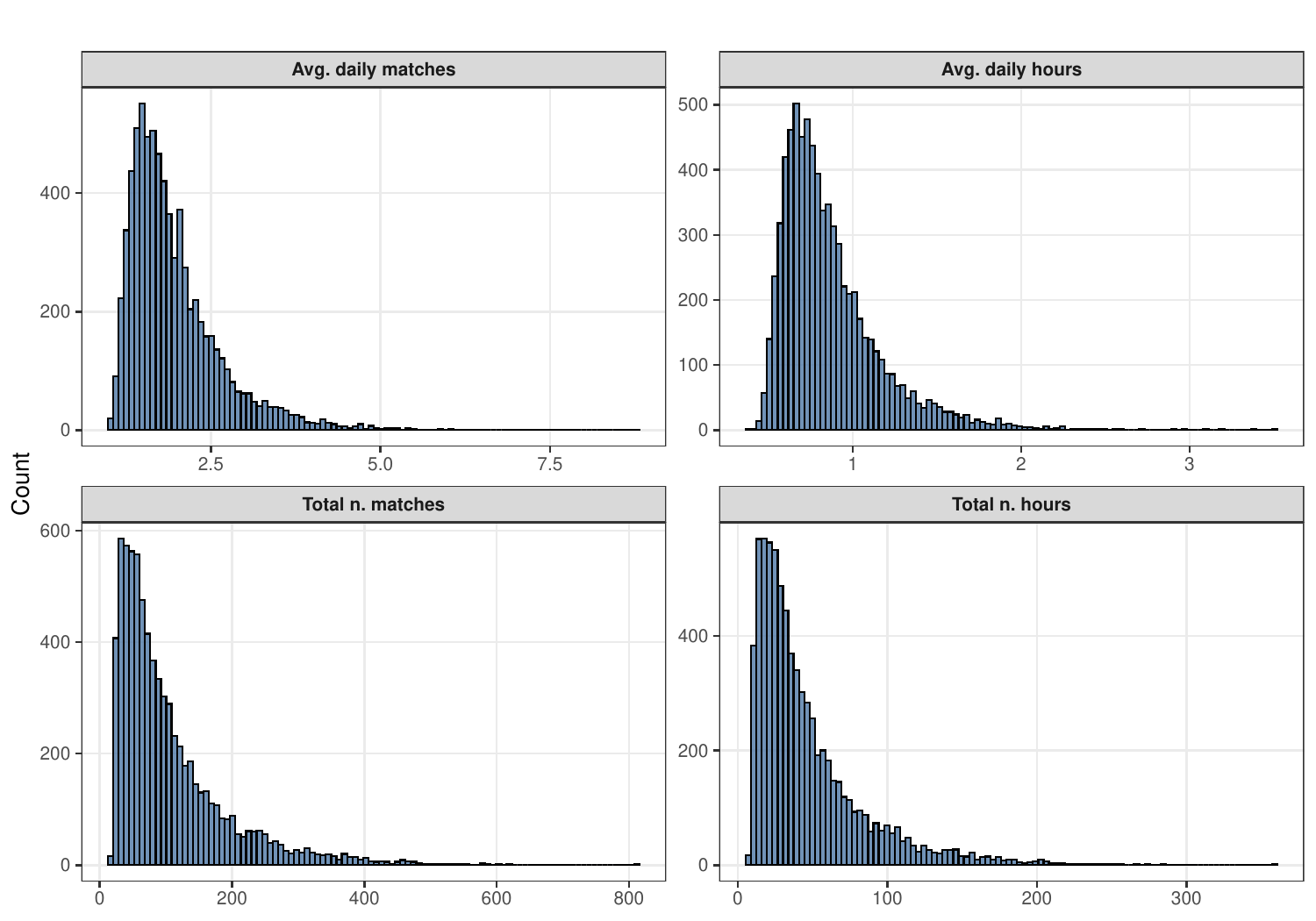}
    \caption{Histograms of players’ total and average daily activity, measured in both number of matches played and hours spent in-game.}
    \label{fig_player_activity}
\end{figure}

\begin{figure}[H]
    \centering
    \includegraphics[scale=0.5]{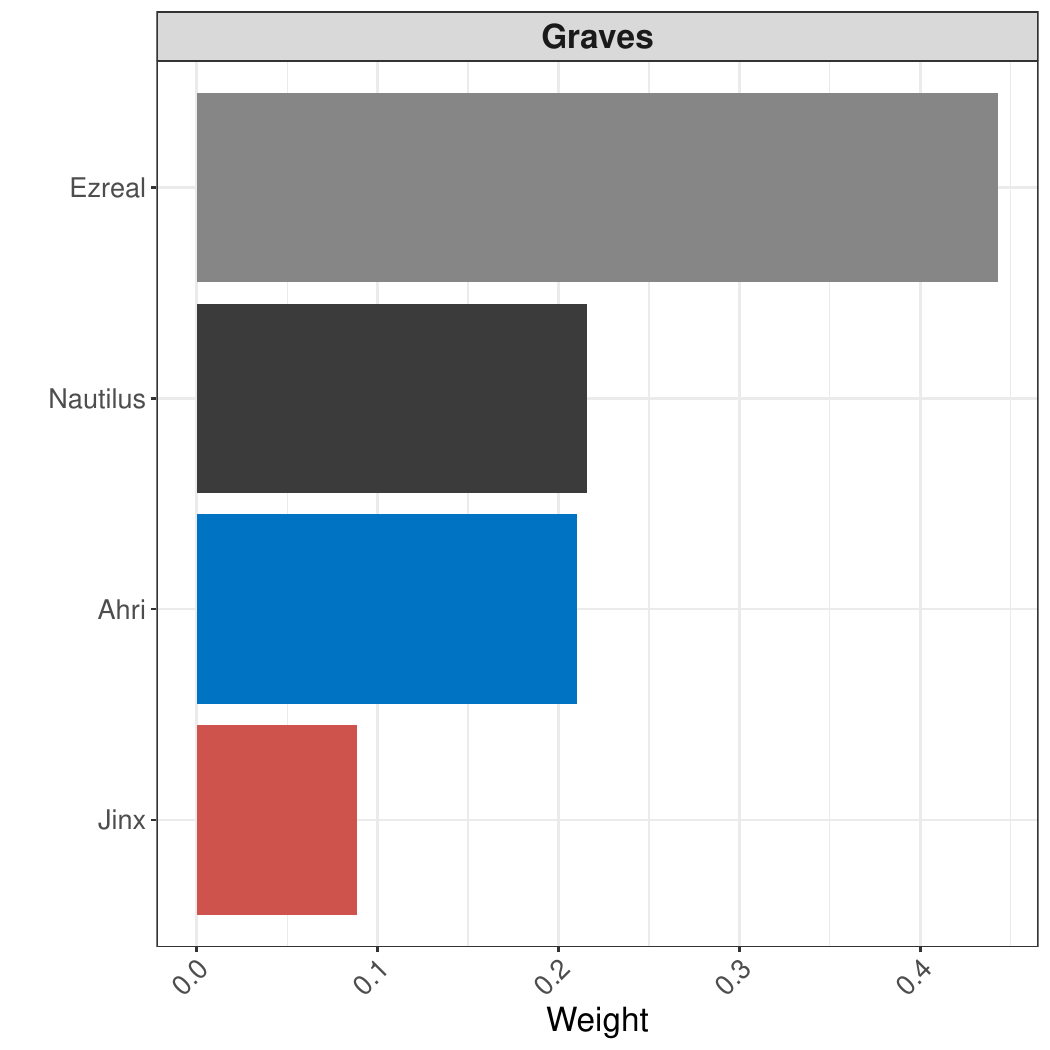}
    \caption{Identities and contributions of characters in the donor pool for the Graves' synthetic control displayed in Figure \ref{fig_graves_pick_rates_pooled}.}
    \label{fig_graves_pick_rates_pooled_weights}
\end{figure}

\begin{figure}[H]
    \centering
    \includegraphics[scale=0.5]{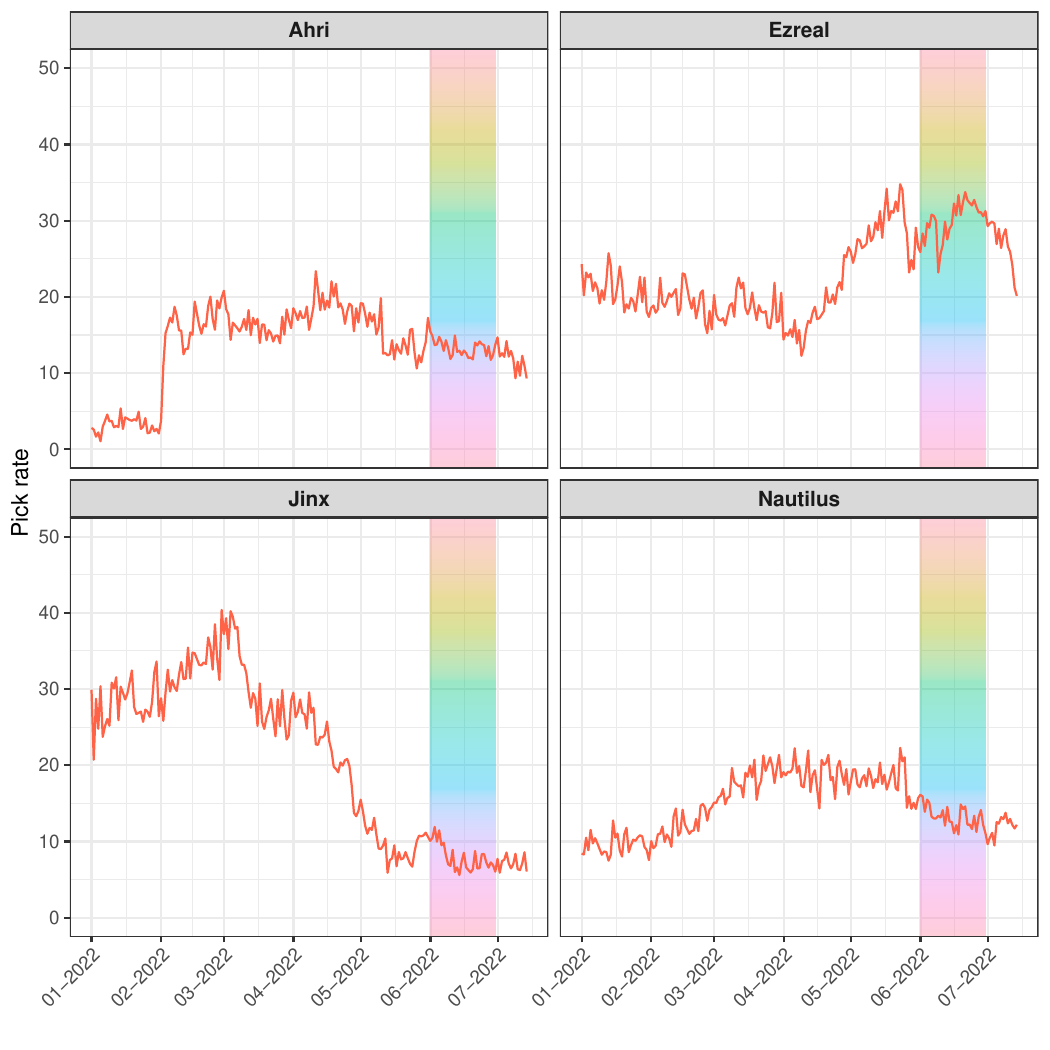}
    \caption{Daily pick rates of the characters contributing to the Graves' synthetic control displayed in Figure \ref{fig_graves_pick_rates_pooled}. The rainbow area highlights LGBT Pride Month.}
    \label{fig_graves_pick_rates_pooled_donors}
\end{figure}

\begin{figure}[H]
    \centering
    \includegraphics[scale=0.5]{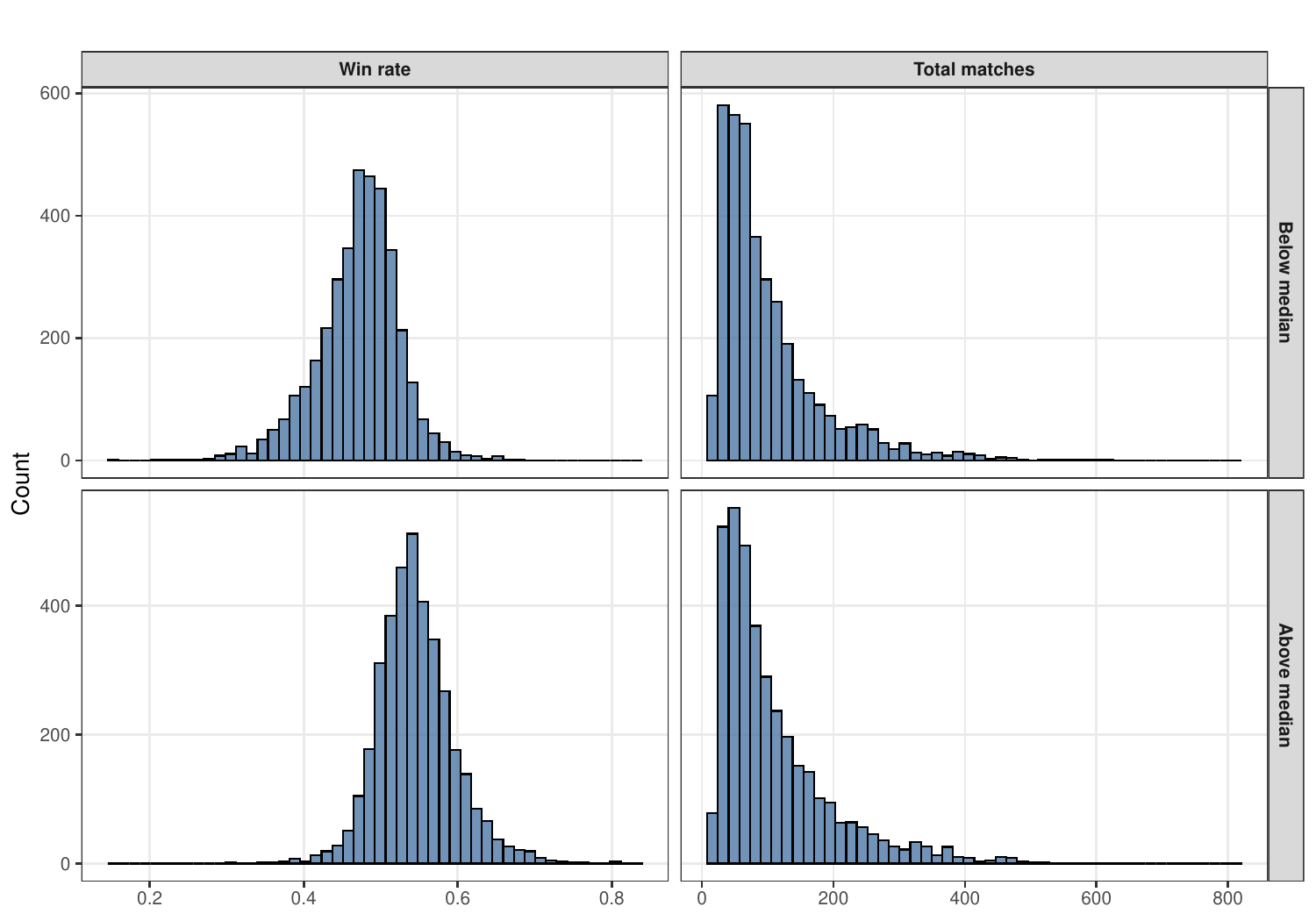}
    \caption{Distribution of overall win rates and total matches played for players classified as above or below the median pre-treatment win rate.}
    \label{fig_player_skills_median}
\end{figure}

\begin{figure}[H]
    \centering
    \includegraphics[scale=0.5]{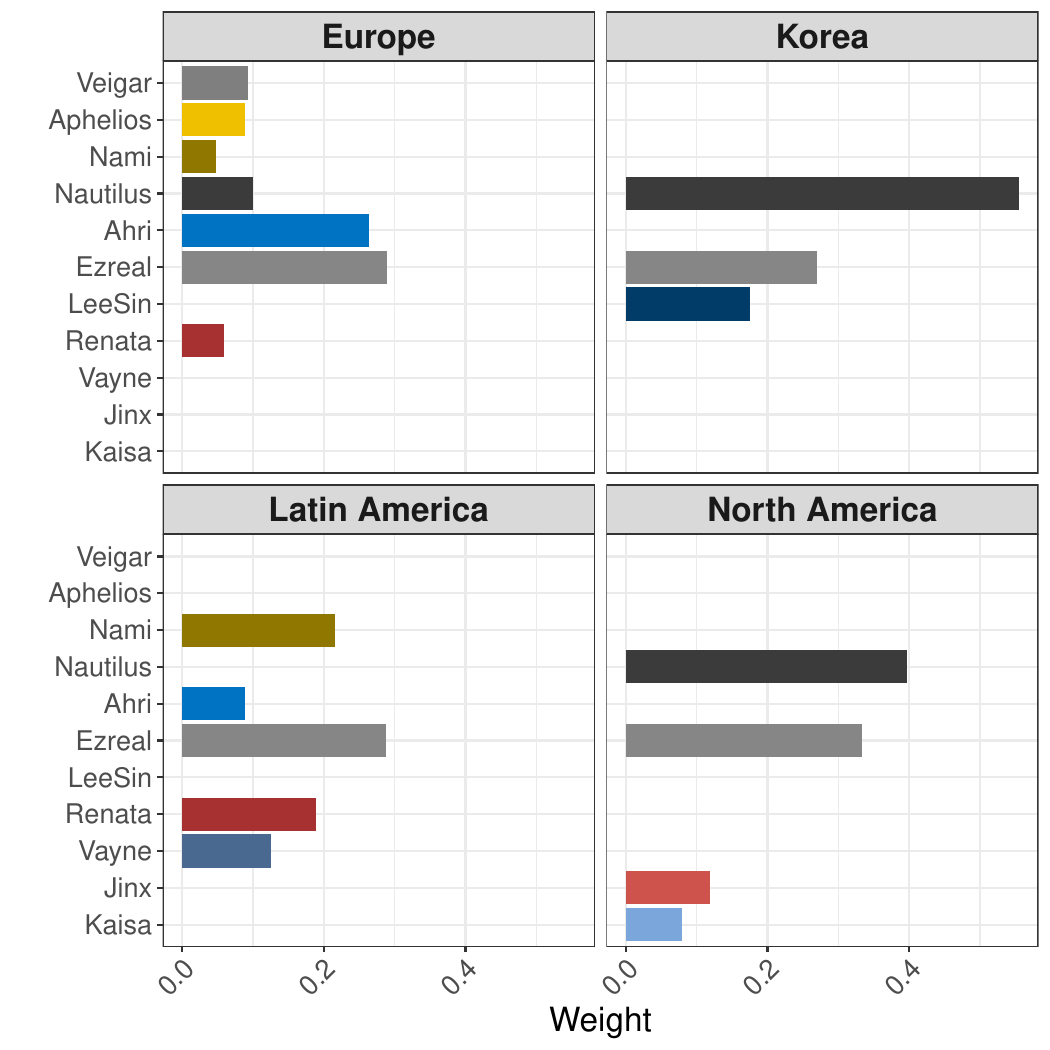}
    \caption{Identities and contributions of characters in the donor pool for the Graves' synthetic controls displayed in Figure \ref{fig_graves_pick_rates_regional}.}
    \label{fig_graves_pick_rates_regional_weights}
\end{figure}

\begin{figure}[H]
    \centering
    \includegraphics[scale=0.5]{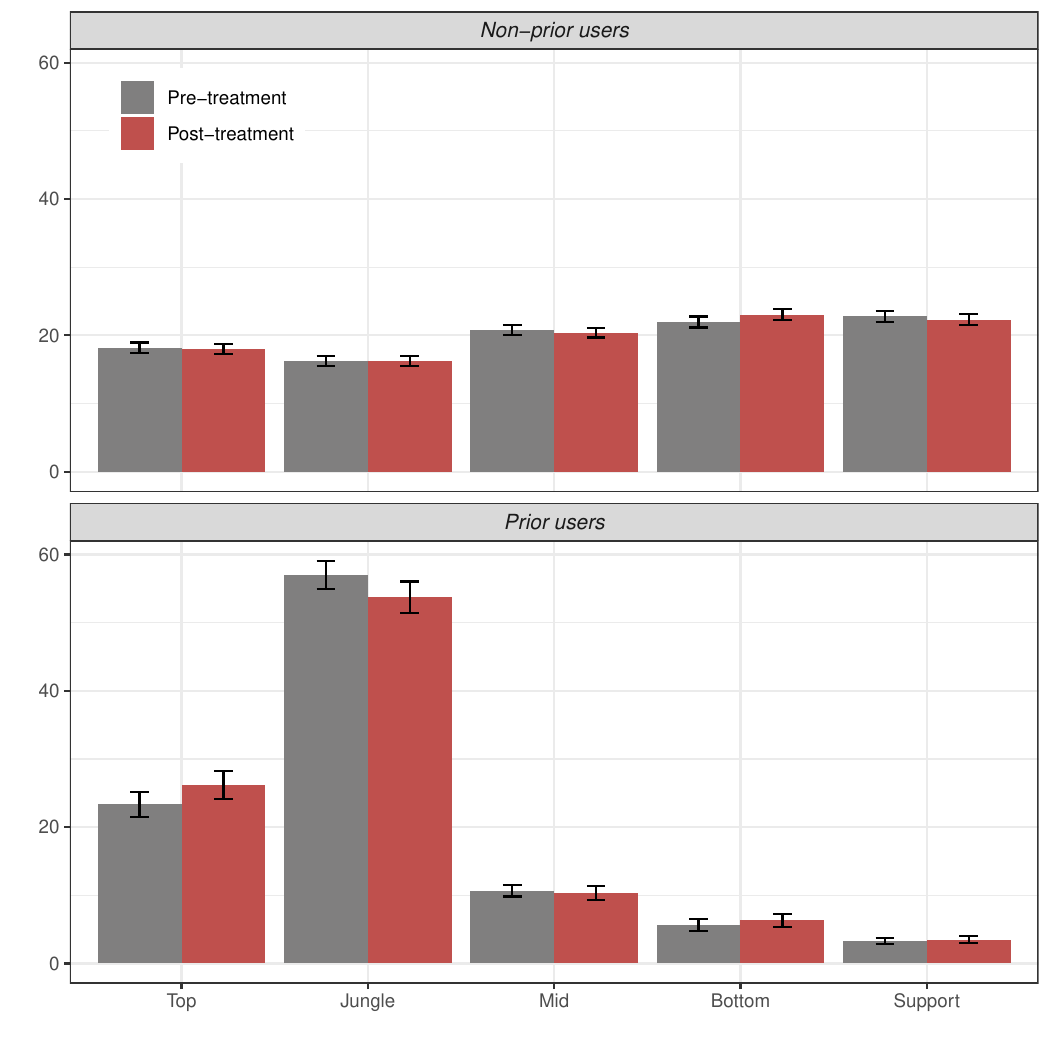}
    \caption{Shares of matches played in each role before and after the coming-out event. Players are divided into two groups based on their preferences for Graves before his disclosure.}
    \label{fig_players_position}
\end{figure}

\begin{figure}[H]
    \centering
    \includegraphics[scale=0.5]{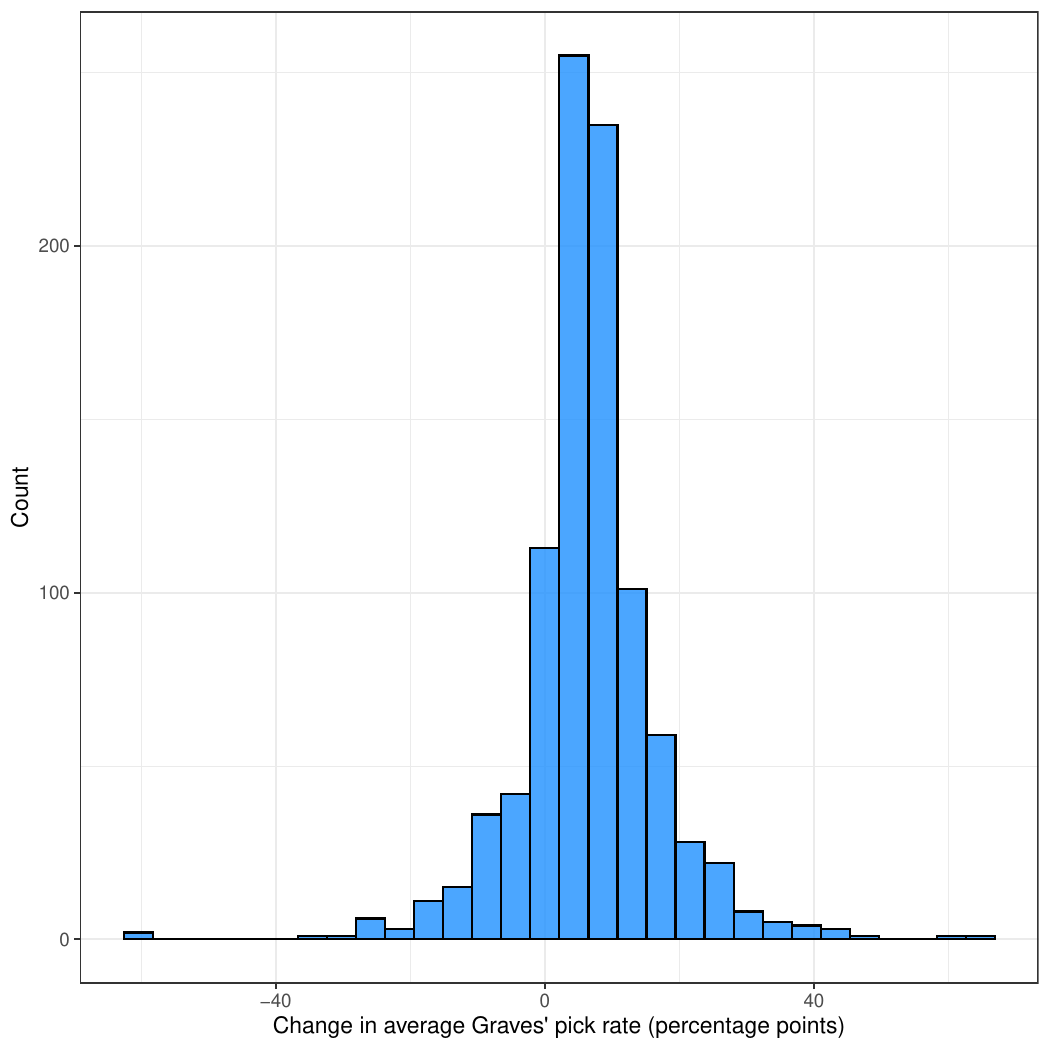}
    \caption{Distribution of changes in average Graves’ pick rates (percentage points) between the pre- and post-disclosure periods among prior users. Negative values indicate increases in pick rates, while positive values indicate reductions.}
    \label{fig_players_graves_reduction}
\end{figure}

\begin{figure}[H]
    \centering
    \includegraphics[scale=0.5]{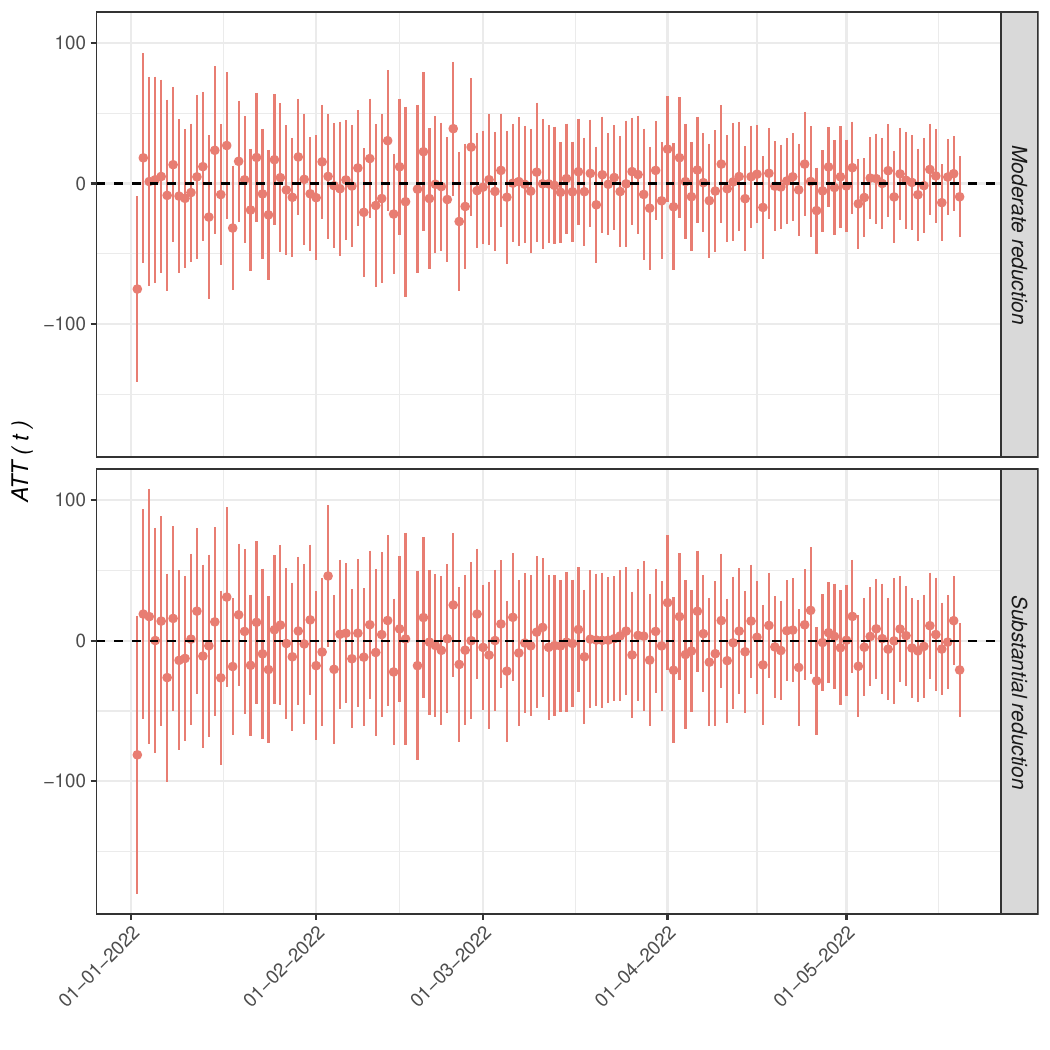}
    \caption{Point estimates and simultaneous $95\%$ confidence bands allowing for clustering at the player level for the placebo $ATT ( t )$. Each row corresponds to a different version of the treatment.}
    \label{fig_did_player_performance_results_pre}
\end{figure}

\begin{figure}[H]
    \centering
    \includegraphics[scale=0.5]{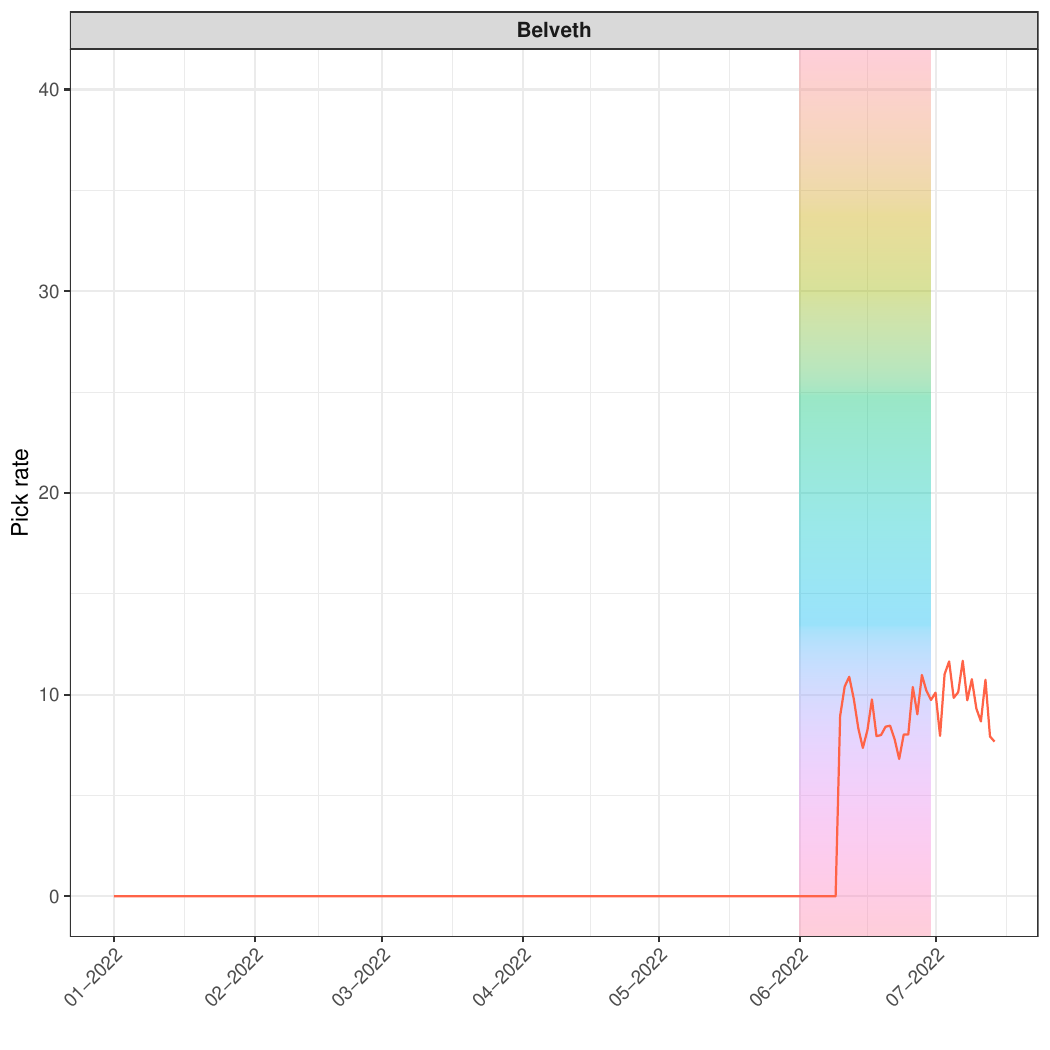}
    \caption{Bel'Veth's daily pick rates. The rainbow area highlights LGBT Pride Month.}
    \label{fig_belveth_pick_rates}
\end{figure}

\begingroup
  \setlength{\tabcolsep}{8pt}
  \renewcommand{\arraystretch}{1.1}
  \begin{table}[H]
    \centering
    \caption{Most popular characters.}
    \begin{adjustbox}{width = 1\textwidth}
    \begin{tabular}{@{\extracolsep{5pt}}l c c c c c c c c c c }
      \\[-1.8ex]\hline
      \hline \\[-1.8ex]
      & \multicolumn{5}{c}{\textit{Pre-treatment}} & \multicolumn{5}{c}{\textit{Post-treatment}} \\ \cmidrule{2-6} \cmidrule{7-11}
      & \textit{Top} & \textit{Jungle} & \textit{Mid} & \textit{Bottom} & \textit{Support} & \textit{Top} & \textit{Jungle} & \textit{Mid} & \textit{Bottom} & \textit{Support} \\
      \addlinespace[2pt]
      \hline \\[-1.8ex] 

      \textbf{1} & Irelia & LeeSin & Ahri & Jinx & Karma & Gangplank & Viego & Yone & Ezreal & Renata \\ 
                 & (12.007) & (19.132) & (13.62) & (24.5) & (19.522) & (12.455) & (19.419) & (17.587) & (29.02) & (16.43) \\ 
      \textbf{2} & Camille & Graves & Akali & Ezreal & Nautilus & Fiora & MonkeyKing & Sylas & Zeri & Karma \\ 
                 & (10.4) & (18.557) & (12.816) & (21.155) & (15.302) & (10.569) & (15.064) & (13.177) & (19.716) & (16.239) \\ 
      \textbf{3} & Jayce & Viego & Yone & Jhin & Lulu & Aatrox & LeeSin & Ahri & Twitch & Yuumi \\ 
                 & (9.756) & (17.982) & (12.197) & (20.746) & (14.322) & (9.451) & (12.78) & (12.83) & (14.751) & (14.215) \\ 
      \textbf{4} & Fiora & Diana & Yasuo & Kaisa & Nami & Irelia & Taliyah & Yasuo & Jhin & Senna \\ 
                 & (8.777) & (13.771) & (10.611) & (17.174) & (11.781) & (9.407) & (12.322) & (10.855) & (14.531) & (13.715) \\ 
      \textbf{5} & Aatrox & Khazix & Viktor & Lucian & Pyke & Kayle & Graves & Viktor & Kaisa & Lulu \\ 
                 & (8.616) & (9.448) & (10.525) & (12.505) & (11.401) & (8.51) & (12.319) & (10.347) & (13.317) & (13.274) \\ 

      \addlinespace[3pt]
      \\[-1.8ex]\hline
      \hline \\[-1.8ex]
      \end{tabular}
      \end{adjustbox}
      \footnotesize
      \renewcommand{\baselineskip}{11pt}
      \textit{Notes.} Most popular characters by role based on average pick rates (displayed in parenthesis)
      \label{table_most_played_champions}
    \end{table}
\endgroup

\begingroup
  \setlength{\tabcolsep}{8pt}
  \renewcommand{\arraystretch}{1.1}
  \begin{table}[H]
    \centering
    \caption{Regional results.}
    \begin{adjustbox}{width = 1\textwidth}
    \begin{tabular}{@{\extracolsep{5pt}}l c c c c c }
      \\[-1.8ex]\hline
      \hline \\[-1.8ex]
      & \multicolumn{2}{c}{\textit{Synthetic Controls}} & \multicolumn{2}{c}{\textit{Regularized Synthetic Controls}} \\ \cmidrule{2-3} \cmidrule{4-5} 
      & (1) & (2) & (3) & (4) \\
      & All characters & Only non-substitutes & All characters & Only non-substitutes \\
      \addlinespace[2pt]
      \hline \\[-1.8ex] 

      \multicolumn{5}{l}{\textbf{\small Panel 1: \textit{Europe}}} \\
      $\hat{\tau}$ &  -8.416 & -9.968 & -6.819 & -7.170 \\
      $p\text{-value}$ & 0.004 & 0.021 & 0.008 & 0.120 \\
      $95\%$ CI & [-14.102, -2.730] & [-18.462, -1.475] & [-11.890, -1.749] & [-16.393,  2.040] \\
      N. Donors & 7 & 7 & 17 & 15 \\
      RMSE & 3.317 & 3.605 & 3.450 & 3.659 \\
      Pre-treatment average & 14.62 & 14.62 & 14.62 & 14.62 \\ \cmidrule{1-5} 

      \multicolumn{5}{l}{\textbf{\small Panel 2: \textit{Korea}}} \\
      $\hat{\tau}$ & -10.637 & -9.963 & -9.461 & -7.300 \\
      $p\text{-value}$ & 0.003 & 0.086 & 0.011 & 0.250 \\
      $95\%$ CI & [-17.649, -3.625] & [-21.353,  1.428] & [-16.791, -2.132] & [-19.743,  5.138] \\
      N. Donors & 3 & 2 & 5 & 4 \\
      RMSE & 8.373 & 8.793 & 8.810 & 9.070 \\
      Pre-treatment average & 31.409 & 31.409 & 31.409 & 31.409 \\ \cmidrule{1-5} 

      \multicolumn{5}{l}{\textbf{\small Panel 3: \textit{Latin America}}} \\
      $\hat{\tau}$ &  -5.188 & -5.225 & -3.407 & -3.200 \\
      $p\text{-value}$ & 0.019 & 0.150 & 0.125 & 0.290 \\
      $95\%$ CI & [ -9.517, -0.858] & [-12.345,  1.895] & [ -7.762,  0.948] & [ -9.235,  2.820] \\
      N. Donors & 5 & 5 & 12 & 11 \\
      RMSE & 3.901 & 3.927 & 3.967 & 4.027 \\
      Pre-treatment average & 16.229 & 16.229 & 16.229 & 16.229 \\ \cmidrule{1-5} 

      \multicolumn{5}{l}{\textbf{\small Panel 4: \textit{North America}}} \\
      $\hat{\tau}$ &   0.597 &  0.597 &  1.587 &  1.830 \\
      $p\text{-value}$ & 0.814 & 0.877 & 0.585 & 0.660 \\
      $95\%$ CI & [ -4.382,  5.576] & [ -6.961,  8.155] & [ -4.109,  7.283] & [ -6.395, 10.074] \\
      N. Donors & 4 & 4 & 6 & 6 \\
      RMSE & 6.237 & 6.237 & 6.351 & 6.382 \\
      Pre-treatment average & 22.281 & 22.281 & 22.281 & 22.281 \\ 

      \addlinespace[3pt]
      \\[-1.8ex]\hline
      \hline \\[-1.8ex]
      \end{tabular}
      \end{adjustbox}
      \footnotesize
      \renewcommand{\baselineskip}{11pt}
      \textit{Notes.} Point estimates, $p$-values, and $95\%$ confidence intervals for $\hat{\tau}$. Additionally, the number of donors receiving a non-zero weight and the pre-treatment root mean squared error are displayed. Each panel reports the results obtained using only matches from a particular macro-region. Each column corresponds to a different specification, with the specifications differing solely in the employed estimator and donor pool composition.
      \label{table_estimation_results_regional}
    \end{table}
\endgroup

\section{Methods} 
\label{app_methodology}

\setcounter{equation}{0}
\setcounter{table}{0}
\setcounter{figure}{0}

\noindent A simple comparison of Graves’ pick rates before and after the disclosure may not accurately reflect the impact of the coming-out event on players' preferences for that character, as other factors could have changed during that period. To address this issue, we construct a synthetic control unit \parencite[see, e.g.,][]{abadie2003economic, abadie2010synthetic, abadie2015comparative, abadie2021using, abadie2022synthetic} by weighting other characters to approximate the pick rates of Graves before the disclosure. This method allows us to isolate the effects of the coming-out event on players' revealed preferences for Graves and gain insight into how these preferences would have behaved in the absence of the disclosure.

Formally, our data set comprises $n = 161$ characters ($i = 1, \dots, n$) observed over $T = 194$ days ($t = 1, \dots, T$). $T^{pre}$ is the length of the period (150 days) before the coming-out event, which occurs at time $T^{pre} + 1$ (i.e., June $1^{st}$, $2022$). For each unit $i$ and time $t$, we denote the observed pick rate as $Y_{i, t}$. We represent the coming out as a binary variable $C_i \in \left\{ 0, 1 \right\}$ equal to one if character $i$ discloses his sexual orientation at time $T^{pre} + 1$. We then posit the existence of two potential pick rates $Y_{i, t}^c$, where one denotes the pick rate in the absence of disclosure ($Y_{i, t}^0$) and the other denotes the pick rate in the presence of disclosure ($Y_{i, t}^1$).\footnote{\ These potential outcomes are based on Rubin’s model for causal inference \parencite{rubin1974estimating}.} 

Without loss of generality, we let the first unit $i = 1$ be Graves. This implies that $C_1 = 1$ and $C_i = 0$ for all $i \neq 1$. Then, for each period $t > T^{pre}$, we define the effect of the coming-out event on players' preferences for Graves as the difference in Graves's potential pick rates at time $t$:

\begin{equation}
    \tau_{t} := Y_{1, t}^1 - Y_{1, t}^0.
    \label{equation_estimand}
\end{equation}  

\noindent Note that we allow the effects to change over time.

Since Graves' sexual orientation has been disclosed after period $T^{pre}$, under a standard SUTVA assumption \parencite[see, e.g.,][]{imbens2015causal} we observe $Y_{1, t} = Y_{1, t}^1$ for all $t>T^{pre}$. Thus, as shown in equation (\ref{equation_estimand}), the challenge in estimating our causal effects of interest is to estimate $Y_{1,t}^0$ for $t>T^{pre}$, i.e., how Graves' pick rates would have evolved in the absence of the disclosure. To this end, we can construct a synthetic control unit that approximates the pick rates of Graves before the coming out. The idea is that if the synthetic control and Graves behave similarly before the disclosure, then the synthetic control can serve as a valid counterfactual. 

The synthetic control unit is characterized by a set of weights, denoted as $\omega := \left( \omega_2, \dots, \omega_{n} \right)$, chosen to align the pre-treatment pick rates of the synthetic unit with those of Graves. This is achieved by solving the following optimization problem \parencite{arkhangelsky2021synthetic}:

\begin{equation}
    \begin{gathered}
        \hat{\omega} = \argmin_{\omega \in \Omega} \ell \left( \omega \right), \\[1ex]
        \ell \left( \omega \right) = \sum_{t = 1}^{T^{pre}} \left( \sum_{i = 2}^n \omega_i Y_{i, t} - Y_{1, t} \right)^2 + \zeta^2 T^{pre} \Vert \omega \Vert_2^2, \quad \Omega = \left\{ \omega \in \R_{+}^{n - 1} : \sum_{i = 2}^n \omega_i = 1 \right\},
    \end{gathered}
    \label{equation_sc_weights}
\end{equation} 

\noindent where the weights are restricted to be non-negative and to sum up to one and a ridge penalty is employed to ensure the uniqueness of the weights. In our main specification, we set the regularization parameter to zero, thus employing a standard synthetic control estimator. As a robustness check, we follow \textcite{arkhangelsky2021synthetic} and set $\zeta = \left( T - T^{pre} \right)^{1/4} \hat{\sigma}$, with $\hat{\sigma}$ denoting the standard deviation of first differences of $Y_{i, t}$ for control units over the pre-treatment period. 

We estimate the counterfactual outcome of Graves as a weighted average of the outcome of the control units:

\begin{equation}
    \widehat{Y}_{1, t}^0 = \sum_{i = 2}^n \hat{\omega}_i Y_{i, t}.
    \label{equation_estimator_counterfactual}
\end{equation}

\noindent Finally, to estimate the causal effects of interest, we compute the differences between Graves' observed pick rates and the synthetic counterfactual for all $t > T^{pre}$:

\begin{equation}
    \hat{\tau}_{t} = Y_{1, t}^1 - \widehat{Y}_{1, t}^0.
    \label{equation_estimator_effects}
\end{equation}  

We summarize the estimated effects by reporting the average treatment effect on players' preferences for Graves, with the averaging carried out over the post-treatment periods:

\begin{equation}
    \hat{\tau} = \frac{1}{T - T^{pre}} \sum_{t = T^{pre} + 1}^T \hat{\tau}_t.
    \label{equation_average_effect_graves}
\end{equation}

\noindent We employ the \open placebo approach" of \textcite{arkhangelsky2021synthetic} to estimate the variance of $\hat{\tau}$. We then use the estimated variance to construct asymptotically valid conventional confidence intervals.\footnote{\ The validity of this placebo approach hinges on a homoskedasticity assumption which requires that treated and control units have the same noise distribution. In general, with only one treated unit, nonparametric variance estimation for treatment effect estimators is typically impossible without a homoskedasticity assumption \parencite{arkhangelsky2021synthetic}.}

\section{Robustness Checks}
\label{app_robust}

\setcounter{equation}{0}
\setcounter{table}{0}
\setcounter{figure}{0}

\noindent We examine the robustness of our main results from Section \ref{subsec_main_results} to alternative estimation strategies and variations in the composition of the donor pool. In particular, we repeat our analysis employing a regularized synthetic control estimator (see Section \ref{app_methodology}) and explore different donor pool configurations focusing on characters from distinct roles. Notably, Graves is predominantly designed for and played in three of the possible roles within a team. Consequently, there is a possibility of spillover effects on other characters mainly played in these positions, as players transitioning away from Graves are likely to switch to these alternatives.\footnote{\ Table \ref{table_summary_stats} and the findings of Section \ref{subsec_players_skills} support this intuition.} To mitigate this potential for spillover effects, we restrict our donor pool to characters that are \open non-substitutes" of Graves, that is, those primarily designed for the remaining two roles.\footnote{\ Graves is predominantly designed for and played in the \textit{top}, \textit{jungle}, and \textit{mid} positions. Therefore, we consider characters designed for and played in the \textit{bottom} and \textit{support} positions as \open non-substitutes."}

\begingroup
  \setlength{\tabcolsep}{8pt}
  \renewcommand{\arraystretch}{1.1}
  \begin{table}[b!]
    \caption{Main results.}
    \centering
    \begin{adjustbox}{width = 1\textwidth}
    \begin{tabular}{@{\extracolsep{5pt}}l c c c c c }
      \\[-1.8ex]\hline
      \hline \\[-1.8ex]
      & \multicolumn{2}{c}{\textit{Synthetic Controls}} & \multicolumn{2}{c}{\textit{Regularized Synthetic Controls}} \\ \cmidrule{2-3} \cmidrule{4-5} 
      & (1) & (2) & (3) & (4) \\
      & All characters & Only non-substitutes & All characters & Only non-substitutes \\
      \addlinespace[2pt]
      \hline \\[-1.8ex] 

      $\hat{\tau}$ & -7.112 & -6.922 & -5.462 & -5.28 \\
      $p\text{-value}$ & 0.005 & 0.071 & 0.027 & 0.17 \\
      $95\%$ CI & [-12.094, -2.130] & [-14.425, 0.580] & [-10.289, -0.636] & [-12.837, 2.269] \\
      N. Donors & 4 & 4 & 6 & 6 \\
      RMSE & 2.734 & 2.875 & 2.770 & 2.877 \\
      Pre-treatment average & 18.557 & 18.557 & 18.557 & 18.557\\ 

      \addlinespace[3pt]
      \\[-1.8ex]\hline
      \hline \\[-1.8ex]
      \end{tabular}
      \end{adjustbox}

      \footnotesize
      \renewcommand{\baselineskip}{11pt}
      \textit{Notes.} Point estimates, $p$-values, and $95\%$ confidence intervals for $\hat{\tau}$. Additionally, the number of donors receiving a non-zero weight and the pre-treatment root mean squared error are displayed. Each column corresponds to a different specification, with the specifications differing solely in the employed estimator and donor pool composition.
      \label{table_estimation_results_main}
    \end{table}
\endgroup

Table \ref{table_estimation_results_main} displays the results. For any donor pool composition, the results are not sensitive to the choice of the regularization parameter, and point estimates are consistently negative. The results are consistent also quantitatively across all specifications, with a decline ranging between $40.98\%$ and $38.17\%$ of the pre-treatment average preferences for Graves. Overall, these results support our main finding of a substantial negative impact of the coming-out event on players' preferences for Graves. 

\begin{figure}[b!]
    \centering
    \includegraphics[scale=0.50]{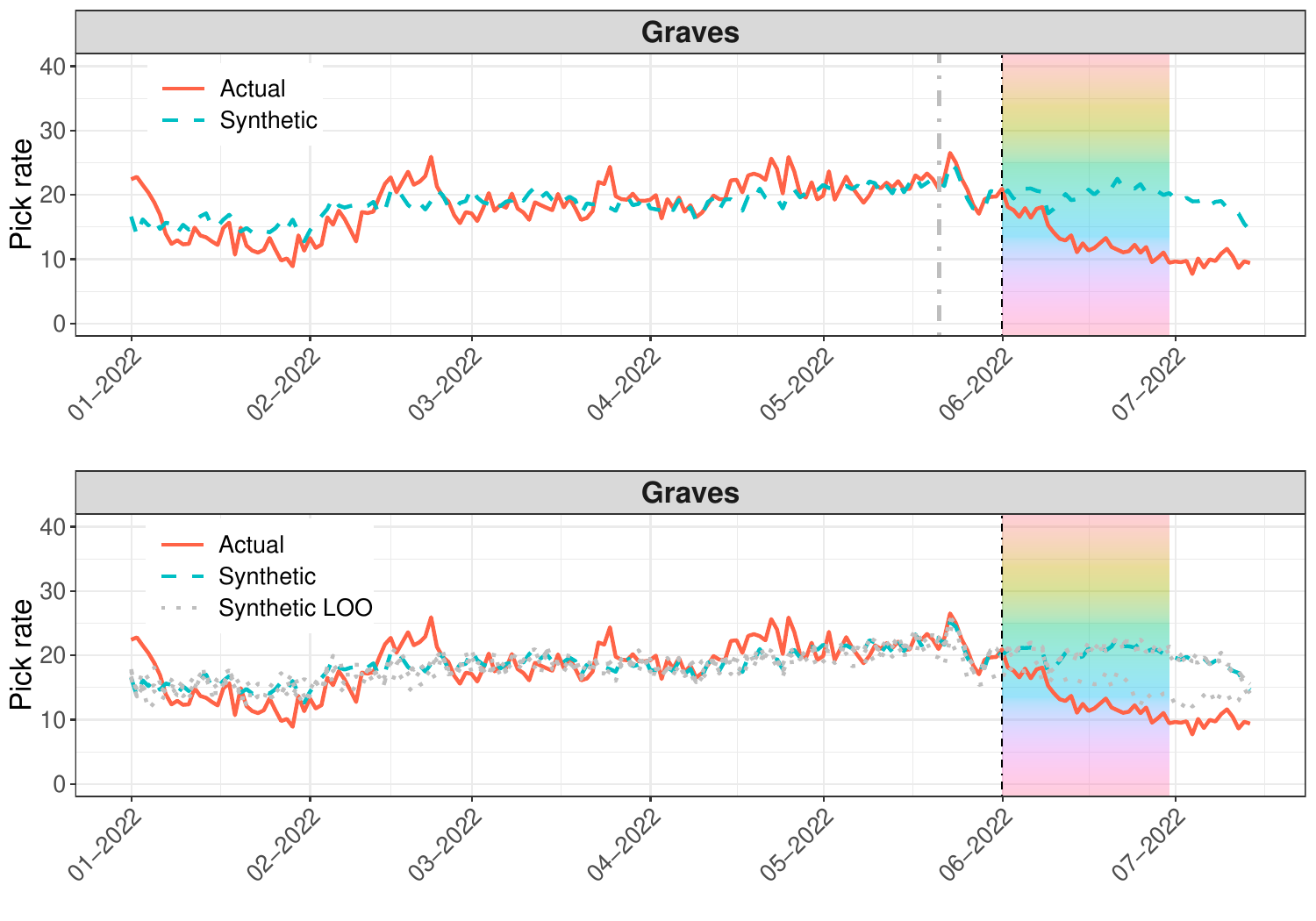}
    \caption{Robustness checks for main results. The upper panel shifts the coming-out event ten days earlier, with the new treatment date denoted by the vertical gray dashed line. The lower panel reports leave-one-out estimates of the synthetic control series, obtained by excluding one of the characters of Figure \ref{fig_graves_pick_rates_pooled_weights} at a time from the donor pool.}
    \label{fig_graves_pick_rates_pooled_robustness}
\end{figure}

We also assess the credibility of the synthetic control estimator by conducting a robustness check that artificially shifts the coming-out event ten days earlier. This backdating exercise allows us to evaluate the estimator's predictive accuracy during a ten-day hold-out period \parencite[see e.g.,][]{abadie2022synthetic}. The upper panel of Figure \ref{fig_graves_pick_rates_pooled_robustness} presents the results of this analysis. We observe three key findings. First, the estimated effects remain qualitatively and quantitatively consistent, confirming a negative and persistent impact of the coming-out event on players' revealed preferences for Graves. Second, the synthetic control estimator demonstrates a good fit during the hold-out period, indicating its ability to accurately capture Graves' usage prior to the disclosure. Third, the actual and the synthetic series begin to diverge on the true day of disclosure, even when the estimator has no knowledge of the actual disclosure date. The absence of estimated effects before the coming-out event also lends support to the plausibility of a no-anticipation assumption \parencite[see e.g.,][]{abadie2021using}. 

We conduct an additional robustness test by performing a leave-one-out exercise, where we repeatedly estimate the synthetic control series by excluding one character with non-zero estimated weights at a time from the donor pool \parencite[see e.g.,][]{abadie2021using}. The lower panel of Figure \ref{fig_graves_pick_rates_pooled_robustness} presents the results of this analysis. Overall, our finding of a negative and persistent impact of the coming-out event on players' preferences for Graves is robust to the exclusion of any particular character. Most of the leave-one-out synthetic series closely align with the main estimate, thus reinforcing the robustness of the main conclusion of our study. One leave-one-out series falls beneath the other synthetic series, suggesting a somewhat reduced, although still negative, impact. However, this series diverges from the actual series in the weeks prior to the treatment, which undermines the reliability of its results.\footnote{\ This series is obtained by excluding the character \textit{Ezreal} from the donor pool. Figure \ref{fig_graves_pick_rates_pooled_weights} in Supplementary Information \ref{app_figures_tables} shows that this character receives the largest weight in our main specification. Therefore, the divergence of this leave-one-out series from the actual series is unsurprising.}

Finally, we implement a within-role placebo design by replicating the synthetic control analysis for the four most frequently selected jungle characters (excluding Graves) in the pre-treatment period. These characters share the same primary role as Graves and are therefore exposed to similar role-specific dynamics and potential demand shifts. The results are reported in Figure \ref{fig_placebo_jungle}. None of these placebo units exhibits a sharp and sustained post-disclosure divergence comparable to that observed for Graves. In particular, we do not observe a synchronized decline in jungle pick rates around June 1, 2022. While some champions display idiosyncratic fluctuations, these movements are not aligned with the disclosure date and are not persistent. Overall, these patterns indicate that the estimated effect for Graves is not driven by broader shocks to jungle demand or role-specific changes, but is instead character-specific.

\begin{figure}[H]
    \centering
    \begin{minipage}[t]{0.48\textwidth}
        \centering
        \includegraphics[width=\textwidth]{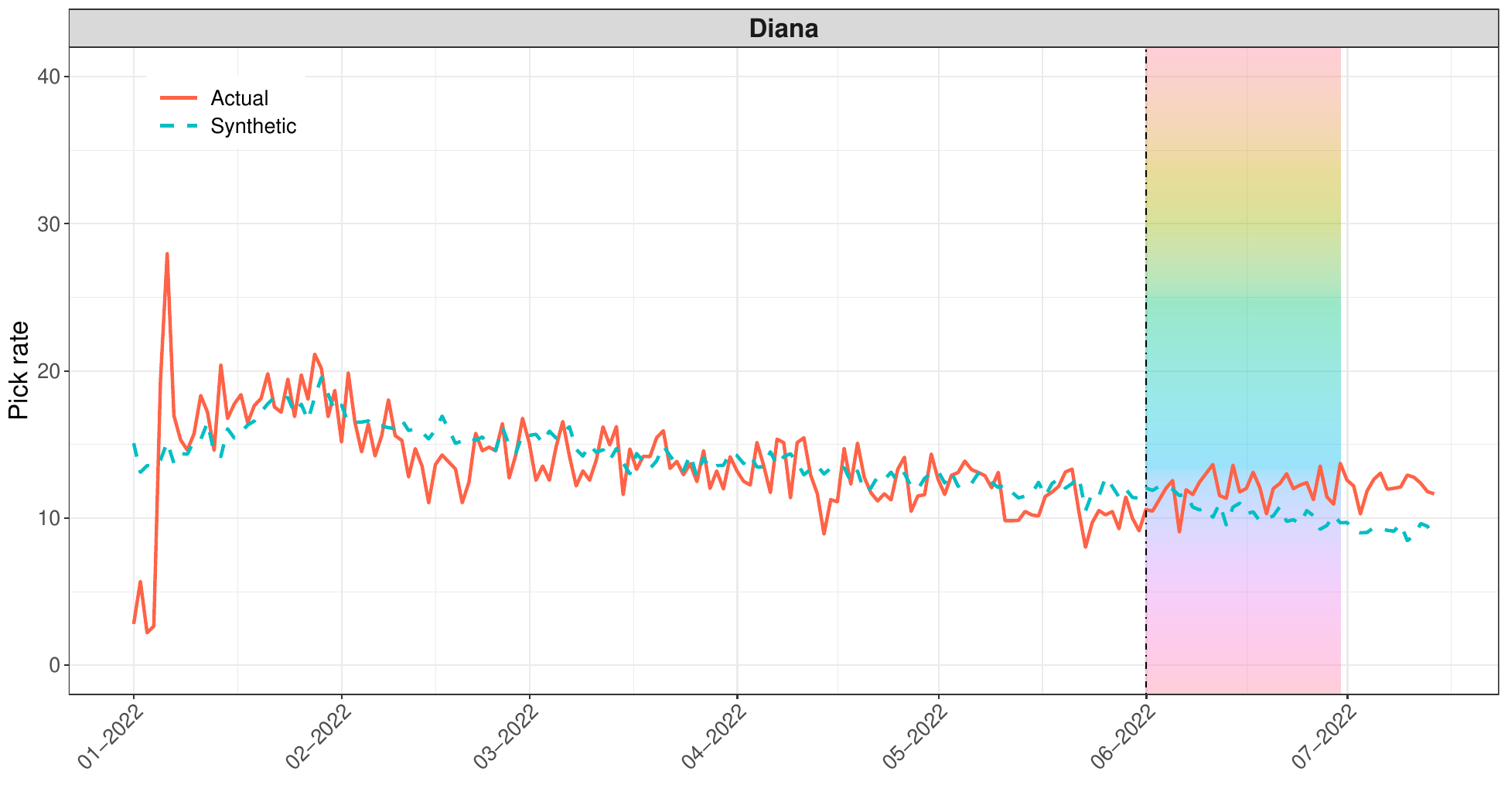}
    \end{minipage}
    \hfill
    \begin{minipage}[t]{0.48\textwidth}
        \centering
        \includegraphics[width=\textwidth]{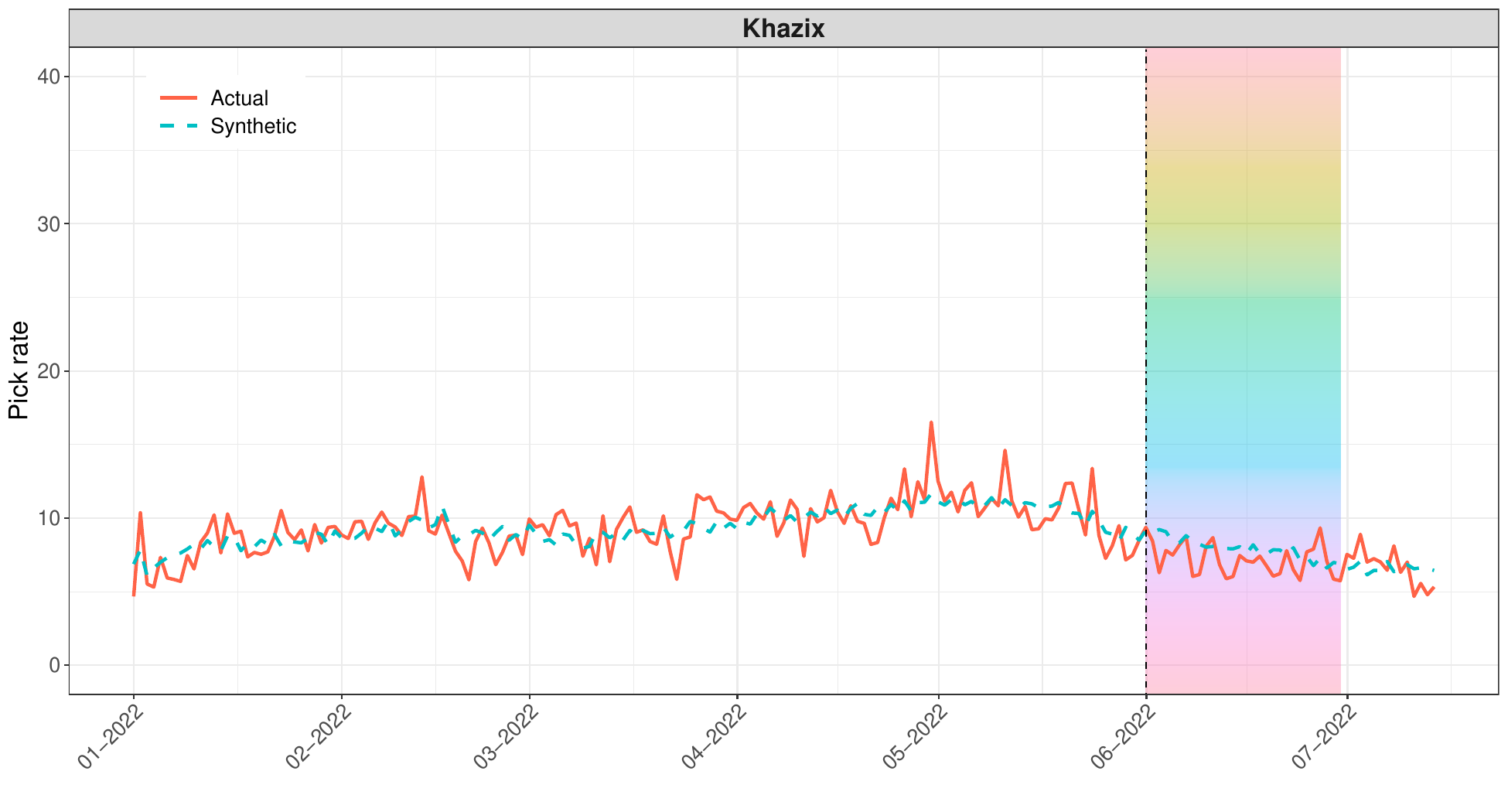}
    \end{minipage}

    \vspace{0.3cm}

    \begin{minipage}[t]{0.48\textwidth}
        \centering
        \includegraphics[width=\textwidth]{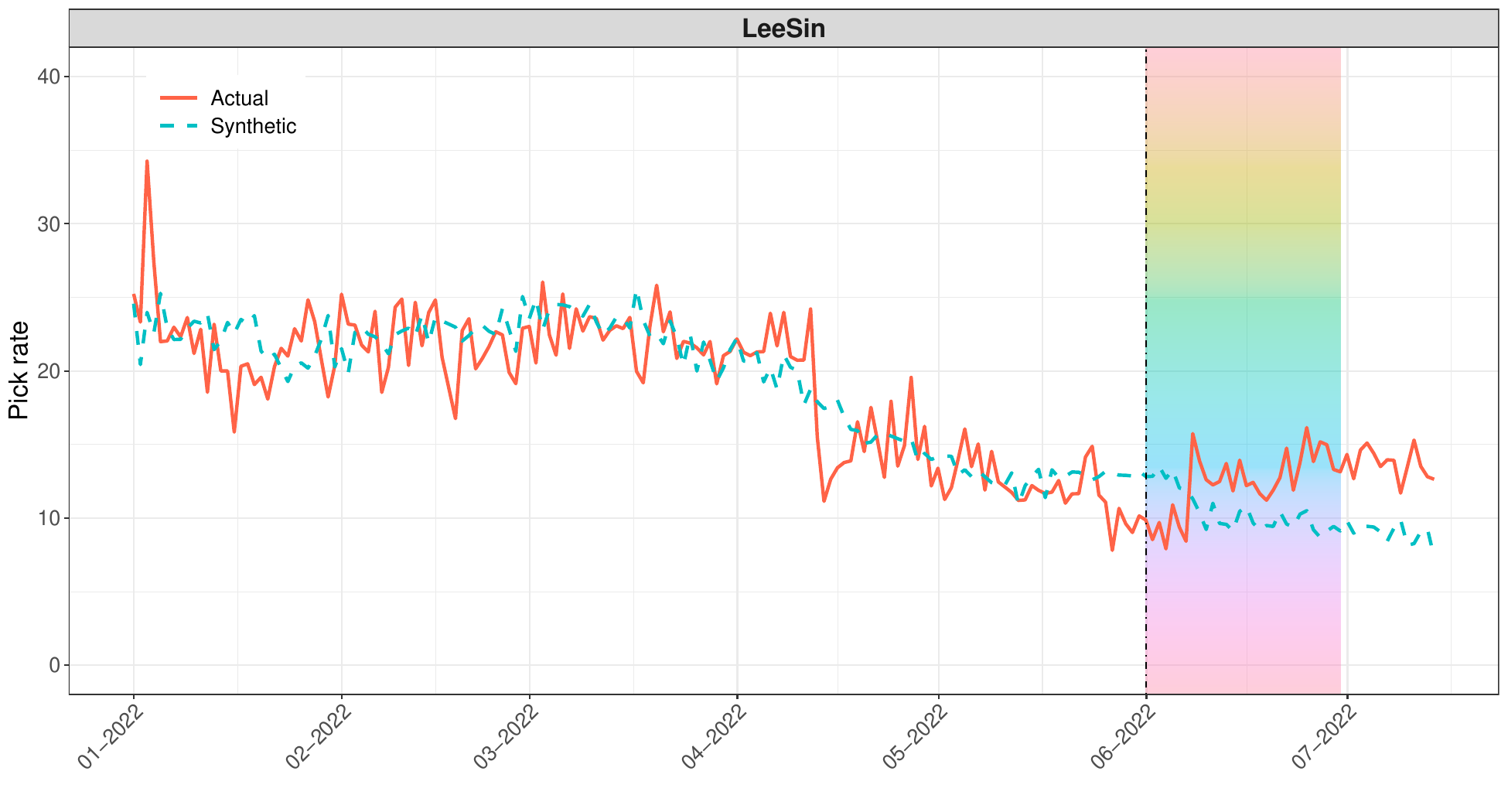}
    \end{minipage}
    \hfill
    \begin{minipage}[t]{0.48\textwidth}
        \centering
        \includegraphics[width=\textwidth]{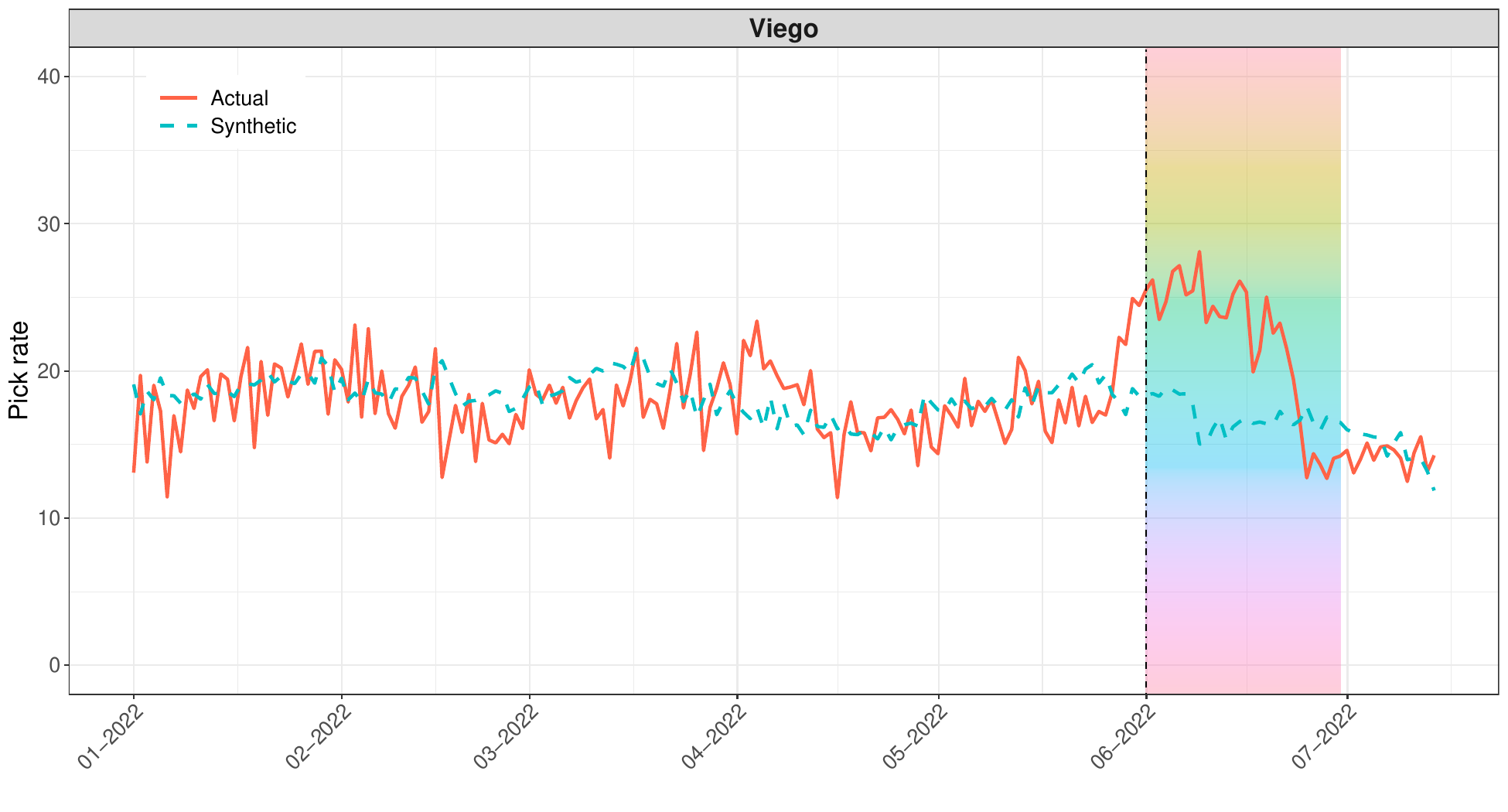}
    \end{minipage}

    \caption{Top four (excluding Graves) jungle characters' daily pick rates and synthetic control estimation results. The dashed vertical line denotes the day of disclosure, and the rainbow area highlights LGBT Pride Month.}
    \label{fig_placebo_jungle}
\end{figure}

\section{Mechanisms}
\label{app_mechanisms}

\setcounter{equation}{0}
\setcounter{table}{0}
\setcounter{figure}{0}

\noindent In Section \ref{sec_methdology_results}, we established evidence of a substantial negative impact of the coming-out event on players' revealed preferences for Graves. However, the players' decision to switch from this character may reflect factors other than a response to the disclosure itself. The objective of this section is to eliminate these alternative channels, thereby strengthening the interpretation that the observed behavior reflects a response to the disclosure rather than confounding gameplay or design factors.

\subsection{Graves' Strength}
\label{subapp_graves_performance}
\noindent Crucial for interpreting the estimated effects as responses to the disclosure is the fact that Graves' strength remained unaffected by the coming-out event, as any change in character relative strengths could explain why players' preferences shift away from Graves. 

To address this concern, we employ the synthetic control estimator described in Section \ref{app_methodology} to examine the potential impact of the coming-out event on Graves’ strength. We measure characters’ strength using daily win rates.

\begin{figure}[b!]
    \centering
    \includegraphics[scale=0.4]{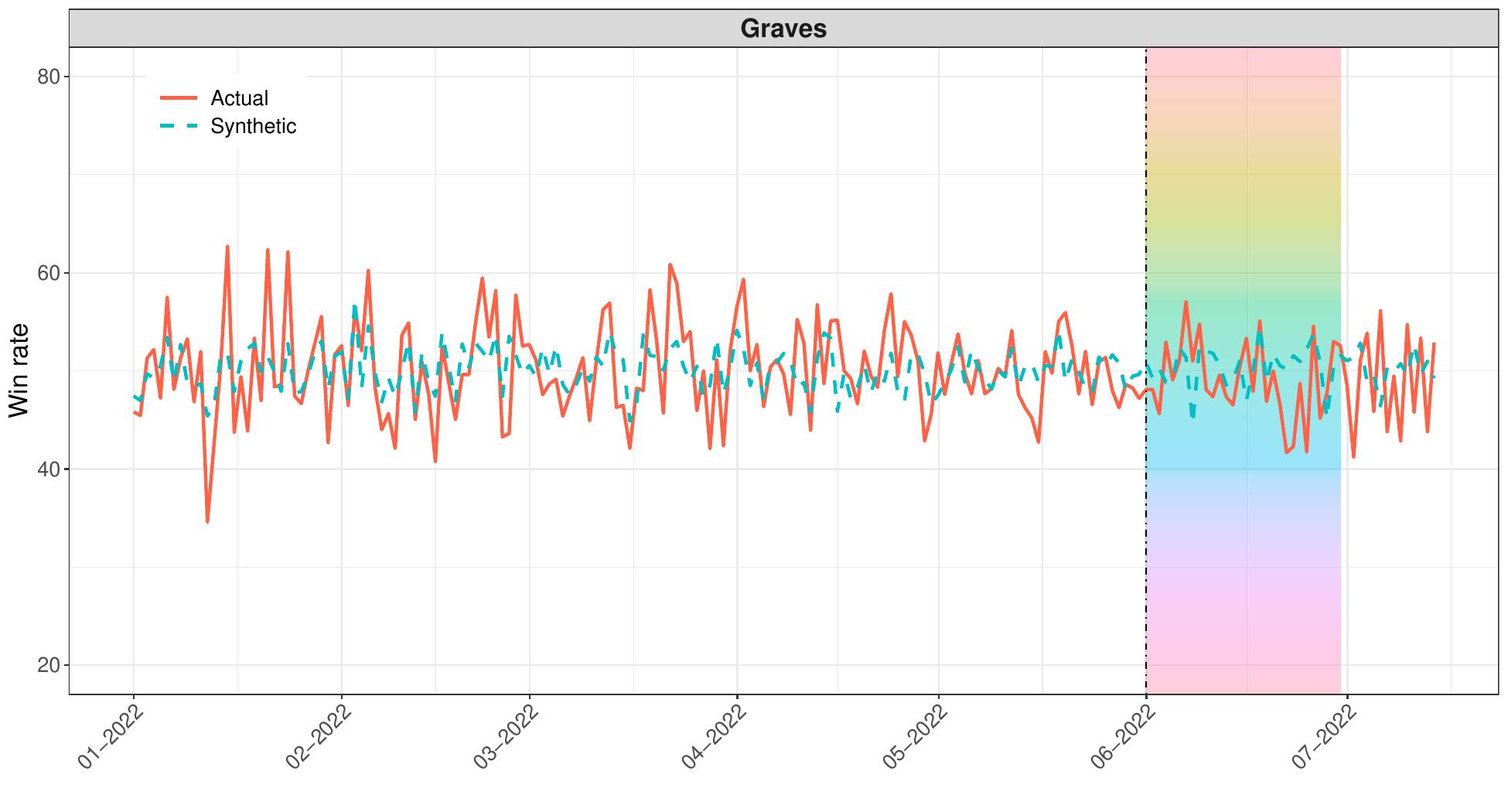}
    \caption{Graves' daily win rates and synthetic control estimation results. The dashed vertical line denotes the day of disclosure, and the rainbow area highlights LGBT Pride Month.}
    \label{fig_graves_performance_pooled}
\end{figure}

Figure \ref{fig_graves_performance_pooled} displays the results. Overall, our analysis reveals that the coming-out event had no impact on Graves' strength. Despite the actual series exhibiting daily fluctuations around the $50\%$ mark, the synthetic control estimator effectively captures its pre-treatment trend, showcasing its ability to predict the counterfactual trend. After the treatment date, the synthetic control estimator continues to align with Graves' win rate trend, confirming that the character's strength was unaffected by the disclosure. The average effect is estimated to be $-1.311$ percentage points (standard error: $4.545$), and the conventional $95\%$ confidence interval encompasses zero, indicating a failure to reject the null hypothesis of no effect. These findings demonstrate that Graves' strength remained unchanged during the coming-out event, dismissing the possibility of a shift in his strength as an explanation for the results of Section \ref{sec_methdology_results}.

Moreover, we note that players have real-time access to detailed information regarding characters' strengths, weaknesses, and overall performance, as numerous websites continuously provide updated data on characters' in-game statistics.\footnote{\ Examples of such websites include \href{https://lolalytics.com/lol/graves/build/}{https://lolalytics.com/lol/graves/build/} and \href{https://www.leagueofgraphs.com/champions/stats/graves}{https://www.leagueofgraphs.com/champions/stats/graves}.} Therefore, players were well-informed that no game-relevant skills or attributes were altered during the treatment period, and they could observe that Graves's strength remained consistent. These factors suggest that the negative impact of the coming-out event estimated in Section \ref{sec_methdology_results} is unlikely to be driven by actual or presumed changes in character relative strengths.

\subsection{Engagement and Performance Across Usage Groups}
\label{subapp_players_skills}

\noindent If the observed decline in Graves’ usage were driven by gameplay considerations rather than the disclosure itself, we would expect systematic differences in engagement or performance across players with different pre-disclosure usage patterns.

To address this concern, we compare engagement and performance outcomes across players with different pre-treatment preferences for Graves. We classify players into two groups based on their pre-disclosure usage of Graves: the first group comprises those who chose Graves in at least $5\%$ of their matches before the coming-out event (henceforth labeled as \textit{prior users}, $n = 953$), while the second group comprises the remaining players (henceforth labeled as \textit{non-prior users}, $n = 6468$). We then examine performance differences both within and between these groups before and after the treatment. 


\begin{figure}[b!]
    \centering
    \includegraphics[scale=0.5]{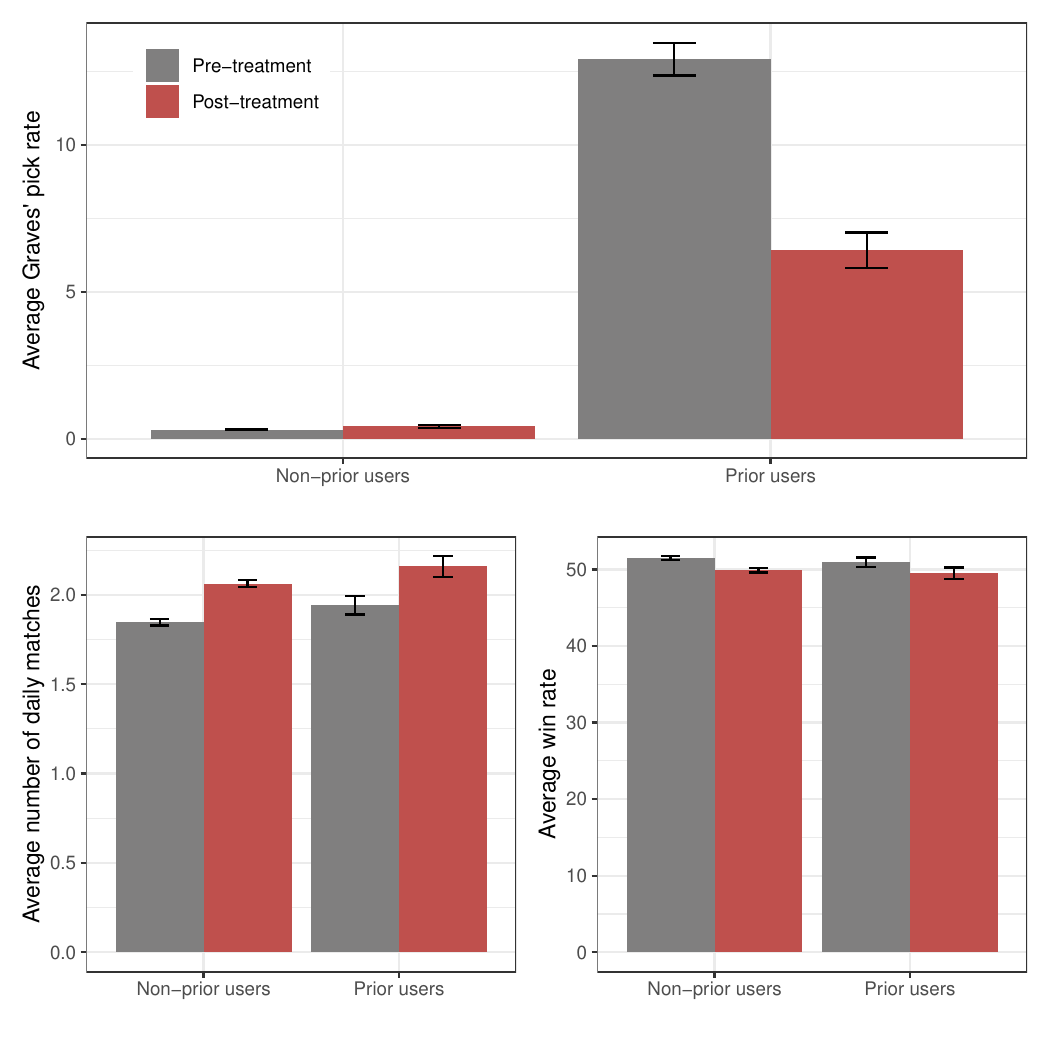}
    \caption{Players' average pick rates for Graves and performance measures. Players are divided into two groups based on their preferences for Graves before his disclosure. The panels display the average pick rates for Graves, number of daily matches, and win rate of each group before and after the coming-out event.}
    \label{fig_players_performance}
\end{figure}

The top panel of Figure \ref{fig_players_performance} displays the average pick rate for Graves among prior and non-prior users before and after the treatment. We observe a sharp decline in pick rates among prior users following the coming-out event, similar in size to the decrease shown in Figure \ref{fig_graves_pick_rates_pooled} and Table \ref{table_summary_stats}. Conversely, non-prior users exhibit a marginal increase in average pick rates post-treatment, although this increase is practically negligible.

In the remaining panels of Figure \ref{fig_players_performance}, we investigate whether prior and non-prior users differ in their in-game behavior. First, the bottom left panel shows the average number of daily matches played by each group. We observe similar engagement levels across groups, both before and after the treatment, indicating that prior and non-prior users tend to play a comparable number of matches per day. Notably, neither group appears to disengage from the game following Graves’ disclosure. On the contrary, the number of daily matches slightly increases post-treatment for both groups, suggesting that players are not leaving the game but instead redirecting their attention to other characters. Supporting this interpretation, Figure~\ref{fig_players_position} in Supplementary Information~\ref{app_figures_tables} shows no meaningful change in role preferences for either group following the event. This pattern is consistent with the idea that players face low switching costs and have access to suitable alternatives, enabling them to express their preferences without disrupting their gameplay. These results align with those displayed in Table \ref{table_summary_stats}.

Second, the bottom right panel displays the players' average win rates. We observe no substantial disparities within or between groups, indicating that differences in pre-disclosure usage of Graves are not associated with differential performance or skill. In particular, the decision to abandon Graves does not coincide with systematic changes in win rates. Overall, these findings provide no evidence that the post-disclosure decline in Graves’ usage is driven by gameplay considerations, thereby strengthening the interpretation that the observed behavior reflects a response to the disclosure rather than changes in in-game incentives.


\subsection{Players' Performance}
\label{subapp_players_performance}
\noindent To ensure that our estimates capture behavioral responses to the disclosure rather than strategic considerations, it is crucial to assess whether shifting away from Graves to other characters impacts players' performance. If there are performance costs, our estimates could be biased toward zero, as players might continue using Graves for strategic considerations. Moreover, if players switch characters primarily for convenience or strategic reasons, our analysis might capture a different phenomenon unrelated to the disclosure.

We employ difference-in-differences identification and estimation strategies to assess the impact of players abandoning Graves on their performance. We gauge players' performance by their daily win rate. Our analysis focuses on the $953$ prior-users of Section \ref{subsec_players_skills}, who are classified into treated or control groups based on their responses to Graves' disclosure. We consider two different definitions of the treatment, sorted by their intensity. In the first version, labeled \textit{moderate reduction}, we classify as treated those players who decreased their average pick rate for Graves following his disclosure by at most $10$ percentage points (the number of treated units is $522$). In the second version, labeled \textit{substantial reduction}, we classify as treated those players who reduced their average pick rate for Graves by $10$ to $100$ percentage points (the number of treated units is $256$).\footnote{\ Figure \ref{fig_players_graves_reduction} in Supplementary Information \ref{app_figures_tables} shows the distribution of changes in average Graves’ pick rates between the pre- and post-disclosure periods among prior users.} In both scenarios, the control group remains the same and consists of the $175$ prior users who did not reduce their average pick rate for Graves following his disclosure.

Under the standard assumptions of parallel trends and no anticipation \parencite[see, e.g.,][]{roth2023s}, we can identify the average treatment effect on the treated (ATT) using observable data. The parallel trend assumption posits that the performance of treated and untreated players would have evolved similarly if Graves' disclosure had not occurred. While we cannot formally test this assumption, the findings of Section \ref{subsec_graves_performance} and Section \ref{subsec_players_skills} provide substantial support for its plausibility.\footnote{\ Moreover, we demonstrate below the absence of pre-treatment differences in trends by reporting placebo estimates of the ATT that are not statistically different from zero. This is often viewed as a natural plausibility check, although even if pre-trends are perfectly parallel, this does not necessarily guarantee the satisfaction of the post-treatment parallel trends assumption \parencite[see, e.g.,][]{roth2023s}.} As for the no anticipation assumption, it stipulates that in the weeks preceding the disclosure, players' performance did not change due to the incoming Graves' disclosure. The plausibility of this assumption was thoroughly discussed in Section \ref{subsec_coming_out_event} and Section \ref{subsec_main_results}.

We implement the approach of \textcite{callaway2021difference} to target the ATT at a particular day $t > T^{pre}$:\footnote{\ The framework outlined in \textcite{callaway2021difference} is broader as it accommodates multiple groups defined by the timing of treatment reception. This enables the identification and estimation of the group-time ATTs, defined as $ ATT \left( g, t \right) := \EX \left[ Y_{i, t} \left( g \right) - Y_{i, t} \left( 0 \right) | G_g = 1 \right]$, where $G_g$ is a binary variable indicating treatment reception in period $g$. However, our data set features a single group, given that all treated players receive the treatment simultaneously (i.e., at Graves' disclosure date). This allows us to simplify notation and focus on the time ATTs in equation (\ref{equation_time_atts}) for the single group we observe.}

\begin{equation}
    ATT ( t ) := \EX [ Y_{i, t} \left( 1 \right) - Y_{i, t} \left( 0 \right) | D_i = 1 ],
    \label{equation_time_atts}
\end{equation}

\noindent where potential outcomes are defined as in Section \ref{app_methodology}, and $D_i$ is a binary variable indicating whether a player is treated or not. Under the assumptions of parallel trends and no anticipation, \textcite{callaway2021difference} show that $ATT ( t )$ can be identified by comparing the change in outcomes between the latest period before the coming-out event and day $t$ experienced by treated players to the change in outcomes experienced by control players.\footnote{\ Formally, \textcite{callaway2021difference} show that $ATT ( t ) = \EX [ Y_{i, t} - Y_{i, T^{pre}} | D_i = 1 ] - \EX [ Y_{i, t} - Y_{i, T^{pre}} | D_i = 0 ]$. Estimation is carried out by replacing expectations with their sample analogs.}

\begin{figure}[b!]
    \centering
    \includegraphics[scale=0.5]{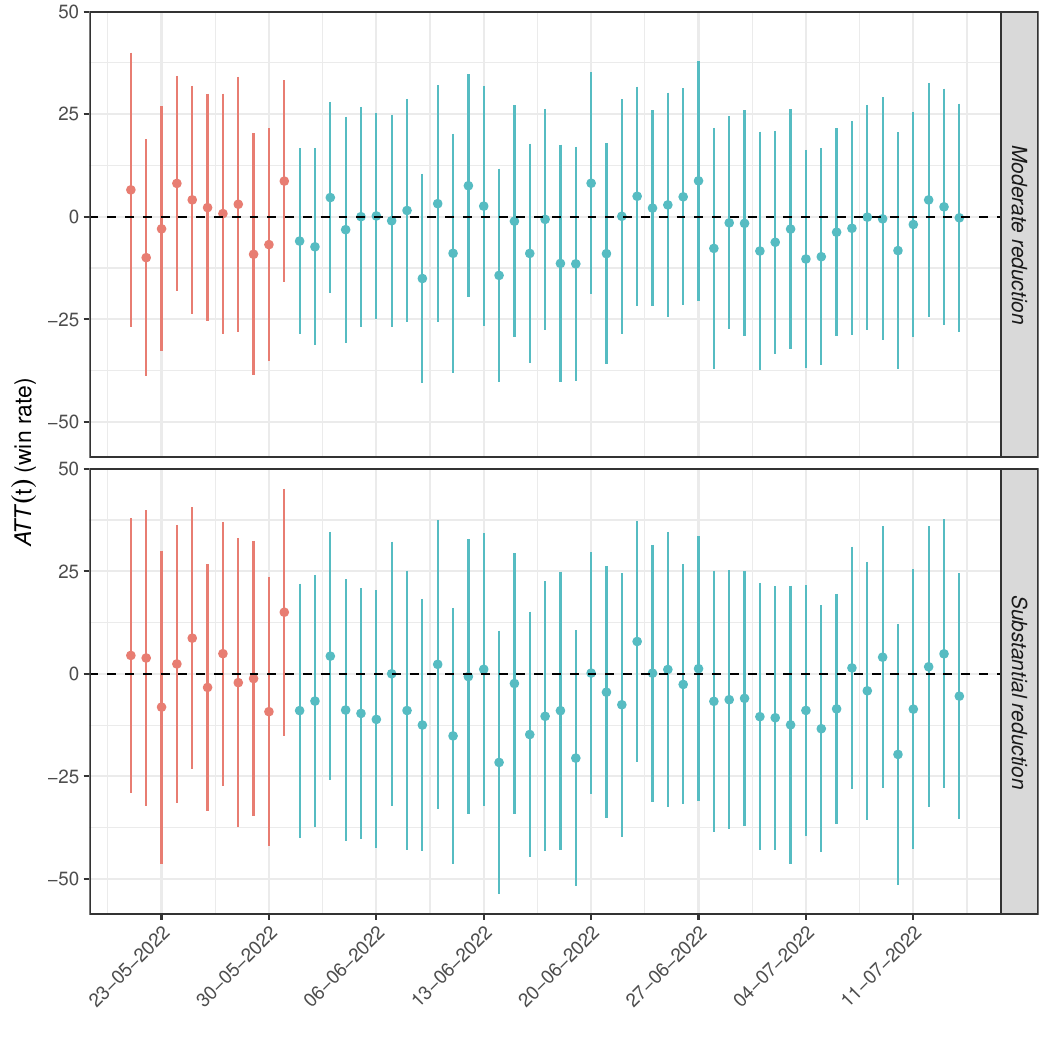}
    \caption{Point estimates and simultaneous $95\%$ confidence bands allowing for clustering at the player level for the $ATT ( t )$. Red lines refer to pre-treatment periods, while blue lines refer to post-treatment periods. Each row corresponds to a different version of the treatment.}
    \label{fig_did_player_performance_results_post}
\end{figure}

Figure \ref{fig_did_player_performance_results_post} displays the point estimates and simultaneous $95\%$ confidence bands for the $ATT ( t )$. Overall, we find that shifting away from Graves to other characters has no impact on players' performance. None of the estimated $ATT ( t )$ is statistically different from zero, suggesting that transitioning to other characters does not result in any performance-related consequences. This finding highlights that the decision to move away from Graves is not influenced by performance considerations. 

Figure \ref{fig_did_player_performance_results_post} further displays placebo estimates of the time ATTs for the ten days before the treatment.\footnote{\ Figure \ref{fig_did_player_performance_results_pre} in Supplementary Information \ref{app_figures_tables} displays the remaining estimated placebo $ATT ( t )$.} As explained above, these estimates are valuable for \open pre-testing" the credibility of the parallel trend assumption \parencite{callaway2021difference}. Notably, all placebo time ATTs in the pre-treatment periods are statistically insignificant, supporting the validity of the parallel trends assumption.

\subsection{New Substitute Character}
\label{subapp_belveth}
\noindent On June $9^{th}$, $2022$, a new character (\textit{Bel'Veth}) was released.\footnote{\ Figure \ref{fig_belveth_pick_rates} in Supplementary Information \ref{app_figures_tables} displays Bel'Veth's daily pick rates.} Since the primary position the character is designed for is the same position as Graves is designed for, it can be considered a close substitute. Therefore, players' decisions to switch away from Graves (after June $9^{th}$) might to some degree be driven by the desire to experiment with the new character and to explore potential competitive advantages, challenging social stigma as the primary explanation behind our main result.

\begin{figure}[b!]
    \centering
    \includegraphics[scale=0.45]{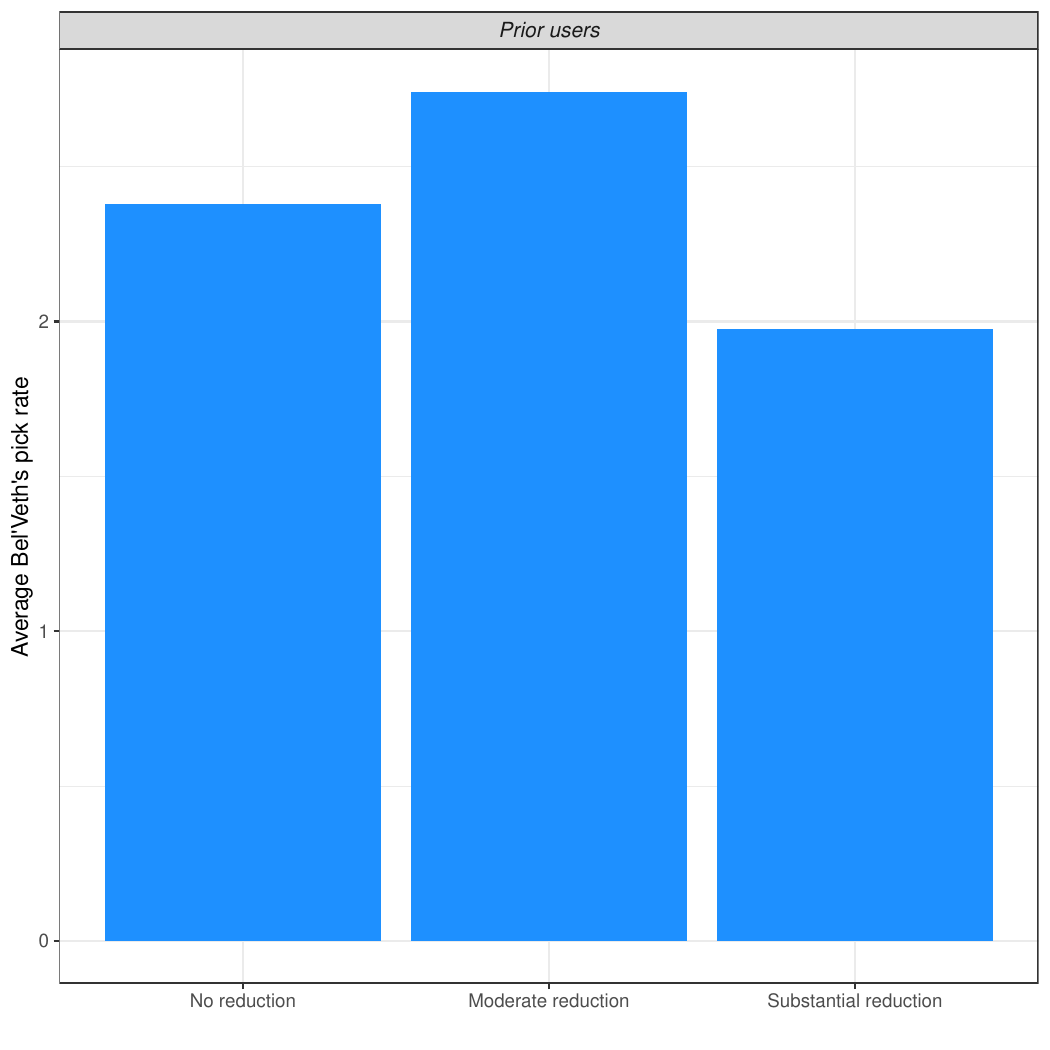}
    \caption{Average pick rate for Bel'Veth after its release among prior users.}
    \label{fig_belveth_barplot}
\end{figure}

If the release serves as the primary explanation for the observed drop in Graves pick rate we would expect a positive correlation between the size of the drop in players' Graves usage and players preferences for the new character. In Figure \ref{fig_belveth_barplot} we show that this is not the case. Figure \ref{fig_belveth_barplot} shows the average pick rate for Bel'Veth for the prior users of Section \ref{subsec_players_skills}, classified as in Section \ref{subsec_players_performance}. We find that those players who feature a substantial reduction in their Graves' pick rates following the coming-out event are less likely to pick Bel'Veth for their matches than players who feature a moderate reduction. Moreover, these players are even less likely to pick Bel'Veth for their matches than players who did not react at all to Graves' disclosure. We therefore argue that the release is unlikely to serve as the primary explanation for the observed main result of the paper.

\subsection{Coming Out versus LGBT Pride Month}
\label{subapp_coming_out_vs_lgbt_month}
\noindent As described in Section \ref{subsec_coming_out_event}, the disclosure of Graves' sexual orientation coincided with the start of LGBT Pride Month. This means that the coming-out event encompasses two \open simultaneous treatments" \parencite[see, e.g.,][]{roller2023differences}, namely the announcement of Graves' homosexuality and the introduction of visual and expressive elements in League of Legends that support the LGBT community. It is therefore plausible that the findings presented in Section \ref{sec_methdology_results} may, to some extent, be influenced by the presence of LGBT Pride Month, which might elicit negative reactions from certain players, leading them to shift their preferences away from LGB characters. While this alternative perspective does not undermine the validity of our identification strategy, it does raise questions about our interpretation of the estimated effects as solely stemming from Graves' disclosure.

In Supplementary Information \ref{app_anatomy_coming_out_event}, we introduce a theoretical framework that formalizes the existence of two simultaneous treatments and discuss the implications for interpretation. Additionally, we outline sufficient assumptions that enable us to separate the impacts of coming out and LGBT Pride Month on players’ preferences for Graves. Here, we provide the main intuitions behind our approach, directing the reader to the supplementary information for technical details.

To examine the potential impact of LGBT Pride Month on players' preferences for LGB characters, we leverage the existence in our data set of other four characters (\textit{Diana}, \textit{Leona}, \textit{Nami}, and \textit{Neeko}) already acknowledged as part of the LGB community before the coming-out event. These characters are subject only to a part of our treatment, specifically being part of the LGB community while LGBT Pride Month is ongoing, whereas Graves experiences both the disclosure of his sexual orientation and LGBT Pride Month. 

\begin{figure}[t!]
    \centering
    \includegraphics[scale=0.4]{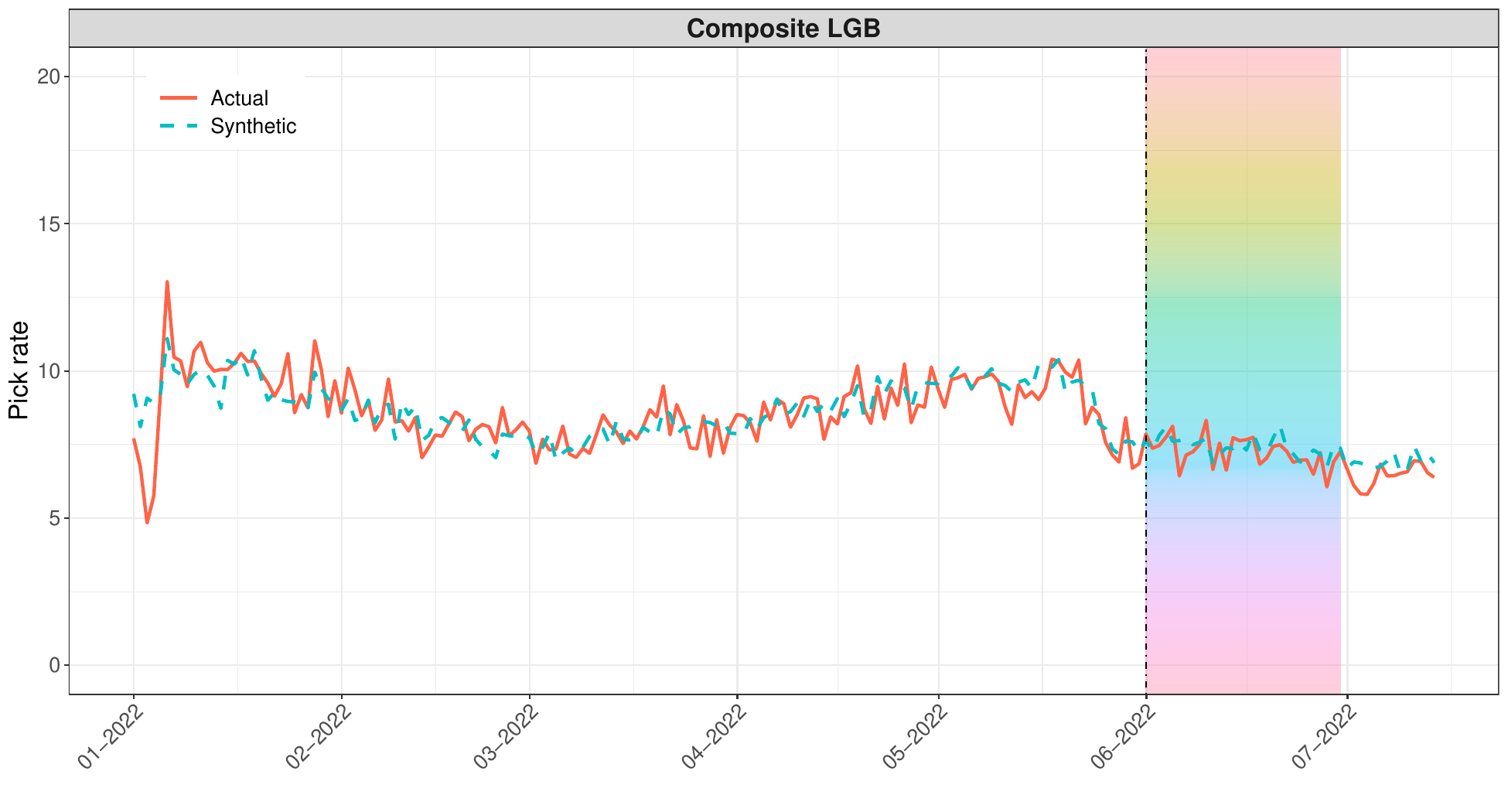}
    \caption{Composite LGB unit's daily pick rates and synthetic control estimation results. The dashed vertical line denotes the day of disclosure, and the rainbow area highlights LGBT Pride Month.}
    \label{fig_composite_lgb_pick_rates_pooled}
\end{figure}

We create a composite LGB unit by averaging the pick rates of Diana, Leona, Nami, and Neeko and employ the synthetic control estimator described in Section \ref{app_methodology} to estimate the effect of LGBT Pride Month on players’ preferences for LGB characters. Then, under the assumption that the influence of LGBT Pride Month is uniform across all LGB characters, we can compare the results with those obtained for Graves to separate the impacts of coming out and LGBT Pride Month on players' preferences for Graves. Intuitively, if the estimated impact of LGBT Pride Month on players' preferences for LGB characters is small relative to the estimated impact of the coming-out event on players' preferences for Graves, this suggests that the findings of Section \ref{sec_methdology_results} must be primarily attributed to Graves' disclosure.

Figure \ref{fig_composite_lgb_pick_rates_pooled} displays the actual and the synthetic pick rate series for the composite LGB unit. Overall, our analysis suggests that LGBT Pride Month had no impact on players' preferences for LGB characters. Before the treatment, the synthetic control estimator closely aligns with the actual series, providing support for the estimator's ability to predict the counterfactual series. After the treatment date, the synthetic control estimator continues to align with the actual series, confirming that the players' preferences for LGB characters were unaffected by LGBT Pride Month. The average effect is estimated to be $-0.672$ percentage points (standard error: $2.806$), and the conventional $95\%$ confidence interval encompasses zero, indicating a failure to reject the null hypothesis of no effect. Under the homogeneity assumption discussed above, these findings support the interpretation that the estimated effects presented in Section \ref{sec_methdology_results} are primarily driven by Graves' disclosure rather than being influenced by the broader context of LGBT Pride Month.

\section{Twisted Fate}
\label{app_why_not_tf}

\setcounter{equation}{0}
\setcounter{table}{0}
\setcounter{figure}{0}

\noindent In the League of Legends universe, Graves, described as a \open gruff-looking, broad-shouldered, middle-aged man," forms a criminal partnership with Twisted Fate, who is characterized as \open a tall, handsome male with tanned skin, trimmed beard, and long dark hair." Together, they engage in various illicit activities until, in the midst of a heist, Graves finds himself captured. This event leads Graves to perceive a betrayal by Twisted Fate, prompting a pursuit of revenge upon his escape. However, the duo ultimately decides to reconcile their differences and resumes their collaboration.\footnote{\ The complete background of Graves and Twisted Fate is available at \href{https://leagueoflegends.fandom.com/wiki/Graves}{https://leagueoflegends.fandom.com/wiki/Graves} and \href{https://leagueoflegends.fandom.com/wiki/Twisted_Fate}{https://leagueoflegends.fandom.com/wiki/Twisted\_Fate}.}

Notably, one of the initially considered narrative concepts for Graves and Twisted Fate involved them being married or ex-lovers. Although this particular aspect was discarded, the general narrative retained the notion of \open palpable sexual tension" between the two characters. 

The story unveiling Graves' sexual orientation (see Section \ref{subsec_coming_out_event}) also subtly hints at Twisted Fate's pansexuality, although this is not explicitly stated. Perhaps, the most notable passage that alludes to this is:

\begin{quote}
    \textit{No matter the size, shape, make, or model, none can resist the charms of Tobias Felix. I have conned hundreds—nay, thousands—of dew-eyed tourists across the whole of this vast and gullible land.} \hfill (Twisted Fate)
\end{quote}

We investigate whether this implied revelation has captured the players' attention in Figure \ref{fig_google_interest_time_graves_twisted_fate}, illustrating the Google search interest for the queries \textit{\open Graves gay"} and \textit{\open Twisted Fate gay."} Throughout $2022$, we observe approximately no interest in the latter query, with a small spike occurring during the week of the coming-out event, amounting to less than half of the spike associated with Graves. Furthermore, the search interest for Twisted Fate is always lower than that for Graves, suggesting the greater popularity of Graves among players.\footnote{\ This is also suggested by Graves's pick rates being approximately 3 to 4 times higher than those of Twisted Fate, as displayed in Figure \ref{fig_graves_twisted_fate_pick_rates}.} These results underscore the relatively low attention directed towards Twisted Fate from players, who were primarily focused on Graves and the explicit establishment of his sexual orientation. As a result, we concentrate our analysis on Graves and his disclosure for a more credible identification of the effects of coming out.

\begin{figure}[H]
    \centering
    \includegraphics[scale=0.4]{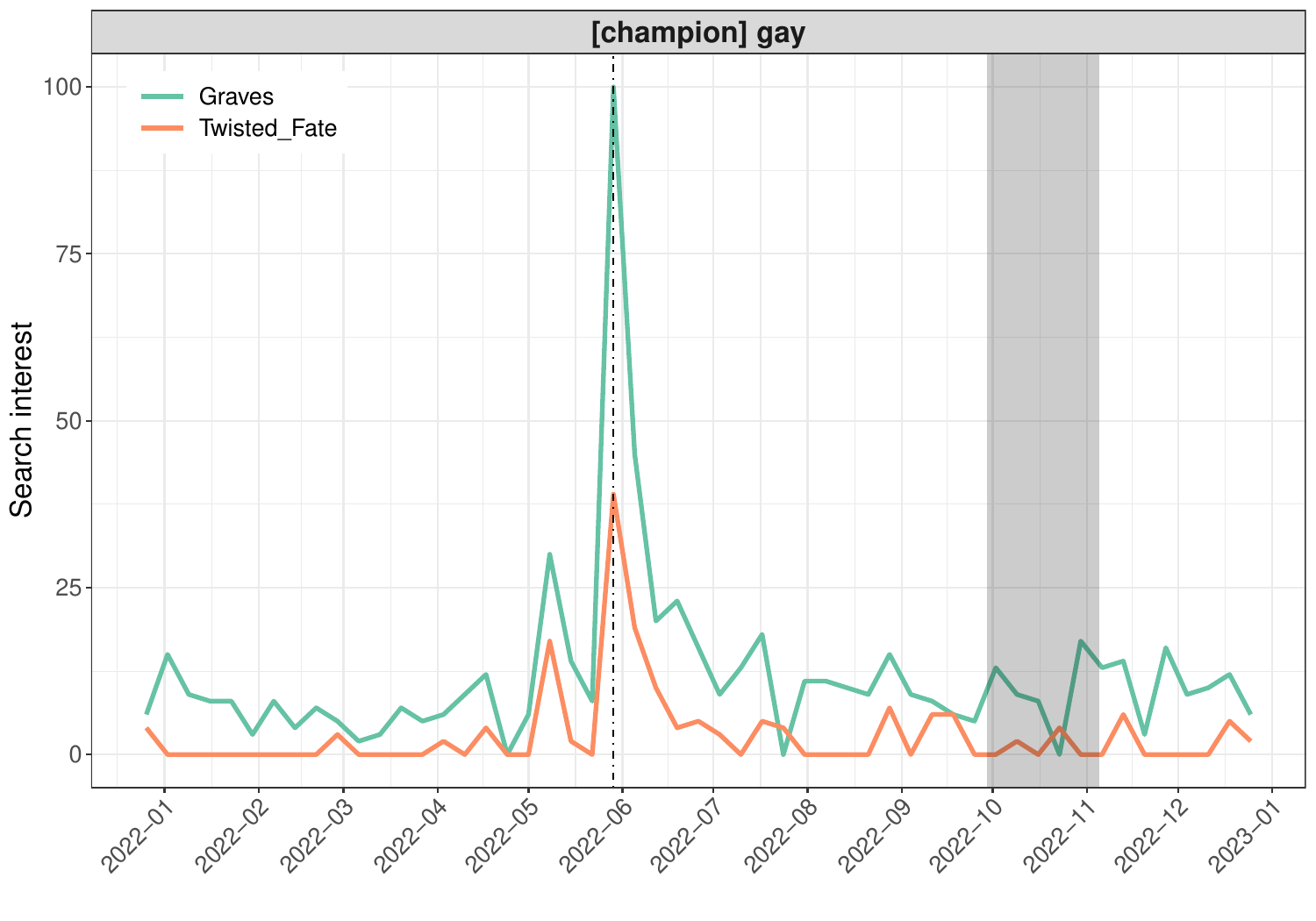}
    \caption{Google search interest over time for the queries \textit{\open Graves gay"} and \textit{\open Twisted Fate gay."} The dashed vertical line denotes the week of disclosure, and the shaded area highlights the League of Legends World Championship.}
    \label{fig_google_interest_time_graves_twisted_fate}
\end{figure}  

\begin{figure}[H]
    \centering
    \includegraphics[scale=0.4]{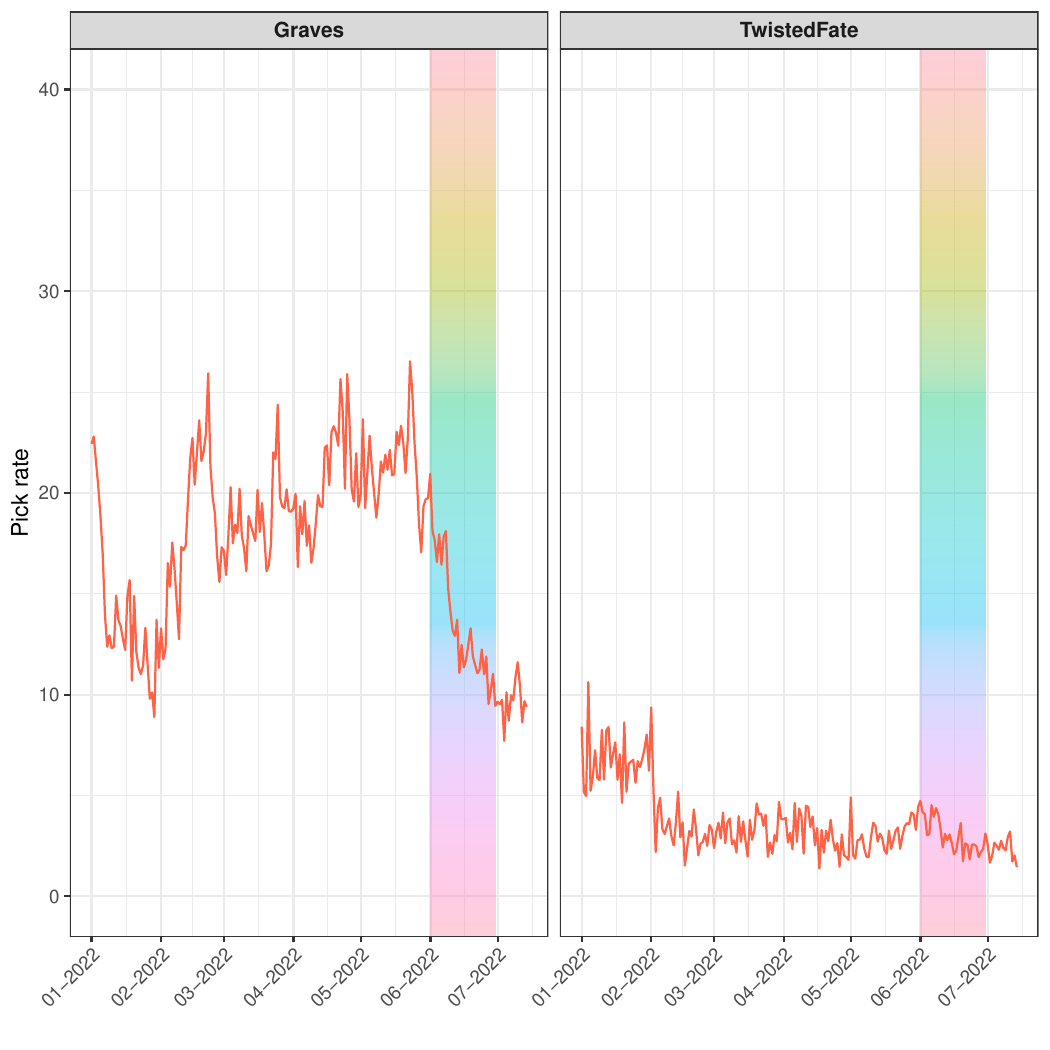}
    \caption{Graves and Twisted Fate's daily pick rates. The rainbow area highlights LGBT Pride Month.}
    \label{fig_graves_twisted_fate_pick_rates}
\end{figure}

\section{Anatomy of the Coming-Out Event}
\label{app_anatomy_coming_out_event}

\setcounter{equation}{0}
\setcounter{table}{0}
\setcounter{figure}{0}

\noindent In this section, we discuss how the existence of two treatments---the disclosure of Graves' sexual orientation and the start of LGBT Pride Month---occurring at the same time may affect the interpretation of the main findings of Section \ref{subsec_main_results}. The notation follows that used in Section \ref{app_methodology}. The results of the analysis are detailed in Section \ref{subsec_coming_out_vs_lgbt_month}.

In the next subsection, we introduce the framework that formalizes the existence of two \open simultaneous treatments." We then outline sufficient assumptions that enable us to separate the impacts of coming out and LGBT Pride Month on players' preferences for Graves.

\subsection{Simultaneous Treatments}
\label{subapp_simultaneous_treatments}
\noindent As described in Section \ref{subsec_coming_out_event}, the disclosure of Graves' sexual orientation coincided with the start of LGBT Pride Month. This means that the coming-out event encompasses two treatments occurring at the same time, namely the announcement of Graves' homosexuality and the introduction of visual and expressive elements in League of Legends that support the LGBT community.\footnote{\ See, e.g., \textcite{roller2023differences} for a discussion on \open simultaneous treatments" and methodologies for disentangling their effects under a Difference-in-Differences identification strategy.} 

We recognize the potential influence of LGBT Pride Month on players' preferences for characters by introducing the binary variable $L_i \in \left\{ 0, 1 \right\}$ to represent character $i$'s inclusion in the LGB community no later than $T^{pre} + 1$. Consequently, we observe three distinct groups of units: the first group includes only Graves, with $C_i = L_i = 1$; the second group includes only Diana, Leona, Nami, and Neeko, with $C_i = 0$ and $L_i = 1$; and the third group includes all other characters, with $C_i = L_i = 0$.\footnote{\ Neglecting the presence of two simultaneous treatments and treating them as a single treatment does not invalidate the results of Section \ref{subsec_main_results}. It primarily affects their interpretation, which, without further investigation, could only be attributed to the combined effects of simultaneously receiving both treatments $C_i$ and $L_i$ - referred to as the \textit{coming-out event} in the main body of the paper.} 

To explicitly account for the influence of the two treatments $C_i$ and $L_i$, we define the potential pick rates as $Y_{i, t}^{c, l}$. Then, for each period $t > T^{pre}$, the effect of the coming-out event on players' preferences for Graves in (\ref{equation_estimand}) corresponds to:

\begin{equation}
    \tau_t = Y_{1, t}^{1, 1} - Y_{1, t}^{0, 0}
    \label{equation_effect_bundled}
\end{equation}

\noindent Equation (\ref{equation_effect_bundled}) shows why we need to be cautious in interpreting the estimated effects of Section \ref{subsec_main_results} as solely stemming from the disclosure of Graves' sexual orientation. Under an extended version of the SUTVA assumption (see Section \ref{subapp_separating_simultaneous_treatments}), we observe $Y_{1, t} = Y_{1, t}^{1, 1}$ for all $t > T^{pre}$, and the estimator in (\ref{equation_estimator_counterfactual}) effectively targets the counterfactual series $Y_{1, t}^{0, 0}$. Consequently, the estimated effects presented in Section \ref{subsec_main_results} encompass the combined impacts of both disclosing Graves' sexual orientation and his affiliation with the LGB community during LGBT Pride Month. This can be formalized as follows:

\begin{equation}
    \begin{split}
        \tau_t & = Y_{1, t}^{1, 1} - Y_{1, t}^{0, 0} \\
        & = \underbrace{\left[ Y_{1, t}^{1, 1} - Y_{1, t}^{0, 1} \right]}_{:= \tau_{t}^C} + \underbrace{\left[ Y_{1, t}^{0, 1} - Y_{1, t}^{0, 0} \right]}_{:= \tau_{t}^L} \\
    \end{split}
    \label{equation_decomposition}
\end{equation}

\noindent with $\tau_{t}^C$ representing the effects of the disclosure on players' preferences for Graves, and $\tau_{t}^L$ representing the effects of being part of the LGB community during LGBT Pride Month on players' preferences for Graves.

\subsection{Separating Simultaneous Treatment Effects}
\label{subapp_separating_simultaneous_treatments}
\noindent The decomposition in (\ref{equation_decomposition}) offers a strategy to disentangle the effects of the two treatments $C_i$ and $L_i$ for Graves. If we can successfully estimate the two counterfactual series $Y_{1, t}^{0, 1}$ and $Y_{1, t}^{0, 0}$, then we would be able to construct estimates $\hat{\tau}_{t}^C = Y_{1, t}^{1, 1} - \widehat{Y}_{1, t}^{0, 1}$ and $\hat{\tau}_{t}^L = \widehat{Y}_{1, t}^{0, 1} - \widehat{Y}_{1, t}^{0, 0}$ of $\tau_{t}^C$ and $\tau_{t}^L$, respectively. This would allow us to quantify the extent to which LGBT Pride Month drives the main findings of Section \ref{subsec_main_results}.

To this end, we assume an extended version of the SUTVA that accommodates the existence of two different treatments.

\begin{assumption}
    (SUTVA): $Y_{i, t} = Y_{i, t}^{1, 1} C_i L_i + Y_{i, t}^{0, 1} \left[ 1 - C_i \right] L_i + Y_{i, t}^{0, 0} \left[ 1 - C_i \right] \left[ 1 - L_i \right]$
    \label{assumption_sutva}
\end{assumption}

\noindent Under Assumption \ref{assumption_sutva}, we can estimate the counterfactual series $Y_{1, t}^{0, 0}$ by constructing a synthetic control unit that approximates the pick rates of Graves before the coming-out event as in Section \ref{app_methodology}. Thus, as shown in (\ref{equation_decomposition}), the challenge in disentangling our causal effects of interest is to estimate $Y_{1, t}^{0, 1}$ for $t > T^{pre}$, i.e., how Graves’ pick rates would have evolved if Graves were already part of the LGB community prior to the 2022 LGBT Pride Month. 

Having a sufficient number of LGB characters other than Graves (that is, sufficient units such as $C_i = 0$ and $L_i = 1$) would enable us to estimate the counterfactual series $Y_{1, t}^{0, 1}$ through standard synthetic control methods. However, since we only have four such characters in our data set, this approach is infeasible. 

One way out is to estimate the impact of LGBT Pride Month on players' preferences for LGB characters and compare the results with those obtained for Graves. If the influence of LGBT Pride Month is uniform across all LGB characters, this strategy provides insight into the role of LGBT Pride Month in driving the main findings of Section \ref{subsec_main_results}. 

To achieve this, we create a composite LGB unit by averaging the pick rates of all characters such as $C_i = 0$ and $L_i = 1$ (namely, Diana, Leona, Nami, and Neeko), denoting this unit as character $j$ without loss of generality. Then, for each period $t > T^{pre}$, we define the effect of LGBT Pride Month on players' preferences for LGB characters as the difference in character $j$'s potential pick rates at time $t$:

\begin{equation}
    \gamma_t^L := Y_{j, t}^{0, 1} - Y_{j, t}^{0, 0}
    \label{equation_effect_lgb}
\end{equation}

\noindent Under Assumption (\ref{assumption_sutva}), we observe $Y_{j, t} = Y_{j, t}^{0, 1}$ for all $t > T^{pre}$, and we can estimate the counterfactual series $Y_{j, t}^{0, 0}$ by constructing a synthetic control unit that approximates the pick rates of character $j$ before the beginning of the 2022 LGBT Pride Month. We can then estimate $\gamma_t^L$ by computing the differences between character $j$’s observed pick rates and the synthetic counterfactual for all $t > T^{pre}$:

\begin{equation}
    \hat{\gamma}_t^L = Y_{j, t}^{0, 1} - \widehat{Y}_{j, t}^{0, 0}
    \label{equation_estimated_effect_lgb}
\end{equation}

Finally, we introduce a homogeneity assumption that leverages the estimates $\hat{\gamma}_t^L$ to provide an interpretation for the estimates $\hat{\tau}_t$ presented in Section \ref{subsec_main_results}:

\begin{assumption}
    (Effect Homogeneity): $\tau_{t}^L = \gamma_{t}^L$ for all $t > T^{pre}$.
    \label{assumption_same_effects}
\end{assumption}

\noindent Under Assumption \ref{assumption_same_effects}, the relationship $\tau_t^C = \tau_t - \gamma_t^L$ holds. Thus, if the estimated effects of LGBT Pride Month on players' preferences for LGB characters are small relative to the estimated effects of the coming-out event on players' preferences for Graves, this suggests that the findings of Section \ref{subsec_main_results} must be primarily attributed to Graves' disclosure.

\end{appendices}


\onehalfspacing
\newrefcontext[sorting = nty]
\newpage
\printbibliography

@article{rubin1974estimating,
  title={Estimating causal effects of treatments in randomized and nonrandomized studies},
  author={Rubin, Donald B.},
  journal={Journal of Educational Psychology},
  volume={66},
  number={5},
  pages={688–-701},
  year={1974}
}

@article{ayres1995race,
  title={Race and gender discrimination in bargaining for a new car},
  author={Ayres, Ian and Siegelman, Peter},
  journal={American Economic Review},
  pages={304--321},
  year={1995},
  publisher={JSTOR}
}

@article{badgett1995wage,
  title={The wage effects of sexual orientation discrimination},
  author={Badgett, MV Lee},
  journal={ILR Review},
  year={1995},
  volume={48},
  number={4},
  pages={726--739}
}

@article{meyer1995minority,
  title={Minority stress and mental health in gay men},
  author={Meyer, Ilan H},
  journal={Journal of Health and Social Behavior},
  pages={38--56},
  year={1995},
  publisher={JSTOR}
}

@article{akerlof2000economics,
  title={Economics and identity},
  author={Akerlof, George A and Kranton, Rachel E},
  journal={The Quarterly Journal of Economics},
  volume={115},
  number={3},
  pages={715--753},
  year={2000},
  publisher={MIT Press}
}

@article{correll2002police,
  title={The police officer's dilemma: using ethnicity to disambiguate potentially threatening individuals.},
  author={Correll, Joshua and Park, Bernadette and Judd, Charles M and Wittenbrink, Bernd},
  journal={Journal of personality and social psychology},
  volume={83},
  number={6},
  pages={1314},
  year={2002},
  publisher={American Psychological Association}
}

@article{abadie2003economic,
  title={The economic costs of conflict: A case study of the Basque Country},
  author={Abadie, Alberto and Gardeazabal, Javier},
  journal={American Economic Review},
  volume={93},
  number={1},
  pages={113--132},
  year={2003},
  publisher={American Economic Association}
}

@article{weichselbaumer2003sexual,
  title={Sexual orientation discrimination in hiring},
  author={Weichselbaumer, Doris},
  journal={Labour economics},
  volume={10},
  number={6},
  pages={629--642},
  year={2003},
  publisher={Elsevier}
}

@article{meyer2003prejudice,
  title={Prejudice, social stress, and mental health in lesbian, gay, and bisexual populations: conceptual issues and research evidence.},
  author={Meyer, Ilan H},
  journal={Psychological Bulletin},
  year={2003},
  volume={129},
  number={5},
  pages={674--697},
}

@article{drydakis2009sexual,
  title={Sexual orientation discrimination in the labour market},
  author={Drydakis, Nick},
  journal={Labour Economics},
  volume={16},
  number={4},
  pages={364--372},
  year={2009},
  publisher={Elsevier}
}

@article{abadie2010synthetic,
  title={Synthetic control methods for comparative case studies: Estimating the effect of California’s tobacco control program},
  author={Abadie, Alberto and Diamond, Alexis and Hainmueller, Jens},
  journal={Journal of the American Statistical Association},
  volume={105},
  number={490},
  pages={493--505},
  year={2010},
  publisher={Taylor \& Francis}
}

@article{oreopoulos2011,
  Author = {Oreopoulos, Philip},
  Title = {Why Do Skilled Immigrants Struggle in the Labor Market? A Field Experiment with Thirteen Thousand Resumes},
  Journal = {American Economic Journal: Economic Policy},
  Volume = {3},
  Number = {4},
  Year = {2011},
  Pages = {148-71}
}

@article{Tilcsik2011,
  author = {Tilcsik, Andr\'{a}s},
  title = {Pride and Prejudice: Employment Discrimination against Openly Gay Men in the United States},
  journal = {American Journal of Sociology},
  volume = {117},
  number = {2},
  pages = {586-626},
  year = {2011},
}

@article{aguiar2012recent,
  title={Recent developments in the economics of time use},
  author={Aguiar, Mark and Hurst, Erik and Karabarbounis, Loukas},
  journal={Annual Review of Economics},
  volume={4},
  number={1},
  pages={373--397},
  year={2012},
  publisher={Annual Reviews}
}

@article{kuhn2013gender,
  title={Gender discrimination in job ads: Evidence from china},
  author={Kuhn, Peter and Shen, Kailing},
  journal={The Quarterly Journal of Economics},
  volume={128},
  number={1},
  pages={287--336},
  year={2013}
}

@article{ahmed2013gay,
  title={Are gay men and lesbians discriminated against in the hiring process?},
  author={Ahmed, Ali M and Andersson, Lina and Hammarstedt, Mats},
  journal={Southern Economic Journal},
  volume={79},
  number={3},
  pages={565--585},
  year={2013},
  publisher={Wiley Online Library}
}

@article{drydakis2014sexual,
  title={Sexual orientation discrimination in the Cypriot labour market. Distastes or uncertainty?},
  author={Drydakis, Nick},
  journal={International Journal of Manpower},
  volume={35},
  number={5},
  pages={720--744},
  year={2014}
}

@article{abadie2015comparative,
  title={Comparative politics and the synthetic control method},
  author={Abadie, Alberto and Diamond, Alexis and Hainmueller, Jens},
  journal={American Journal of Political Science},
  volume={59},
  number={2},
  pages={495--510},
  year={2015},
  publisher={Wiley Online Library}
}

@article{borowiecki2015video,
  title={Video games playing: A substitute for cultural consumptions?},
  author={Borowiecki, Karol J and Prieto-Rodriguez, Juan},
  journal={Journal of Cultural Economics},
  volume={39},
  number={3},
  pages={239--258},
  year={2015},
  publisher={Springer}
}

@article{boden2016bullying,
  title={Bullying victimization in adolescence and psychotic symptomatology in adulthood: Evidence from a 35-year study},
  author={Boden, Joseph M and van Stockum, Saskia and Horwood, L John and Fergusson, David M},
  journal={Psychological Medicine},
  volume={46},
  number={6},
  pages={1311--1320},
  year={2016},
  publisher={Cambridge University Press}
}

@article{broockman2016,
author = {David Broockman  and Joshua Kalla },
title = {Durably reducing transphobia: A field experiment on door-to-door canvassing},
journal = {Science},
volume = {352},
number = {6282},
pages = {220-224},
year = {2016}
}

@incollection{borowiecki2017cultural,
  title={The cultural value and variety of playing video games},
  author={Borowiecki, Karol J and Prieto-Rodriguez, Juan},
  booktitle={Enhancing Participation in the Arts in the EU: Challenges and Methods},
  pages={323--336},
  year={2017},
  publisher={Springer}
}

@article{bharadwaj2017mental,
  title={Mental health stigma},
  author={Bharadwaj, Prashant and Pai, Mallesh M and Suziedelyte, Agne},
  journal={Economics Letters},
  volume={159},
  pages={57--60},
  year={2017},
  publisher={Elsevier}
}

@article{carpenter2017does,
  title={Does it get better? Recent estimates of sexual orientation and earnings in the United States},
  author={Carpenter, Christopher S and Eppink, Samuel T},
  journal={Southern Economic Journal},
  year={2017},
  volume={84},
  number={2},
  pages={426--441}
}

@article{coffman2017size,
  title={The size of the LGBT population and the magnitude of antigay sentiment are substantially underestimated},
  author={Coffman, Katherine B and Coffman, Lucas C and Ericson, Keith M Marzilli},
  journal={Management Science},
  volume={63},
  number={10},
  pages={3168--3186},
  year={2017},
  publisher={INFORMS}
}

@article{pope2018awareness,
author = {Pope, Devin G. and Price, Joseph and Wolfers, Justin},
title = {Awareness Reduces Racial Bias},
journal = {Management Science},
volume = {64},
number = {11},
pages = {4988-4995},
year = {2018}
}

@article{Pope2011,
Author = {Pope, Devin G. and Schweitzer, Maurice E.},
Title = {Is Tiger Woods Loss Averse? Persistent Bias in the Face of Experience, Competition, and High Stakes},
Journal = {American Economic Review},
Volume = {101},
Number = {1},
Year = {2011},
Pages = {129–57}}

@article{Price2010,
    author = {Price, Joseph and Wolfers, Justin},
    title = {Racial Discrimination Among NBA Referees},
    journal = {The Quarterly Journal of Economics},
    volume = {125},
    number = {4},
    pages = {1859-1887},
    year = {2010},
    month = {11}
}

@article{Brox2025,
  author={Brox, Enzo and Goller, Daniel},
  title={Tournaments, Contestant Heterogeneity and Performance},
  journal={Journal of Political Economy Microeconomics},
  year={2025}
}

@article{borowiecki2018did,
  title={Did you really take a hit? Understanding how video games playing affects individuals},
  author={Borowiecki, Karol J and Bakhshi, Hasan},
  journal={Research in Economics},
  volume={72},
  number={2},
  pages={313--326},
  year={2018},
  publisher={Elsevier}
}

@article{burn2018not,
  title={Not all laws are created equal: Legal differences in state non-discrimination laws and the impact of LGBT employment protections},
  author={Burn, Ian},
  year={2018},
  journal={Journal of Labor Research},
  volume={39},
  number={4},
  pages={462--497}
}

@article{geijtenbeek2018penalty,
  title = {Is there a penalty for registered women? Is there a premium for registered men? Evidence from a sample of transsexual workers},
  journal = {European Economic Review},
  volume = {109},
  pages = {334-347},
  year = {2018},
  author = {Lydia Geijtenbeek and Erik Plug}
}

@article{parshakov2018diversity,
  title={Is diversity good or bad? Evidence from eSports teams analysis},
  author={Parshakov, Petr and Coates, Dennis and Zavertiaeva, Marina},
  journal={Applied Economics},
  volume={50},
  number={47},
  pages={5064--5075},
  year={2018},
  publisher={Taylor \& Francis}
}

@article{ofosu2019same,
  title={Same-sex marriage legalization associated with reduced implicit and explicit antigay bias},
  author={Ofosu, Eugene K and Chambers, Michelle K and Chen, Jacqueline M and Hehman, Eric},
  journal={Proceedings of the National Academy of Sciences},
  volume={116},
  number={18},
  pages={8846--8851},
  year={2019},
  publisher={National Acad Sciences}
}

@inproceedings{wang2019personality,
  title={Personality and behavior in role-based online games},
  author={Wang, Zhao and Sapienza, Anna and Culotta, Aron and Ferrara, Emilio},
  booktitle={2019 IEEE Conference on Games (CoG)},
  pages={1--8},
  year={2019},
  organization={IEEE}
}

@article{sansone2019pink,
  title = {Pink work: Same-sex marriage, employment and discrimination},
  journal = {Journal of Public Economics},
  volume = {180},
  pages = {104086},
  year = {2019},
  author = {Dario Sansone}
}

@article{burn2020relationship,
  title={The relationship between prejudice and wage penalties for gay men in the United States},
  author={Burn, Ian},
  journal={ILR Review},
  volume={73},
  number={3},
  pages={650--675},
  year={2020},
  publisher={SAGE Publications Sage CA: Los Angeles, CA}
}

@article{pachankis2020sexual,
  title={Sexual orientation concealment and mental health: A conceptual and meta-analytic review.},
  author={Pachankis, John E and Mahon, Conor P and Jackson, Skyler D and Fetzner, Benjamin K and Br{\"a}nstr{\"o}m, Richard},
  journal={Psychological Bulletin},
  year={2020},
  volume={146},
  number={10},
  pages={831--871}
}

@article{tang2020investigating,
  title={Investigating sexual harassment in online video games: How personality and context factors are related to toxic sexual behaviors against fellow players},
  author={Tang, Wai Yen and Reer, Felix and Quandt, Thorsten},
  journal={Aggressive Behavior},
  volume={46},
  number={1},
  pages={127--135},
  year={2020},
  publisher={Wiley Online Library}
}

@article{callaway2021difference,
  title={Difference-in-differences with multiple time periods},
  author={Callaway, Brantly and Sant’Anna, Pedro HC},
  journal={Journal of Econometrics},
  volume={225},
  number={2},
  pages={200--230},
  year={2021},
  publisher={Elsevier}
}

@article{abadie2021using,
  title={Using synthetic controls: Feasibility, data requirements, and methodological aspects},
  author={Abadie, Alberto},
  journal={Journal of Economic Literature},
  year={2021},
  volume={59},
  number={2},
  pages={391--425}
}

@article{arkhangelsky2021synthetic,
  title={Synthetic difference-in-differences},
  author={Arkhangelsky, Dmitry and Athey, Susan and Hirshberg, David A and Imbens, Guido W and Wager, Stefan},
  journal={American Economic Review},
  volume={111},
  number={12},
  pages={4088--4118},
  year={2021}
}

@article{badgett2021lgbtq,
  title={LGBTQ economics},
  author={Badgett, MV and Carpenter, Christopher S and Sansone, Dario},
  journal={Journal of Economic Perspectives},
  volume={35},
  number={2},
  pages={141--170},
  year={2021}
}

@article{martell2021labor,
  title={Labor market differentials estimated with researcher-inferred and self-identified sexual orientation},
  author={Martell, Michael E},
  journal={Economics Letters},
  volume={205},
  pages={109959},
  year={2021}
}

@article{carpenter2022economic,
  title={Economic outcomes for transgender people and other gender minorities in the United States: First estimates from a nationally representative sample},
  author={Carpenter, Christopher S and Lee, Maxine J and Nettuno, Laura},
  journal={Southern Economic Journal},
  volume={89},
  number={2},
  pages={280--304},
  year={2022}
}

@article{kudashvili2022minorities,
  title={Minorities’ strategic response to discrimination: Experimental evidence},
  author={Kudashvili, Nikoloz and Lergetporer, Philipp},
  journal={Journal of Public Economics},
  volume={208},
  pages={104630},
  year={2022}
}

@article{deal2022bound,
  title={Bound by Bostock: The effect of policies on attitudes},
  author={Deal, Cameron},
  journal={Economics Letters},
  volume={217},
  pages={110656},
  year={2022},
  publisher={Elsevier}
}

@article{arnold2022measuring,
  title={Measuring racial discrimination in bail decisions},
  author={Arnold, David and Dobbie, Will and Hull, Peter},
  journal={American Economic Review},
  volume={112},
  number={9},
  pages={2992--3038},
  year={2022}
}

@article{parshakov2023lgbtq,
  title={Do LGBTQ-Supportive Corporate Policies Affect Consumer Behavior? Evidence from the Video Game Industry},
  author={Parshakov, Petr and Naidenova, Iuliia and Gomez-Gonzalez, Carlos and Nesseler, Cornel},
  journal={Journal of business ethics},
  volume={187},
  number={3},
  pages={421--432},
  year={2023}
}

@article{oh2023does,
  title={Does identity affect labor supply?},
  author={Oh, Suanna},
  journal={American Economic Review},
  volume={113},
  number={8},
  pages={2055--2083},
  year={2023}
}

@inproceedings{deal2023heterogeneity,
  title={Heterogeneity in Attitude Responses: Evidence from Bostock v. Clayton County},
  author={Deal, Cameron},
  booktitle={AEA Papers and Proceedings},
  volume={113},
  pages={546--550},
  year={2023},
  organization={American Economic Association 2014 Broadway, Suite 305, Nashville, TN 37203}
}

@article{aksoy2023sexual,
  title={Sexual identity, gender, and anticipated discrimination in prosocial behavior},
  author={Aksoy, Billur and Chadd, Ian and Koh, Boon Han},
  journal={European Economic Review},
  volume={154},
  pages={104427},
  year={2023},
  publisher={Elsevier}
}

@article{roth2023s,
  title = {What’s trending in difference-in-differences? A synthesis of the recent econometrics literature},
  journal = {Journal of Econometrics},
  volume = {235},
  number = {2},
  pages = {2218-2244},
  year = {2023},
  author = {Jonathan Roth and Pedro H.C. Sant’Anna and Alyssa Bilinski and John Poe}
}

@article{dell2023super,
  title={Super Mario Meets AI: Experimental Effects of Automation and Skills on Team Performance and Coordination},
  author={Dell'Acqua, Fabrizio and Kogut, Bruce and Perkowski, Patryk},
  journal={Review of Economics and Statistics},
  pages={Forthcoming},
  year={2023},
  publisher={MIT Press One Rogers Street, Cambridge, MA 02142-1209, USA journals-info~…}
}

@article{badgett2023review,
  title={A Review of the Economics of Sexual Orientation and Gender Identity},
  author={Badgett, MV and Carpenter, Christopher S and Lee, Maxine J and Sansone, Dario},
  journal={Journal of Economic Literature},
  pages={Forthcoming},
  year={2023}
}

@article{carpenter2023orient,
  title = {Sexual Orientation and Earnings in New Zealand},
  journal = {Economics Letters},
  pages = {111493},
  year = {2023},
  author = {Christopher S. Carpenter and Kabir Dasgupta and Alexander Plum},
}

@article{gillin2024attitudes,
  title={Attitudes toward sexual orientation and gender identity in online multiplayer gaming spaces},
  author={Gillin, Laura E and Signorella, Margaret L},
  journal={Psychological Reports},
  volume={127},
  number={6},
  pages={3066--3088},
  year={2024},
  publisher={SAGE Publications Sage CA: Los Angeles, CA}
}

@techreport{gandhi2024beliefs,
  title={Beliefs that Entertain},
  author={Gandhi, Ashvin and Giuliano, Paola and Guan, Eric and Keefer, Quinn and McDonald, Chase and Pagel, Michaela and Tasoff, Joshua},
  year={2024},
  institution={National Bureau of Economic Research}
}

@article{tampellini2024pride,
  title = {Latin American pride: Labor market outcomes of sexual minorities in Brazil},
  journal = {Journal of Development Economics},
  volume = {167},
  pages = {103239},
  year = {2024},
  author = {João Tampellini},
}

@article{marques2024positive,
  title={Positive behaviour interventions in online gaming: a systematic review of strategies applied in other environments},
  author={Marques, Tiago Garrido and Schumann, Sandy and Mariconti, Enrico},
  journal={Crime Science},
  volume={13},
  number={1},
  pages={14},
  year={2024},
  publisher={Springer}
}

@article{ham2024102503,
  title = {How accurately are household surveys measuring the LGBT population in Colombia? Evidence from a list experiment},
  journal = {Labour Economics},
  volume = {87},
  pages = {102503},
  year = {2024},
  author = {Andrés Ham and Ángela Guarín and Juanita Ruiz}
}

@article{palacios2025beautiful,
  author={Palacios-Huerta, Ignacio},
  title={The Beautiful Dataset},
  journal={Journal of Economic Literature},
  note={Forthcoming},
  year={2025}
}

@article{mcpherson2001birds,
  title = {Birds of a Feather: Homophily in Social Networks},
  author = {McPherson, Miller and Smith-Lovin, Lynn and Cook, James M.},
  journal = {Annual Review of Sociology},
  volume = {27},
  number = {1},
  pages = {415--444},
  year = {2001}
}

@article{currarini2009economic,
  title = {An Economic Model of Friendship: Homophily, Minorities, and Segregation},
  author = {Currarini, Sergio and Jackson, Matthew O. and Pin, Paolo},
  journal = {Quarterly Journal of Economics},
  volume = {124},
  number = {3},
  pages = {1003--1045},
  year = {2009}
}

@article{bekes2025cultural,
  title={Cultural homophily and collaboration in superstar teams},
  author={B{\'e}k{\'e}s, G{\'a}bor and Ottaviano, Gianmarco IP},
  journal={Management Science},
  volume={71},
  number={10},
  pages={8149--8168},
  year={2025},
  publisher={INFORMS}
}

@article{oecd2019,
  title={Society at a Glance 2019},
  author={OECD},
  journal={OECD Social Indicators},
  publisher= {Paris. OECD},
  year={2019}
}

@article{delhommer2020effect,
  title={Effect of state and local sexual orientation anti-discrimination laws on labor market differentials},
  author={Delhommer, Scott},
  journal={Available at SSRN 3625193},
  year={2020}
}

@article{seror2021legalized,
  title={Legalized Same-Sex Marriage and Coming Out in America: Evidence from Catholic Seminaries},
  author={Seror, Avner and Ticku, Rohit},
  year={2021}
}

@article{abadie2022synthetic,
  title={Synthetic controls in action},
  author={Abadie, Alberto and Vives-i-Bastida, Jaume},
  journal={arXiv preprint arXiv:2203.06279},
  year={2022}
}

@TechReport{roller2023differences,
  author={Roller, Marcus and Steinberg, Daniel},
  title={{Differences-in-Differences with multiple treatments under control}},
  year=2023,
  institution={Universitaet Bern, Departement Volkswirtschaft - CRED},
  type={Diskussionsschriften},
  number={credresearchpaper41}
}

@article{grodmadzki2026,
  title = {\#IamLGBT: Social networks and coming out},
  journal = {European Economic Review},
  volume = {183},
  pages = {105216},
  year = {2026},
  author = {Jan Gromadzki and Przemysław Siemaszko}
}

@book{becker1957economics,
  title={The economics of discrimination},
  author={Becker, Gary S},
  year={1957},
  publisher={University of Chicago Press}
}

@book{imbens2015causal, 
  author={Imbens, Guido W. and Rubin, Donald B.}, 
  title={Causal Inference for Statistics, Social, and Biomedical Sciences: An Introduction},
  publisher={Cambridge University Press}, 
  year={2015}
}

@article{bertrand2017field,
  title={Field experiments on discrimination},
  author={Bertrand, Marianne and Duflo, Esther},
  journal={Handbook of economic field experiments},
  volume={1},
  pages={309--393},
  year={2017},
  publisher={Elsevier}
}

@book{badgett2020economic,
  title={The economic case for LGBT equality: Why fair and equal treatment benefits us all},
  author={Badgett, MV Lee},
  year={2020},
  publisher={Beacon Press}
}

\end{document}